# Complete inelastic transparency of time-modulated resonant photonic circuits

M. Sumetsky

Aston Institute of Photonic Technologies, Aston University, Birmingham B4 7ET, UK

**Abstract**

Photonic circuits modulated in time can convert the input light frequency $\omega_0$ shifting it by multiples of the modulation frequency $\omega_p$ and, in certain cases, amplify the total input light power. Of special interest are photonic circuits employing microwave capacitors, which *instantaneously* modulate photonic waveguides with frequency $\omega_p \ll \omega_0$. While the amplification of light is negligible in such circuits, ideally, frequency conversion can be completed with the conservation of the light amplitude. Therefore, similar to the elastically transparent photonic structures (i.e., structures conserving both the light amplitude and frequency), we can say that a photonic circuit parametrically modulated in time exhibits *complete inelastic transparency* if a wave enters the structure with frequency $\omega_0$ and exits it with a different frequency and the same amplitude. Here, we develop an approach that allows us to introduce and investigate a broad class of time-modulated photonic circuits exhibiting complete inelastic transparency. Light enters these circuits with a resonant frequency $\omega_0$, cascades between their $N$ eigenstates separated by the modulation frequency $\omega_p$, and exits with frequency $\omega_0 + (N-1)\omega_p$ and the output amplitude close to the input amplitude. As examples, we consider circuits of ring microresonators and SNAP microresonators.

# I. INTRODUCTION

A monochromatic wave (optical, acoustic, electronic, etc.) propagating through a microscopic structure is transformed by elastic and inelastic interactions with the structure material. In photonics and electronics, waves commonly enter and exit a microstructure through waveguides, which can be connected to the microstructure directly or coupled to it evanescently. Provided that the propagation is elastic, i.e., proceeds with minimal losses and accurately conserved wave frequency, the problem of *complete transparency* of a structure arises. We say that such a *stationary structure* is completely transparent for a monochromatic wave with frequency $\omega_{in}$ if this wave enters the structure along one waveguide and exits it along another one with the same frequency $\omega_{out} = \omega_{in}$ and unity amplitude, $|S_{in,out}| = 1$ (Fig. 1(a)). Besides trivial cases, such as propagation along a low-loss optical waveguide, the simplest commonly known structures exhibiting complete elastic transparency are those described by the resonant tunneling through a quantum well surrounded by two barriers – the problem presented in many textbooks on quantum mechanics (see, e.g., [1, 2]). More general systems include quantum microstructures [3-8], heterostructures in electronics [9, 10], arrangements of scatterers in an optical waveguide [11, 12], and resonant microphotonic circuits [13-19].

In contrast to stationary structures, the wave propagation in a *time-modulated structures* can be amplified or attenuated depending on whether the modulation supplies or removes energy from the wave. For instance, a refractive-index modulation with frequency $\omega_p$ close to frequency $2\omega_{in}$ can lead to parametric gain or loss described by Floquet theory [20, 21]. For a relatively *small modulation frequency* considered in this paper, $\omega_p \ll \omega_{in}$, amplification can still occur if the modulation is performed by a traveling wave [22-25]. However, realizing this process in optical systems is challenging since it requires the dispersionless and low-loss propagation of light over a wide transmission bandwidth [22-25]. It has recently been shown that amplification of light by an acoustic wave can be achieved in a racetrack optical resonator within a reduced bandwidth [26]. Yet, the determined conditions of amplification, though relaxed, remain challenging.

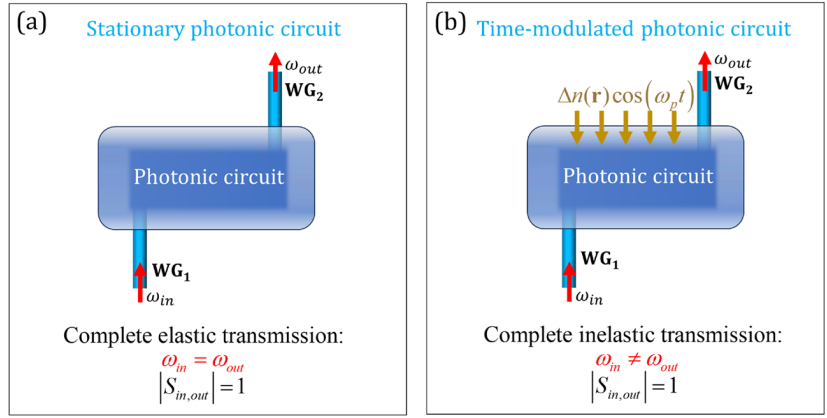

FIG. 1. Illustration of (a) stationary and (b) time-modulated photonic circuits.

Beyond amplification, one may ask whether a slowly varying time-modulated microscopic system could, like a stationary one, achieve *complete inelastic transparency*. Generally, we can say that a system exhibits complete inelastic transparency (i.e., possesses one or more *transparent inelastic channels*) if a wave entering the system with frequency $\omega_{in}$ exits it with frequency $\omega_{out} \neq \omega_{in}$ and unity amplitude $|S_{in,out}| = 1$ (Fig. 1(b)). The positive answer to this question for a particular case of two time-modulated coupled quantum wells (equivalent to two coupled microresonators in photonics) with two states was given three decades ago [27], where the effect of complete inelastic resonant transparency (CIRT) was demonstrated theoretically.

In optics and photonics addressed in the present paper, there is a significant difference between the effects of complete elastic and inelastic transparency. In the elastic case, it is generally possible to design a multichannel, strongly reflecting system – microscopic but not necessarily resonant – that enables complete transparency along selected channels (see [11, 12, 19] and references therein). On the contrary, a system, which has microscopic dimensions and possesses completely transparent inelastic channels, should necessarily be resonant. Indeed, the temporal parametric variations of the medium causing optical frequency conversion are commonly very weak. For this reason, the propagation length of an optical wave, which accomplishes the complete frequency conversion in a realistic photonic waveguide, has to be sufficiently long [28]. To make the interaction area *microscopic* and, simultaneously, the inelastic transmission amplitude close to unity, the wave should experience both multiple reflections and resonant propagation.

Recently, the effect of CIRT has found essential applications in photonics. The authors of Refs. [29, 30] experimentally demonstrated photonic circuits of optical microresonators with periodically modulated parameters, which can potentially act as complete frequency converters and frequency beam splitters. In particular, the transparency of the inelastic transmission channel through a system of two coupled microring resonators with a conversion efficiency of up to 80% has been achieved in Ref. [30]. In the same paper, the concept of CIRT of a resonant photonic circuit having a finite number $N$ of eigenfrequencies equally spaced by the modulation frequency $\omega_p$, generating the concept of two-level based CIRT [27], was introduced. Light propagating through such circuit cascades between its $N$ eigenstates and can exit it with unity amplitude and the frequency shifted from its original frequency by $(N-1)\omega_p$.

It is of great interest to explore and generalize the concept of the CIRT further. In particular, it is interesting to expand the conditions enabling the complete transparency of systems for the elastic wave propagation investigated in quantum mechanics [1-8], photonics [13-19], and acoustics [31, 32] to the inelastic case.

## II. STATEMENT OF THE PROBLEM AND MAJOR RESULTS

In this paper, we investigate and optimize the effect of CIRT for a broad range of resonant photonic circuits modulated by a relatively small instantaneous refractive index variation changing in space and time as $\Delta n(\mathbf{r})\cos(\omega_p t)$ with $\omega_p \ll \omega_{in}$. We show that, in the absence of losses, propagation of light in such circuits is described by a *unitary S-matrix* and, therefore, amplification of light is insignificant. Indeed, then in the ideal case of CIRT, we have $|S_{in,out}| = 1$ and the relative power amplification or attenuation is defined by the ratio $\Delta P = |\omega_{out} - \omega_{in}|/\omega_{in} \sim \omega_p \ll \omega_{in}$. For example, for the characteristic modulation frequency $\omega_p \sim 10 - 100$ GHz, and light frequency $\omega_{in} \sim 200$ THz, we have $\Delta P \lesssim 10^{-3}$. Consequently, the amplitude of CIRT $|S_{in,out}|$ and relative output power can reach unity for the negligible transmission loss only.

The simplest microscopic optical system exhibiting CIRT illustrated in Fig. 2(a1) is similar to the electronic system of two coupled quantum wells described in Ref. [27]. It consists of two microresonators, $MR_1$ and $MR_2$, weakly coupled to each other and to the input and output waveguides, $WG_1$ and $WG_2$. The resonant eigenfrequencies and eigenstates of uncoupled $MR_m$ are $\omega_m^{(0)}$ and $\Psi_m^{(0)}$, which become $\omega_n$ and $\Psi_n$ after the microresonator coupling is taken into account (Figs. 2(a2) and (a3)). It is assumed that the refractive index of microresonators is modulated with frequency $\omega_p \cong |\omega_1 - \omega_2|$. In the absence of losses, the CIRT of this system may be realized for $\omega_{in} = \omega_1$ and $\omega_{out} = \omega_2$. Provided that $MR_1$ and $MR_2$ are identical and, their compound eigenstates $\Psi_n$ are expressed through their individual eigenstates $\Psi_m^{(0)}$ as $\Psi_{1,2} = 2^{-1/2}\left(\Psi_1^{(0)} \pm \Psi_2^{(0)}\right)$. Then, coupling of $\Psi_1$ and $\Psi_2$ to each of the waveguides are the same and we have $|S_{in,out}| = |S_{in,in}|$. For this reason, due to the S-matrix's unitarity, the CIRT condition $|S_{in,out}| = 1$ is impossible. To solve the problem, $MR_1$ and $MR_2$ should be slightly dissimilar, so that the distribution of $\Psi_1$ is primarily localized near $WG_1$ and the distribution of $\Psi_2$ is primarily localized near $WG_2$ (Fig.2(a3)). However, the microresonator asymmetry, required to minimize the magnitude and, thus, the leakage of the eigenstate $\Psi_n$ into waveguide $MR_{2-n}$, leads to the vanishing effect of modulation which is proportional to $\langle\Psi_1|\Delta n|\Psi_2\rangle$. Thus, approaching the CIRT in the considered case requires a practically challenging enhancement of the modulation amplitude.

A solution to the above problem was proposed in Ref. [30], where the inelastic transparency of a photonic circuit including a modulated racetrack resonator was considered both experimentally and theoretically. The idea of Ref. [30] is illustrated in Fig. 2(b). To limit the inelastic transitions to a finite number $N$ of eigenstates of the racetrack resonator $MR_1$ ($N = 4$ in Fig. 2(b)) with constant free spectral range (FSR) $\Delta\omega$, the authors of Ref. [30] coupled this resonator to a smaller $MR_2$ having the $N$ times larger FSR (Figs. 2(b1) and 2(b2)). As proposed in Ref. [30], to exclude leakage of intermediate racetrack resonator eigenstates $\Psi_2, \Psi_3, \ldots, \Psi_{N-1}$ into waveguides (Fig. 2(b3), we can resonantly couple $\Psi_1$ to the input $WG_1$ and $\Psi_N$ to the output $WG_2$ through additional microresonators $MR_3$ and $MR_4$ (Fig 2(b1)).

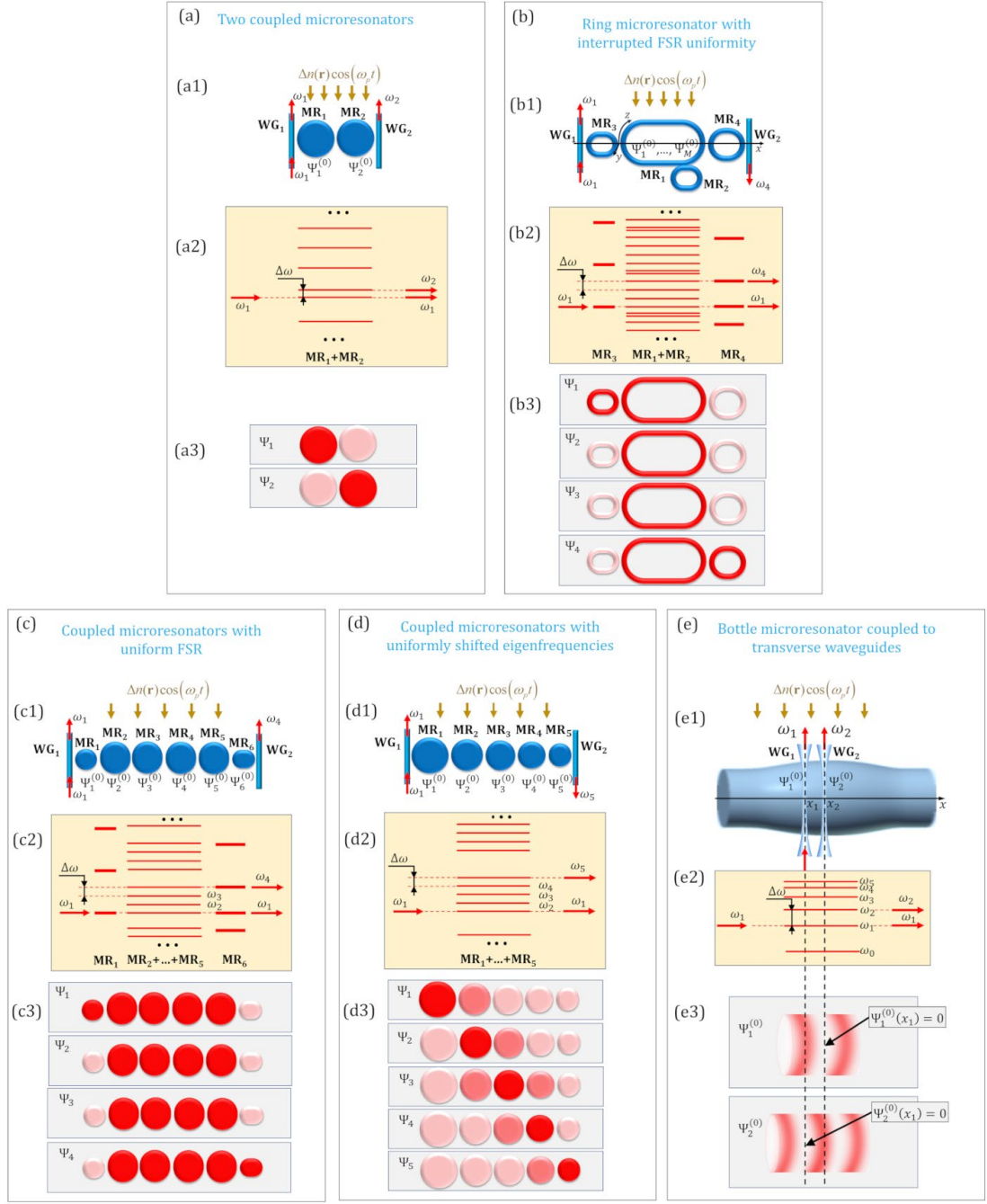

FIG. 2. (a) Two microresonators coupled to each other and to the input and output waveguides. (b) Racetrack resonator coupled to waveguides through microresonators resonantly transmitting selected eigenstates. (c) A series of equal microresonators coupled to each other and to waveguides, through which microresonators resonantly transmit selected eigenstates. (d) A series of microresonators with uniformly shifted eigenfrequencies coupled to each other and to waveguides. (e) Bottle microresonator coupled to transverse input and output waveguides.

Fig. 2(c1) shows a photonic circuit consisting of modulated successively coupled microresonators. In this circuit, a finite sequence of eigenstates $\Psi_1,\ldots,\Psi_N$ with constant FSR $\Delta\omega$ ($N = 4$ in Fig. 2(c)) can be designed by adjusting the inter-resonator couplings (Fig. 2(c2)). Additional microresonators ($MR_1$ and $MR_6$ in Fig. 2(c) similar to $MR_1$ and $MR_4$ in Fig. 2(b)) are included to suppress the unwanted leakages.

A method to avoid the additional resonators, which enable the exclusive coupling of the input and output waveguides to a given eigenstate of the photonic circuit, is illustrated in Fig. 2(d). The sequence of coupled microresonators shown in this figure have gradually decreasing dimensions and constant FSR $\Delta\omega$. The distribution of the density of eigenstates $\Psi_1,\ldots,\Psi_N$ of this circuit is more expanded than in the case of two coupled microresonators shown in Fig. 2(a). This allows us to optimize the spatial distribution of modulation more effectively without its significant enhancement as in the case of two coupled microresonators.

Remarkably, unlike the photonic circuit composed of two coupled microresonators shown in Fig. 2(a), the three-dimensional configuration of a bottle microresonator coupled to the input-output waveguides (Fig. 2(e)) enables complete suppression of eigenstate leakage in an exceptionally simple manner. We suggest that the considered bottle resonator is a SNAP microresonator, i.e., it has a very shallow profile with a nanoscale effective radius variation [33, 34]. The axial distribution of whispering gallery eigenstates of a SNAP bottle microresonator has a characteristic length of 10-100 μm [33] convenient for selective coupling of the transverse input-output waveguides (Fig. 2(e1)). Fig. 2(e2) shows the eigenfrequencies $\omega_n$ of this resonator corresponding to the same azimuthal and radial quantum numbers and different axial quantum numbers $n = 1,2,\ldots,5$. We assume that the input frequency $\omega_{in}$ is close to the frequency $\omega_1$ of the bottle eigenstate $\Psi_1^{(0)}$ and the output frequency $\omega_1 + \omega_p$ is close to the frequency $\omega_2$ of the bottle eigenstate $\Psi_2^{(0)}$. The spatial distribution of these eigenstates is illustrated in Fig. 2(e3). We choose the position of the input waveguide $WG_1$ at the node of $\Psi_2^{(0)}$ so that it is also located close to the antinode of $\Psi_1^{(0)}$. Respectfully, we choose the position of the output waveguide $WG_2$ at the node of $\Psi_1^{(0)}$ so that it is also located close to the antinode of $\Psi_2^{(0)}$. In doing this, we ensure that light with frequency $\omega_{in} \cong \omega_1$ is completely uncoupled and cannot leak to $WG_2$ and light with frequency $\omega_{out} = \omega_{in} + \omega_p$ is completely uncoupled and cannot leak to $WG_2$.

Below, we develop a theoretical approach describing both elastic and inelastic resonant propagation of light in time-modulated photonic circuits. By mapping the problem of inelastic frequency conversion to an equivalent elastic scattering problem in a compound configuration space that includes the degrees of freedom of the modulation, we show that the Mahaux-Weidenmüller formalism that governs stationary resonant transport [35-38] can be applied to describe CIRT. We show that under the conditions of negligible losses, the S-matrix of a time-modulated circuit remains unitary for the compound configuration of photonic and modulation states.

Based on the developed formalism, we identify the general condition of CIRT and demonstrate that it can be fulfilled by adjusting only two experimentally accessible parameters: the modulation-induced interstate coupling and the input frequency, while other circuit parameters remain fixed. We describe a broad family of time-modulated resonant photonic circuits, ranging from coupled microrings to SNAP bottle microresonators, in which light cascades coherently between $N$ eigenstates with equidistance eigenfrequencies and exits with nearly unchanged amplitude.

In Section III, we develop a general formalism that describes the interaction between resonant photonic circuits and oscillations, derive the general condition for CIRT, and introduce an approach for optimizing the spatial distribution of modulation. In Sections IV and V, we consider systems with two and three resonant states, illustrating the effect of CIRT and its sensitivity to detuning, coupling asymmetry, and loss. In Section VI, we extend the analysis to an arbitrary number of resonant eigenstates. In Sections VI and VII we present specific realizations of CIRT in photonic circuits within experimentally feasible parameters including coupled microrings, Kac-type microresonator chains, as well as SNAP bottle and semi-parabolic microresonators.

## III. GENERAL FORMALISM

Here, we present a general formalism that allows us to introduce and analyze a broad class of resonant microscopic systems, enabling the effects of complete resonant elastic and inelastic transparency. The idea of our approach follows from an observation that the problem of *inelastic* propagation of a wave (e.g., of a photon) $\Psi(\mathbf{r}, t)$ interacting with medium which parameters are changing in time can be reformulated as a problem of *elastic* propagation of a wave $\Sigma(\mathbf{r}, \mathbf{u})\exp(i\Omega t)$ in an expanded configuration space $(\mathbf{r}, \mathbf{u})$. Here $\mathbf{u}$ includes the coordinates of particles, quasiparticles, and collective excitations (e.g., phonons or plasmons) in a medium interacting with the propagating wave. Consequently, frequency $\Omega$ is proportional to the *total energy of the formed closed system* which is now conserved in the process of propagation.

The primary goal of this paper is to introduce and analyze photonic circuits that enable CIRT. However, the elastic resonant transmission circuits, which are designed to support CIRT, will also be considered and analyzed. Furthermore, the circuits supporting inelastic resonant propagation of light are reduced to the stationary resonant circuits provided that the separation between their eigenfrequencies vanishes together with the modulation frequency $\omega_p$.

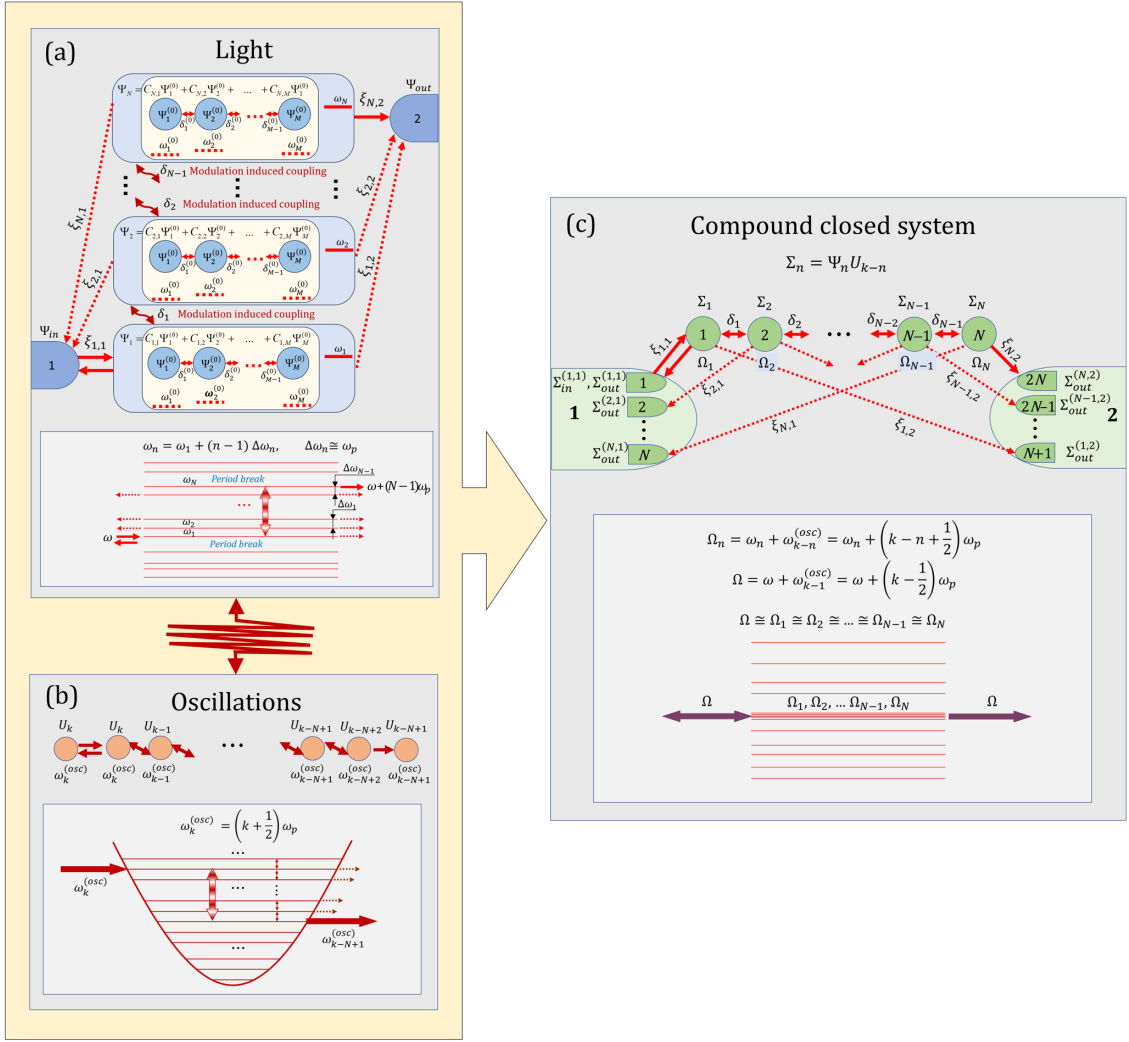

FIG. 3 (a) A photonic circuit composed of weakly coupled resonant states $\Psi_m^{(0)}$ possessing a series of equally spaced eigenfrequencies $\omega_n$ corresponding to the circuit eigenstates $\Psi_n$. (b) Eigenstates $U_k$ of a harmonic oscillator with eigenfrequencies $\left(k + \frac{1}{2}\right)\omega_p$. (c) Compound system of a photonic circuit coupled to a harmonic oscillator.

Photonic circuits considered in this paper are modeled as the circuits of series of weakly coupled microresonators. We distinguish between the full eigenstates of the photonic circuits in the absence of modulation $\Psi_n(\mathbf{r})\exp(-i\omega_n t)$, $n = 1, 2, ..., N$, with eigenfrequencies $\omega_n$ and eigenstates of independent $M$ microresonators $\Psi_m^{(0)}(\mathbf{r})\exp(-i\omega_m^{(0)} t)$, $m = 1, 2, ..., M$ with eigenfrequencies $\omega_m^{(0)}$ comprising these circuits as illustrated in Fig. 3(a). In the absence of modulation-induced couplings $\delta_n$, states $\Psi_n(\mathbf{r})$ are orthogonal to each other. Similarly, in the absence of couplings between microresonators $\delta_m^{(0)}$, their eigenstates $\Psi_m^{(0)}(\mathbf{r})$ do not overlap and, thus, are orthogonal as well.

Below, we describe the elastic and inelastic resonant propagation of light through the introduced photonic circuits. The circuit eigenstates $\Psi_n(\mathbf{r})$ are expressed through eigenstates of individual microresonators by linear relations determined by a matrix $\mathbf{C} = \{C_{m,n}\}$ (Fig. 3(a)). This matrix, together with eigenfrequencies $\omega_n$, are expressed through eigenfrequencies $\omega_m^{(0)}$ of individual microresonators and couplings $\delta_m^{(0)}$ between them by the equations given below in Section IIIB. In particular, if all $\delta_m^{(0)} = 0$ then we simply have $\Psi_n(\mathbf{r}) = \Psi_n^{(0)}(\mathbf{r})$ and $\omega_n = \omega_n^{(0)}$. For the elastic propagation, we assume that the spacings between the considered group of eigenfrequencies $\omega_m^{(0)}$ is small compared to the spacing of this group to other eigenfrequencies of microresonators. In the case of inelastic propagation, we assume that the spacing between the circuit eigenfrequencies $\Delta\omega_n = \omega_{n+1} - \omega_n$ is close to the modulation frequency $\omega_p$. In the latter case, without loss of degeneracy, we assume $\omega_{n+1} > \omega_n$, i.e., consider the frequency up-conversion.

## A. Description of the model

We begin with the introduction of a general model of photonic circuits considered in this paper, which enables us to describe inelastic propagation through time-modulated resonant photonic circuits in a manner analogous to elastic propagation. In an ideal system, we suggest that only two of the considered eigenstates, $\Psi_1(\mathbf{r}) \exp(-i\omega_1 t)$ and $\Psi_N(\mathbf{r})\exp(-i\omega_N t)$, are coupled respectively to the waves $\Psi_{in}^{(1,1)}(\mathbf{r}) \exp(-i\omega_{in} t)$ and $\Psi_{out}^{(N,2)}(\mathbf{r})\exp(-i\omega_{out} t)$ of the input and output waveguides. The spatial extension of states $\Psi_n(\mathbf{r})$ may cause additional leakage to the channels $\Psi_{out}^{(n,1)}(\mathbf{r})\exp(-i\omega_{out} t)$ and $\Psi_{out}^{(n,2)}(\mathbf{r})\exp(-i\omega_{out} t)$ of the input waveguide 1 and output waveguide 2 as indicated by dashed red arrows in Fig. 3(a). In our design of photonic microresonator circuits below, we minimize this additional coupling to have the amplitude of the inelastic transition with frequency conversion $\omega_1 \to \omega_2 = \omega_1 + (N-1)\omega_p$ close to unity.

In Fig. 3(b), we show the states of a harmonic oscillator, $U_k(u)\exp(-ik\omega_p t)$, $U_{k-1}(u)\exp(-i(k-1)\omega_p t)$, ..., $U_{k-N}(u) \exp(-i(k-N)\omega_p t)$ with eigenfrequencies $\omega_k^{(osc)} = \left(k + \frac{1}{2}\right)\omega_p$. This oscillator models the modulation of the system's parameters with frequency $\omega_p$. For large quantum numbers $k$ of the oscillator's eigenstates, its effect is equivalent to the periodic parametric modulation in time. We assume that the time modulation induces coupling between adjacent states $\Psi_n(r)\exp(-ik\omega_n t)$ and $\Psi_{n+1}(r)\exp(-ik\omega_{n+1} t)$ only. As noted above, to enable the resonant transmission, the spacing between all eigenfrequencies $\omega_n$ of adjacent states $\Psi_n(\mathbf{r})\exp(-ik\omega_n t)$ is set close to the modulation frequency, $\Delta\omega = \omega_{n+1} - \omega_n \cong \omega_p$.

In Fig. 3(c), the states of the optical system and of the harmonic oscillator are joined into compound states $\Sigma_n(\mathbf{r}, u) \exp(-i\Omega_n t) = \Psi_n(\mathbf{r})U_{k-n-1}(u) \exp(-i\Omega_n t)$. Each of these states has the total eigenfrequency $\Omega_n = \omega_n + \left(k - n + \frac{1}{2}\right)\omega_p$ and satisfy the resonant condition $\Omega_1 \cong \Omega_2 \cong \cdots \cong \Omega_N$. We introduce $N$ input channels $\Sigma_{in}^{(n,1)}(\mathbf{r}, u) \exp(-i\Omega t) = \Psi_{in}^{(n,1)}(\mathbf{r})U_k(u) \exp(-i\Omega t)$, $n = 1, 2, ..., N$, and the corresponding output channels $\Sigma_{out}^{(n,1)}(\mathbf{r}, u) \exp(-i\Omega t) = \Psi_{out}^{(n,1)}(\mathbf{r})U_k(u) \exp(-i\Omega t)$ in the input waveguide and $N$ output channels $\Sigma_{out}^{(n,2)}(\mathbf{r}, u)\exp(-i\Omega t) = \Psi_{out}^{(n,2)}(\mathbf{r})U_{k-N+1}(u) \exp(-i\Omega t)$ in the output waveguide. Here the oscillator component of frequency of states $\Sigma_{in}^{(n,1)}$ and $\Sigma_{out}^{(n,1)}$ for $n \leq N$ is assumed to be equal to $\left(k - n + \frac{1}{2}\right)\omega_p$, so that these states are resonantly coupled with states $\Sigma_n$ only. Respectively, the light frequency component of the input waves $\Sigma_{in}^{(n,1)}$ with frequency $\Omega$ is equal to $\omega$, so that $\Omega = \omega + \left(k - \frac{1}{2}\right)\omega_p$ is close to $\Omega_n$. Similar to the elastic transmission, all channels are assumed to be normalized to the unity fluxes along the waveguides. Ideally, to enable the CIRT along a single inelastic channel of this system, the input state $\Sigma_{in}^{(1,1)}$ should be coupled to the state $\Sigma_1$ only and the output state $\Sigma_{out}^{(N,2)}$ should be coupled to state $\Sigma_N$ only, while coupling to other output channels from states $\Sigma_n$ indicated by red dashed arrows should vanish.

Reduction of a system of optical states (Fig. 3(a)) interacting with parametric oscillations (Fig. 3(b)), to a stationary system of weakly coupled compound states in configuration space $(r, u)$ (Fig. 3(c)) clarifies and significantly simplifies the description of inelastic resonant propagation. In particular, in full analogy to the elastic resonant propagation [3, 4, 13, 19, 39], it becomes obvious from Fig. 3(c) that the transmission amplitude (now inelastic) of a system of resonant quantum wells connected in series to each other and to the waveguides (i.e., in the absence of couplings indicated by dashed red arrows) can approach unity for vanishing material losses. Furthermore, resonant systems with a more complex topology than that shown in Fig. 3(a), which are transparent at resonant frequencies, can be used to design associated time-dependent systems exhibiting inelastic transparent channels.

### B. The Mahaux-Weidenmüller formalism: a stationary photonic circuit

Photonic circuits considered in this paper are modelled as the circuits of weakly coupled microresonators. Eigenstates $\Psi_n(\mathbf{r})$ of such photonic circuits can be presented as linear combinations of eigenstates $\Psi_m^{(0)}(\mathbf{r})$ of microresonators they include. The resonant transmission and reflection amplitudes of waves propagating through *a stationary circuit of weakly coupled resonant states* $\Psi_m^{(0)}(\mathbf{r})$ shown in Fig. 4 can be determined by the Mahaux-Weidenmüller (MW) equation [35-38], the matrix generalization of the well-known Breit-Wigner formula [1, 2]. The MW formalism was used previously for the description of elastic resonant propagation and, in particular, for the analysis of complete transmission and reflection of complex resonant systems in Refs. [13, 37, 38, 39]. In particular, Ref. [38] applied this formalism to investigate inelastic processes in cavity quantum electrodynamic systems.

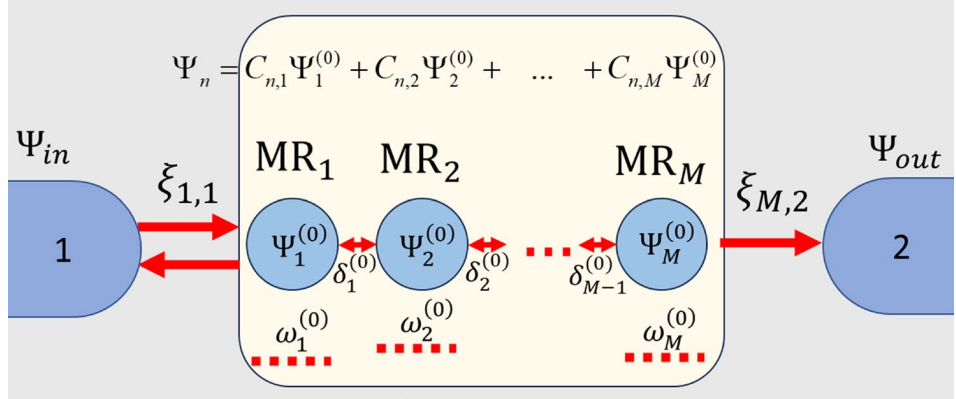

FIG. 4. A stationary photonic circuit possessing eigenstates $\Psi_n$ composed of $M$ weakly coupled resonant states $\Psi_m^{(0)}$.

We consider photonic circuits of $M$ microresonators, $MR_1$, $MR_2$, …, $MR_M$, which are coupled in series to each other and to waveguides $WG_1$ and $WG_2$ as illustrated in Fig. 4. The transmission and reflection amplitudes of propagation between input and output waveguides of these circuits are determined by the S-matrix given by the MW equation (see more general expressions in [37, 38]):

$$\mathbf{S}(\omega) = \mathbf{I}_2 - i\xi^\dagger \left[ \omega \mathbf{I}_M - \mathbf{\Omega}^{(0)} + \frac{i}{2}\left(\mathbf{\Gamma} + \xi\xi^\dagger\right) \right]^{-1} \xi,$$

$$\mathbf{\Omega}^{(0)} = \begin{Bmatrix} \omega_1^{(0)} & \delta_1^{(0)} & 0 & \cdots & 0 & 0 \\ \delta_1^{(0)*} & \omega_2^{(0)} & \delta_2^{(0)} & \cdots & 0 & 0 \\ 0 & \delta_2^{(0)*} & \omega_3^{(0)} & \cdots & 0 & 0 \\ \cdots & \cdots & \cdots & \cdots & \cdots & \cdots \\ 0 & 0 & 0 & \cdots & \omega_{M-1}^{(0)} & \delta_{M-1}^{(0)} \\ 0 & 0 & 0 & \cdots & \delta_{M-1}^{(0)*} & \omega_M^{(0)} \end{Bmatrix}, \quad \mathbf{\Gamma} = \begin{Bmatrix} \gamma_1 & 0 & \cdots & 0 & 0 \\ 0 & \gamma_2 & \cdots & 0 & 0 \\ 0 & 0 & \cdots & 0 & 0 \\ \cdots & \cdots & \cdots & \cdots & \cdots \\ 0 & 0 & \cdots & \gamma_{M-1} & 0 \\ 0 & 0 & \cdots & 0 & \gamma_M \end{Bmatrix}, \quad \xi = \begin{Bmatrix} \xi_{1,1} & 0 \\ 0 & 0 \\ 0 & 0 \\ 0 & 0 \\ \cdots & \cdots \\ 0 & \xi_{M,2} \end{Bmatrix}. \quad (1)$$

where $\mathbf{I}_M$ is the $M \times M$ unity matrix, $\omega_m^{(0)}$ and $\gamma_m$ are the eigenfrequency and loss of MR$_m$ assumed standing alone, $\delta_m^{(0)}$ is coupling between MR$_m$ and MR$_{m+1}$, and $\xi_{1,1}$ and $\xi_{M,2}$ are couplings between MR$_1$ and WG$_1$ and between MR$_M$ and WG$_2$, respectfully.

Provided that $\Psi_m^{(0)}$ are normalized, the normalized eigenstates $\Psi_n(\mathbf{r})$ of the described photonic circuit uncoupled from the waveguides are determined as

$$\Psi_n = \sum_{m=1}^M C_{nm} \Psi_m^{(0)}, \quad \sum_{m=1}^M |C_{nm}|^2 = 1, \quad (2)$$

where the coefficients $C_{nm}$ and eigenfrequencies $\omega_n$ of $\Psi_n$ are determined by equations

$$\left(\omega_n \mathbf{I}_N - \mathbf{\Omega}^{(0)} + \tfrac{i}{2}\mathbf{\Gamma}\right)\mathbf{C}_n = \mathbf{0}, \quad \det\left(\omega_n \mathbf{I}_N - \mathbf{\Omega}^{(0)} + \tfrac{i}{2}\mathbf{\Gamma}\right) = 0, \quad \mathbf{C}_n = \begin{Bmatrix} C_{1n} \\ C_{2n} \\ \cdots \\ C_{Nn} \end{Bmatrix}. \quad (3)$$

The normalization condition for $C_{nm}$ in Eq. (2) together with Eq. (3) allow to determine all coefficients $C_{nm}$.

### C. Condition of complete resonant transparency

Provided that losses $\gamma_m$ are negligible and $\xi_{1,1}$ and $\xi_{M,2}$ are the only nonzero couplings to waveguides, the S-matrix defined by Eq. (1) is unitary and its elements satisfy the conservation of energy relation $|S_{1,1}(\omega)|^2 + |S_{1,2}(\omega)|^2 = 1$. Therefore, the condition of complete resonant transparency, $|S_{1,2}(\omega)|^2 = 1$, is equivalent to:

$$S_{1,1}(\omega) = 0. \quad (4)$$

The complex-valued Eq. (4) is equivalent to *two equations* establishing the required relations between $2M + 1$ real parameters introduced in Eq. (1). Consequently, there is a wide variety of circuit parameters satisfying the condition of complete resonant transparency. In the presence of losses, Eq. (4) manifests the condition of zero reflection that corresponds to the maximum possible inelastic transmission. Several examples of circuits that enable complete elastic resonant transparency will be considered below as particular cases of time-modulated circuits that enable CIRT.

### D. Photonic circuit with two resonant states

For $M = 2$ Eq. (1) is reduced to

$$S_{1,2}(\omega) = \frac{\xi_{1,1}\xi_{2,2}\delta_1^{(0)}}{D(\omega)}, \quad S_{1,1}(\omega) = 1 - \frac{i\xi_{1,1}^2\left(\omega - \omega_2 + \omega_p + \frac{i}{2}(\gamma_2 + \xi_{2,2}^2)\right)}{D(\omega)},$$

$$D(\omega) = \left(\omega - \omega_1 + \frac{i}{2}(\gamma_1 + \xi_{1,1}^2)\right)\left(\omega - \omega_2 + \omega_p + \frac{i}{2}(\gamma_2 + \xi_{2,2}^2)\right) - \left(\delta_1^{(0)}\right)^2. \tag{5}$$

Here, the expression for $S_{1,1}(\omega)$ coincides with that found in Ref. [27] and, in the absence of the second waveguide, WG$_2$ (i.e., for $\xi_{2,2} = 0$), the expression for $S_{1,1}(\omega)$ coincides with that given in the Supplementary Information of Ref. [29]. The condition of zero reflection, Eq. (4), now yields:

$$\left(\delta_1^{(0)}\right)^2 = \left(\xi_{1,1}^2 - \gamma_1\right)\left(\xi_{2,2}^2 + \gamma_2\right)\left[\frac{1}{4} + \frac{(\omega_1 - \omega_2)^2}{\left(\gamma_1 + \gamma_2 - \xi_{1,1}^2 + \xi_{2,2}^2\right)^2}\right]. \tag{6}$$

Under this condition, the frequency $\omega_0$ corresponding to zero reflection, $S_{1,1}(\omega_0) = 0$, is

$$\omega_0 = \frac{\omega_1(\gamma_2 + \xi_{2,2}^2) + \omega_2(\gamma_1 - \xi_{1,1}^2)}{\gamma_1 + \gamma_2 - \xi_{1,1}^2 + \xi_{2,2}^2} \tag{7}$$

Consequently, we find from Eqs. (5)-(7):

$$S_{1,1}(\omega_0) = 0, \quad S_{1,2}(\omega_0) = \frac{1 - \dfrac{\gamma_1}{\xi_{1,1}^2}}{1 + \dfrac{\gamma_2}{\xi_{2,2}^2}}. \tag{8}$$

As expected, we have $|S_{1,2}(\omega_0)| = 1$ in the absence of losses and $|S_{1,2}(\omega_0)| < 1$ in their presence. For pumped photonic circuits, the signs of $\gamma_1$ and $\gamma_2$ can be positive and negative, so that the absolute value of the transmission amplitude $S_{1,2}(\omega_0)$ can be smaller than, equal to, or greater than unity. In particular, it follows from Eq. (8) that the condition of complete transparency, $|S_{1,2}(\omega_0)| = 1$ is satisfied if

$$\gamma_1\xi_{2,2}^2 + \gamma_2\xi_{1,1}^2 = 0. \tag{9}$$

### E. The Mahaux-Weidenmüller formalism: a time-modulated photonic circuit

In application to the problems of inelastic transmission of light through the time-modulated photonic circuits considered below, the MW formalism is introduced as follows. We assume that eigenstates $\Sigma_n$ are coupled in series as illustrated in Fig. 3(c) and that only single-quantum transitions between adjacent states with acquiring or loosing frequency $\omega_p$ are possible, i.e., that the coupling with oscillations is relatively small. In our model, states $\Sigma_n$ are coupled to the input-output wave channels of waveguides 1 and 2 with coupling coefficients $\xi_{n,1}$ and $\xi_{n,2}$, respectively, while $\delta_n$ are the coupling parameters between eigenstates $\Sigma_n$ and $\Sigma_{n+1}$ (Fig. 3(c)). In most of the applications considered below, the values delta sub n are proportional to the modulation amplitude, so that these states are uncoupled in the absence of modulation. Then, the MW formalism [35-38] for the $2N \times 2N$ S-matrix of propagation between $2N$ input-output channels $\Sigma_{in}^{(n,1)}, \Sigma_{out}^{(n,1)}$ and $\Sigma_{in}^{(n,2)}, \Sigma_{out}^{(n,2)}, n = 1,2,\ldots,N$, reads:

$$\mathbf{S}(\omega) = \mathbf{I}_{2N} - i\boldsymbol{\xi}^{\dagger}\left[\Omega(\omega)\mathbf{I}_N - \boldsymbol{\Omega} + \frac{i}{2}\left(\boldsymbol{\Gamma} + \boldsymbol{\xi}\boldsymbol{\xi}^{\dagger}\right)\right]^{-1}\boldsymbol{\xi},$$

$$\boldsymbol{\Omega} = \begin{Bmatrix} \Omega_1 & \delta_1 & 0 & \ldots & 0 & 0 \\ \delta_1^* & \Omega_2 & \delta_2 & \ldots & 0 & 0 \\ 0 & \delta_2^* & \Omega_3 & \ldots & 0 & 0 \\ & & & \ldots & & \\ 0 & 0 & 0 & \ldots & \Omega_{N-1} & \delta_{N-1} \\ 0 & 0 & 0 & \ldots & \delta_{N-1}^* & \Omega_N \end{Bmatrix}, \quad \boldsymbol{\Gamma} = \begin{Bmatrix} \gamma_1 & 0 & \ldots & 0 & 0 \\ 0 & \gamma_2 & \ldots & 0 & 0 \\ 0 & 0 & \ldots & 0 & 0 \\ \ldots & \ldots & & \ldots & \ldots \\ 0 & 0 & \ldots & \gamma_{N-1} & 0 \\ 0 & 0 & \ldots & 0 & \gamma_N \end{Bmatrix},$$

$$\Omega(\omega) = \omega + \left(k - \tfrac{1}{2}\right)\omega_p, \quad \Omega_n = \omega_n + \left(k - n + \tfrac{1}{2}\right)\omega_p,$$

$$\boldsymbol{\xi} = \begin{Bmatrix} \xi_{1,1} & 0 & 0 & \ldots & 0 & \xi_{1,2} & 0 & 0 & \ldots & 0 \\ 0 & \xi_{2,1} & 0 & \ldots & 0 & 0 & \xi_{2,2} & 0 & \ldots & 0 \\ 0 & 0 & \xi_{3,1} & \ldots & 0 & 0 & 0 & 0 & \ldots & 0 \\ & & & & \ldots & & & & & \\ 0 & 0 & 0 & \ldots & \xi_{N,1} & 0 & 0 & 0 & \ldots & \xi_{N,2} \end{Bmatrix}. \tag{10}$$

Here $\mathbf{I}_N$ is the $N \times N$ unity matrix, $\boldsymbol{\xi}$ is the $2N \times N$ matrix, and $\boldsymbol{\Omega} - \frac{i}{2}(\boldsymbol{\Gamma} + \boldsymbol{\xi}\boldsymbol{\xi}^{\dagger})$ with a diagonal matrix $\boldsymbol{\xi}\boldsymbol{\xi}^{\dagger}$ is the $N \times N$ matrix which determines the complex-valued eigenfrequencies of uncoupled states. Of major interest below is the transmission amplitude $S_{1,2N}(\omega)$ along the inelastic transmission channel with frequency conversion $\omega \to \omega + (N-1)\omega_p$ between the input-output waveguide 1 and the output waveguide 2, as well as the inelastic amplitude $S_{1,N}(\omega)$ back into the input-output waveguide 1 with the same frequency conversion $\omega \to \omega + (N-1)\omega_p$. The latter amplitude is of primary interest in the case of the absence of waveguide 2, which was investigated previously for special cases of photonic circuits in Refs. [29, 30]. In the latter case, the considered model is significantly simplified since the matrix $\boldsymbol{\xi}$ is reduced to a diagonal $N \times N$ matrix.

In full analogy with Eqs. (2) and (3), the eigenstates $\tilde{\Sigma}_n$ of the described system uncoupled from the waveguides are determined as

$$\tilde{\Sigma}_n = \sum_{m=1}^{N} \tilde{C}_{nm} \Sigma_m, \tag{11}$$

where the coefficients $\tilde{C}_{nm}$ and eigenfrequencies $\tilde{\Omega}_n$ of $\tilde{\Sigma}_n$ are determined by equations

$$\left(\tilde{\Omega}_n \mathbf{I}_N - \boldsymbol{\Omega} + \tfrac{i}{2}\boldsymbol{\Gamma}\right)\tilde{\mathbf{C}}_n = \mathbf{0}, \quad \det\left(\tilde{\Omega}_n \mathbf{I}_N - \boldsymbol{\Omega} + \tfrac{i}{2}\boldsymbol{\Gamma}\right) = 0, \quad \tilde{\mathbf{C}}_n = \begin{Bmatrix} \tilde{C}_{1n} \\ \tilde{C}_{2n} \\ \ldots \\ \tilde{C}_{Nn} \end{Bmatrix}. \tag{12}$$

As in Eqs. (2) and (3), provided that $\Sigma_n$ and $\tilde{\Sigma}_n$ are normalized, we use the normalization $\sum_{m=1}^{N}|\tilde{C}_{nm}|^2 = 1$ together with Eq. (12) to determine all coefficients $\tilde{C}_{nm}$.

### F. Tuning eigenfrequencies and interstate coupling by parametric variation and modulation.

Below, we apply Eqs. (1)-(3) and (11), (12) to parametrically modulated as well as stationary photonic circuits. We assume that the circuit parameters are tunable due to the induced variation in the refractive index and dimensions of the microresonators. For example, the tunable part of the refractive index variation contains the stationary and nonstationary components, $\Delta n_0^{(tun)}(\mathbf{r})$ and $\Delta n_1^{(tun)}(\mathbf{r})\cos(\omega_p t)$, so that

$$n(\mathbf{r}) = n_0 + \Delta n_0^{(tun)}(\mathbf{r}) + \Delta n_1^{(tun)}(\mathbf{r})\cos(\omega_p t). \quad (13)$$

It is instructive for future consideration to present the spatial distribution of the tunable part of the refractive index in frequency units introducing

$$\Delta \omega_0^{(tun)}(\mathbf{r}) = \frac{\omega}{n_0}\Delta n_0^{(tun)}(\mathbf{r}), \quad \Delta \omega_1^{(tun)}(\mathbf{r}) = \frac{\omega}{n_0}\Delta n_1^{(tun)}(\mathbf{r}). \quad (14)$$

Then, perturbation $\Delta \omega_n$ of circuit eigenfrequency $\omega_n$ is found as

$$\Delta \omega_n = \langle \Psi_n | \Delta \omega_0^{(tun)} | \Psi_n \rangle. \quad (15)$$

In turn, the modulation-induced coupling $\delta_n$ between the originally uncoupled states $\Psi_n(\mathbf{r})$ and $\Psi_{n+1}(\mathbf{r})$ is

$$\delta_n = \langle \Psi_n | \Delta \omega_1^{(tun)} | \Psi_{n+1} \rangle. \quad (16)$$

Below we call $\Delta \omega_1^{(tun)}(\mathbf{r})$ as the *spatial distribution of modulation* (SDM).

Experimentally, stationary $\Delta \omega_0^{(tun)}(\mathbf{r})$ and SDM $\Delta \omega_1^{(tun)}(\mathbf{r})$ are used to adjust the eigenfrequencies $\omega_n$ and to arrive at the required $\delta_n$. The refractive index variation induced by the Pockel effect is proportional to the applied electric field $E(\mathbf{r}) = E_0(\mathbf{r}) + E_1(\mathbf{r})\cos(\omega_p t)$, so that $\Delta \omega_0^{(tun)}(\mathbf{r}) \sim E_0(\mathbf{r})$ and $\Delta \omega_1^{(tun)}(\mathbf{r}) \sim E_1(\mathbf{r})$. For the refractive index variation induced by the Kerr effect, we assume $E(\mathbf{r}) = E_0(\mathbf{r})\exp(i\omega^{(Kerr)}t) + E_1(\mathbf{r})\exp(i(\omega^{(Kerr)} + \omega_p)t)$. Then $\Delta \omega_0^{(tun)}(\mathbf{r}) \sim E_1^2(\mathbf{r}) + E_2^2(\mathbf{r})$ and $\Delta \omega_1^{(tun)}(\mathbf{r}) \sim 2E_1(\mathbf{r})E_2(\mathbf{r})$. Exclusion of the eigenfrequency shifts for the Pockel modulation can be set by the requirement $E_0(\mathbf{r}) = 0$. For the Kerr modulation, the same exclusion can be achieved by setting $\Delta \omega_0^{(tun)}(\mathbf{r}) \sim E_1^2(\mathbf{r}) + E_2^2(\mathbf{r}) = const$. Then $\Delta \omega_n$ defined by Eq. (15) vanishes due to the orthogonality of eigenstates $\Psi_n$.

To tune couplings $\delta_n$, the modulation power can be varied as a whole or locally, being applied to separate parts of the system. For example, if $\Psi_n(\mathbf{r})$ represent eigenfunctions of a system of coupled microresonators, then each of these resonators can be modulated together or separately. The magnitude of SDM $\omega_1^{(tun)}(\mathbf{r})$ necessary to introduce the required $\delta_n$ can be minimized by optimizing its spatial distribution [,]. For ring resonators, the Pockel modulation can be introduced by adjacent RF capacitors (see, e.g., [30]). For essentially three dimensional bottle and SNAP microresonators [33], the Kerr modulation can be introduced by resonant excitation of whispering gallery modes (WGMs). The required spatial distribution of $\omega_1^{(tun)}(\mathbf{r})$ can be achieved by excitation of a resonant WGM with the eigenfrequency $\omega^{(Kerr)}$ (or a series of WGMs with eigenfrequencies $\omega^{(Kerr)} + n\Delta\omega$, $n = 1, 2, ..., N$, $\Delta\omega \cong \omega_p$) with appropriate amplitudes. To avoid interference, the frequency $\omega^{(Kerr)}$ should be chosen sufficiently *separated* from the frequency $\omega$ of the relatively weak input light under consideration.

From Eq. (16), the values of $\delta_n$ are proportional to the spatial overlap of adjacent eigenstates $\Psi_{n+1}(\mathbf{r})$ and $\Psi_n(\mathbf{r})$, which, therefore, are required to be as large as possible. On the other hand, the spatial extension of eigenstates may lead to their leakages into waveguides causing significant losses. The design of photonics circuits considered below suggests several approaches to suppress these leakages and, at the same time, to keep the spatial overlap of eigenstates large, as required for their efficient inelastic interactions leading to CIRT.

### G. Condition of CIRT

For a circuit with negligible losses and leakages ($\tilde{\gamma}_n \ll \xi_{1,1}^2, \xi_{N,2}^2, \delta_k$; $\xi_{n,1}^2 \ll \xi_{1,1}^2, \xi_{N,2}^2, \delta_k$, $n = 2, 3, \ldots, N$; and $\xi_{n,2} \ll, \xi_{1,1}^2 \xi_{N,2}^2, \delta_k,$, $n = 1, 2, \ldots, N-1$), the S-matrix in Eq. (10) is unitary and its only nonzero elements are the reflection amplitude $S_{1,1}(\omega)$ and the inelastic transmission amplitude $S_{1,2N}(\omega)$. As in the elastic case, these amplitudes satisfy the relation $|S_{1,1}(\omega)|^2 + |S_{1,2N}(\omega)|^2 = 1$. Therefore, the condition of CIRT $|S_{1,2N}(\omega)|^2 = 1$ is equivalent to the condition similar to Eq. (4) for stationary systems:

$$S_{1,1}(\omega) = 0. \tag{17}$$

Similar to Eq. (4), the complex-valued Eq. (17) is equivalent to *two equations* for $2N + 4$ real system's parameters $\omega, \omega_n, \omega_p, \delta_n, \xi_{1,1}, \xi_{N,2}$, $n = 1, 2, \ldots, N$. In the experimental realization of the photonic circuits enabling CIRT, it is generally challenging to arrive at the exact resonant condition satisfying Eq. (17) by accurate tuning the circuit eigenfrequencies $\omega_n$ and couplings to the waveguides $\xi_{1,1}$ and $\xi_{N,2}$. It is much easier to tune the modulation-induced couplings $\delta_n$ by varying the modulation power. For example, it is straightforward to design a circuit with all couplings $\delta_n$ varied proportionally, i.e., $\delta_n = \alpha_n \delta_1$ where $\alpha_n$ are constants and $\delta_1$ changes with the modulation power. Then, formally, the two real-valued equations following from Eq. (17) can be satisfied by appropriate choosing of *only two parameters* simple to tune such as the *modulation-induced coupling $\delta_1$ and the input light frequency $\omega$*. In a more general problem when both the losses and the leakages are present, Eq. (17) should be complemented by conditions minimizing the transmission along the leaking channels.

### H. Optimization of the spatial distribution of modulation

Since the magnitudes of the stationary variation $\Delta\omega_0^{(tun)}(\mathbf{r})$ and SDM $\Delta\omega_1^{(tun)}(\mathbf{r})$ are limited experimentally, it is important to achieve the required shifts of eigenfrequencies $\omega_n$ and modulation-induced couplings $\delta_n$ (see Eqs. (14) and (16)) with the possible smallest amplitude of modulation and modulation power. Here, we address this problem by distinguishing photonic circuits with $N = 2$ resonant states and with $N > 2$ resonant states. In this paper, we consider situations when couplings $\delta_n$ defined by Eq. (16) are real, which is assumed in the derivations below.

#### 1. $N = 2$

It is intuitively clear from Eq. (16) that the modulation is more effective if $\Delta\omega_1^{(tun)}(\mathbf{r})$ is proportional to the product $\Psi_1(\mathbf{r})\Psi_2^*(\mathbf{r})$. Indeed, the Cauchy–Schwarz inequality $\left(\int \Psi_1(\mathbf{r})\Psi_2^*(\mathbf{r})\Delta\omega_1^{(tun)}(\mathbf{r})d\mathbf{r}\right)^2 \leq \int |\Psi_1(\mathbf{r})\Psi_2(\mathbf{r})|^2 d\mathbf{r} \int \left|\Delta\omega_1^{(tun)}(\mathbf{r})\right|^2 d\mathbf{r}$ becomes an equality for $\Delta\omega_1^{(tun)}(\mathbf{r})$ proportional to $\Psi_1(\mathbf{r})\Psi_2^*(\mathbf{r})$. This fact was used in Ref. [40] to optimize the modulation-induced inelastic transitions in single microresonators. Based on the same argument, we can look for the SDM in the form $\Delta\omega_1^{(tun)}(\mathbf{r}) = A \cdot \Psi_1(\mathbf{r})\Psi_2(\mathbf{r})$ where functions $\Psi_n(\mathbf{r})$ are normalized. Substitution of the latter equation into Eq. (16) yields

$$\Delta\omega_1^{(tun)}(\mathbf{r}) = \delta_1 \frac{\Psi_1(\mathbf{r})\Psi_2^*(\mathbf{r})}{\left\langle (\Psi_1)^2 |(\Psi_2)^2\right\rangle}. \tag{18}$$

## 2. $N > 2$

In analogy to the case $N = 2$, for $N > 2$, we look for the optimum spatial distribution of $\Delta\omega_1^{(tun)}(\mathbf{r})$ in the form

$$\Delta\omega_1^{(tun)}(\mathbf{r}) = \sum_{n=1}^{N-1} A_n \Psi_n(\mathbf{r}) \Psi_{n+1}^*(\mathbf{r}). \tag{19}$$

As follows from Eq. (16), $N - 1$ coefficients $A_n$ in Eq. (19) are determined by the following system of $N - 1$ linear equations:

$$\delta_n = \sum_{k=1}^{N-1} A_k \langle \Psi_n \Psi_k | \Psi_{k+1} \Psi_{n+1} \rangle. \tag{20}$$

For $N = 2$, Eqs. (19) and (20) are reduced to Eq. (18).

### I. Photonic circuits comprising $M$ successively coupled microresonators

We consider photonic circuits constructed of a series of $M$ weakly coupled microresonators, MR$_m$, $m = 1, 2, \ldots M$, with eigenfrequencies $\omega_m^{(0)}$. The eigenstates $\Psi_n(\mathbf{r})$ of this circuit are expressed through though the eigenstates of individual microresonators $\Psi_m^{(0)}$ by Eq. (2):

$$\Psi_n = \sum_{m=1}^{M} C_{nm} \Psi_m^{(0)}. \tag{21}$$

Then we find from Eq. (19):

$$\Delta\omega_1^{(tun)}(\mathbf{r}) = \sum_{m=1}^{M} B_m \left|\Psi_m^{(0)}(\mathbf{r})\right|^2, \quad B_m = \sum_{n=1}^{N-1} A_n C_{n,m} C_{n+1,m}. \tag{22}$$

For each individual microresonator with an eigenstate $\Psi_m^{(0)}$, we introduce the shift of its eigenfrequency $\omega_m^{(0)}$ by perturbation $\Delta\omega_1^{(tun)}(\mathbf{r})$:

$$\Delta\omega_{1,m}^{(0)} = \langle \Psi_m^{(0)} | \Delta\omega_1^{(tun)} | \Psi_m^{(0)} \rangle = B_m \langle (\Psi_m^{(0)})^2 | (\Psi_m^{(0)})^2 \rangle. \tag{23}$$

We can determine the optimal values of these $M$ shifts (and, thus, coefficients $B_m$) as follows. First, we express $N - 1$ of these shifts through the rest of them by solving the system of $N - 1$ linear equations following from Eqs. (16), (21), and (23):

$$\delta_n = \sum_{m=1}^{M} C_{n,m} C_{n+1,m} \Delta\omega_{1,m}^{(0)}. \tag{24}$$

Then, we can find all $\Delta\omega_{1,m}^{(0)}$ by minimizing the sum $\sum_{m=1}^{N} \left(\Delta\omega_{1,m}^{(0)}\right)^2$ varying the rest free parameters.

In the particular case of modulation-induced coupling between only two eigenstates, $\Psi_{n1}(\mathbf{r})$ and $\Psi_{n2}(\mathbf{r})$, Eqs. (22)-(24) yield:

$$\Delta\omega_1^{(tun)}(\mathbf{r}) = \delta_1 \frac{\sum_{m=1}^{M} C_{n1,m} C_{n2,m} \left|\Psi_m^{(0)}(\mathbf{r})\right|^2}{\sum_{m=1}^{M} C_{n1,m}^2 C_{n1,m}^2 \left\langle \left(\Psi_m^{(0)}\right)^2 \middle| \left(\Psi_m^{(0)}\right)^2 \right\rangle}. \qquad (25)$$

In turn, Eq. (24) is reduced to

$$\delta_1 = \sum_{m=1}^{M} C_{n1,m} C_{n2,m} \Delta\omega_{1,m}^{(0)}. \qquad (26)$$

In this case, minimization of this sum over $\Delta\omega_{1,m}^{(0)}$ can be performed using the Cauchy–Schwarz inequality, which suggests that the optimum frequency shifts should be found in the form $\Delta\omega_{1,m}^{(0)} = A C_{n1,m} C_{n2,m}$. The latter equation together with Eq. (26) yields:

$$\Delta\omega_m^{(0)} = \delta_1 \frac{C_{n1,m} C_{n2,m}}{\sum_{m=1}^{M} C_{n1,m}^2 C_{n2,m}^2}. \qquad (27)$$

### J. SDM figure of merit

Below, we consider microphotonic circuits constructed of coupled microresonators, such as ring resonators [41] and SNAP circuits [33], which are designed by varying the nanoscale effective radius of an optical fiber. The eigenstates of individual ring microresonators $\Psi_m^{(0)}$ depend on their waveguide length $z$ as $\exp(i\beta z)$ where $\beta$ is the propagation constant. Therefore, it follows from Eq. (22) that the optimized SDM, provided that it is independent on the transverse coordinates of the ring waveguide, should be constant within each of the microresonators. Then, from Eq. (22) we have for the optimized SDM

$$\left.\Delta\omega_1^{(tun)}(\mathbf{r})\right|_{\substack{\text{within}\\ \text{MR}_m}} = \Delta\omega_{1,m}^{(0)} \qquad (28)$$

where $\Delta\omega_{1,m}^{(0)}$ are determined by the system of linear equations, Eq. (24), as described above. In our major applications we consider equal modulation-induced couplings $\delta_n = \delta_1$. In this case, the figure of merit characterizing the SDM can be defined as

$$\frac{\max\left|\Delta\omega_{1,m}^{(0)}\right|}{\delta_1} \qquad (29)$$

i.e., we are looking for the SDM producing a given $\delta_1$ with the smallest possible SDM magnitude. For the case of propagation through two states of a circuit of two weakly coupled equal ring microresonators, we have $\Psi_{1,2} = 2^{-1/2}\left(\Psi_1^{(0)} \pm \Psi_2^{(0)}\right)$, i.e., from Eq. (21), $C_{1,1} = C_{1,2} = C_{2,1} = 2^{-1/2}$ and $C_{2,2} = -2^{-1/2}$ (see … for details). Then, from Eq. (27), we have $\Delta\omega_{1,m}^{(0)}/\delta_1 = (-1)^m$. We suggest that the figure of merit value $\max\left|\Delta\omega_{1,m}^{(0)}\right|/\delta_1 = 1$ is the smallest possible ratio of the SDM amplitude and $\delta_1$. We show below that the CIRT induced by modulation with that small ratio can be achieved only if the input and output waveguides coincide spatially.

In our further analysis of SNAP microresonator circuits, we assume that the SDM introduced along the optical fiber axis $z$ can be factorized in cylindrical coordinates $(z, \rho, \varphi)$ as $\Delta\omega_1^{(tun)}(\mathbf{r}) =$ $\Delta\omega_1^{(tun)}(z)Q(\rho,\varphi)$. Analogously to Eq. (29), in this case we define the figure of merit as

$$\frac{\max\left|\Delta\omega_1^{(tun)}(z)\right|}{\delta_1} \quad (30)$$

As an example, below we optimize the SDM of a resonant circuit comprising two states of a SNAP bottle microresonator and achieve $\max\left|\Delta\omega_1^{(tun)}(z)\right|/\delta_1 \cong 2$.

Generally, while the optimization of SDM requires larger spatial extension of its eigenstates $\Psi_n$ to increase their overlap, suppression of leakages from these states into waveguides restricts their spatial distribution. Below, we address the problem of the design of photonic circuits under the condition of the possible smallest figure of merit values defined by Eqs. (24) and (25) and minimum leakages.

### K. Leakage of photonic circuit eigenstates into waveguides

We consider photonic circuits constructed of a series of $M$ successively coupled microresonators with individual eigenstates $\Psi_m^{(0)}$, $m = 1, 2, \ldots, M$. In this series, only two edge microresonators with $m = 1$ and $m = M$ are coupled, respectfully, to waveguides 1 and 2 (Figs. 3(a) and 4). However, several of $N$ full eigenstates $\Psi_n$ of this series determined by Eqs. (2) and (3), which support the inelastic resonant transmission, may couple to waveguides and introduce leakages reducing the effect of CIRT. In our consideration below, we require that, among $N$ eigenstates $\Psi_n$, only $\Psi_1$ and $\Psi_N$ have to be coupled to waveguides 1 and 2, while coupling of $\Psi_n$ with $n = 2, 3, \ldots, N$ to waveguide 1 and coupling of $\Psi_n$ with $n = 1, 2, \ldots, N-1$ to waveguide 2 causing leakages are minimized by choosing the appropriate circuit parameters. Remarkably, the relative values of these leakages are simply expressed through coefficients $C_{n,m}$. Indeed, the coupling of state $\Psi_n$ with waveguide 1 is proportional to $\langle\Psi_n|\Psi_{in}^{(1,1)}\rangle$. Since only the eigenstate $\Psi_1^{(0)}$ with $m = 1$ is coupled to waveguide 1, we find from Eq. (21) that $\langle\Psi_n|\Psi_{in}^{(1,1)}\rangle = C_{n,1}\langle\Psi_1^{(0)}|\Psi_{in}^{(1,1)}\rangle$. Therefore, the leakage power from state $\Psi_n$ relative to coupling from $\Psi_1$ to waveguide 1 is $\left(\langle\Psi_n|\Psi_{in}^{(1,1)}\rangle/\langle\Psi_1|\Psi_{in}^{(1,1)}\rangle\right)^2 = (C_{n,1}/C_{1,1})^2$. Similarly, the relative leakage power of state $\Psi_n$ into waveguide 2 is $(C_{n,M}/C_{N,M})^2$.

### IV. PHOTONIC CIRCUITS WITH TWO RESONANT STATES

The effect of CIRT can be better understood by analysis of the simplest systems. For the system of two states illustrated in Fig. 5(a), we have $\Omega_1 = \omega_1 + (k + \frac{1}{2})\omega_p$, and $\Omega_2 = \omega_2 + (k + \frac{3}{2})\omega_p$. Photonic circuits with two eigenstates possessing only a single input-output waveguide [29, 30] present a particular case of the considered system. It can be modelled assuming that waveguides 1 and 2 coincide and, therefore, the leakages $\xi_{12}$ and $\xi_{21}$ are absent.

### A. Lossless photonic circuits without leakages

In the presence of both input-output waveguide 1 and output waveguide 2 and in the absence of resonant leakage channels, $\xi_{12} = \xi_{21} = 0$, and losses, $\gamma_1 = \gamma_2 = 0$, we have $S_{12} = S_{13} = 0$ and the inelastic transmission amplitude $S_{14}$ and elastic reflection amplitude $S_{11}$ are

$$S_{1,4}(\omega) = \frac{\xi_{1,1}\xi_{2,2}\delta_1}{D(\omega)}, \quad S_{11}(\omega) = 1 - \frac{i\xi_{1,1}^2\left(\omega - \omega_2 + \omega_p + \frac{i}{2}\xi_{2,2}^2\right)}{D(\omega)}, \quad (31)$$

$$D(\omega) = \left(\omega - \omega_1 + \frac{i}{2}\xi_{1,1}^2\right)\left(\omega - \omega_2 + \omega_p + \frac{i}{2}\xi_{2,2}^2\right) - \delta_1^2.$$

From this equation and Eq. (17), we find the values of $\omega = \omega_1^{(res)}$ and $\delta_1 = \delta_1^{(res)}$ which ensure the CIRT:

$$\omega_1^{(res)} = \omega_1 + \frac{(\omega_1 - \omega_2 + \omega_p)\xi_{1,1}^2}{\xi_{2,2}^2 - \xi_{1,1}^2}, \tag{32}$$

$$\delta_1^{(res)} = \xi_{1,1}\xi_{2,2}\left(\frac{1}{4} + \frac{(\omega_1 - \omega_2 + \omega_p)^2}{(\xi_{2,2}^2 - \xi_{1,1}^2)^2}\right)^{1/2}. \tag{33}$$

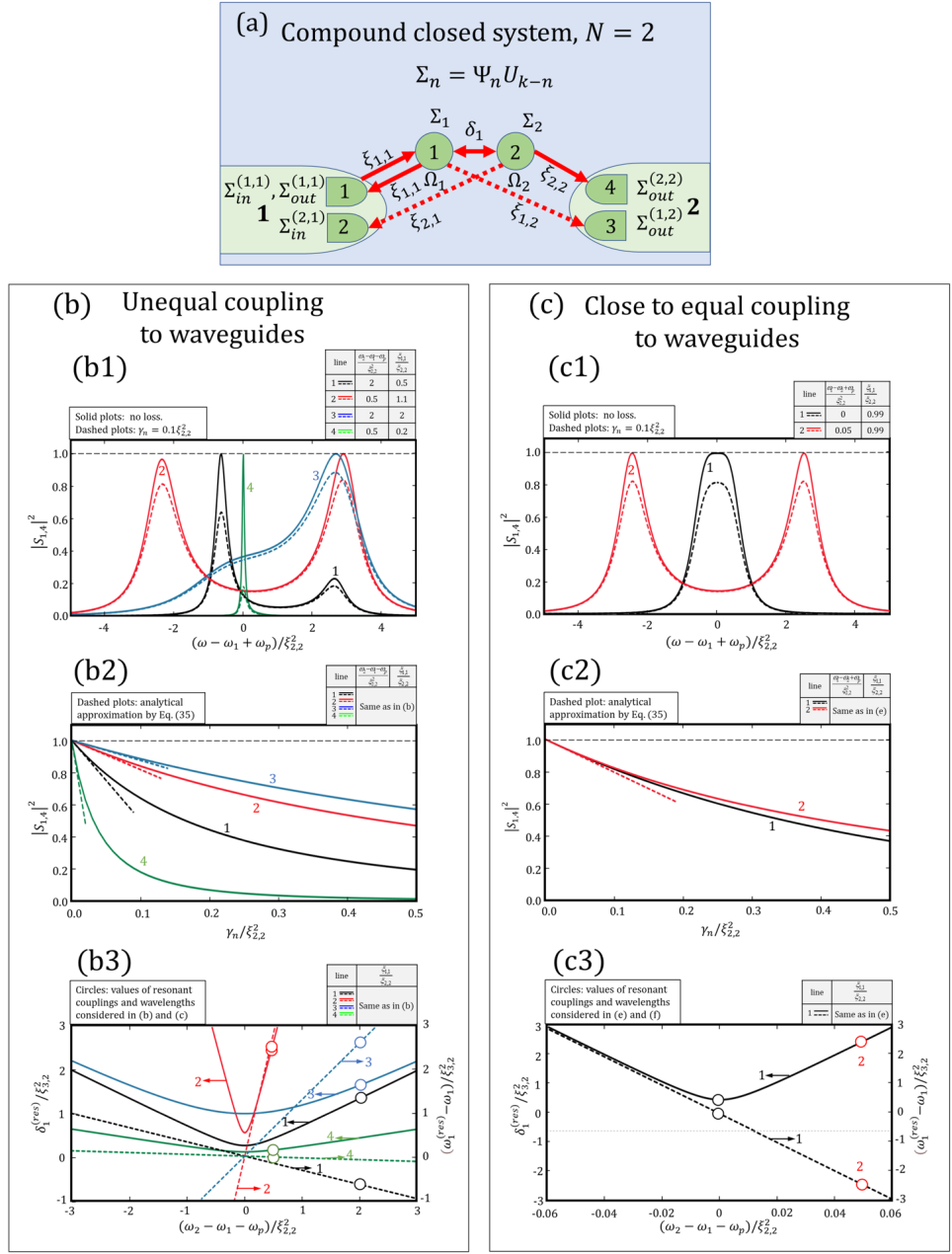

FIG. 5. Characteristic plots of transmission power as a function of frequency for the modulated system of two resonant states and interstate coupling satisfying Eq. (33).

It is seen from these equations that the condition of CIRT does not assume the exact frequency resonance $\Omega_1 = \Omega_2$, i.e., $\omega_2 = \omega_1 + \omega_p$. Formally, this condition can be achieved for any values of couplings to waveguides, $\xi_{11}$ and $\xi_{22}$, system's eigenfrequencies, $\omega_1$ and $\omega_2$, and modulation frequency, $\omega_p$, provided that the frequency $\omega$ and modulation-induced coupling $\delta_1$ are tuned appropriately. In particular, if one of the couplings to waveguides is small compared to another one, e.g., $\xi_{11} \ll \xi_{22}$, then the required for the CIRT modulation-induced coupling $\delta_1^{(res)}$, which is typically limited by several GHz, can be proportionally small. However, reduction of couplings $\xi_{11}$ and $\xi_{22}$ leads to the suppression of resonant transparency due to the enhancement of the effect of material losses (see below).

It follows from Eqs. (32) and (33) that a lossless photonic circuit consisting of $N = 2$ coupled states can have one exact maximum of transmission power $|S_{14}(\omega)|^2$ equal to unity. Generally, a lossless circuit consisting of $N$ resonant states can have $N$ maxima equal to unity and, in particular two such maxima for $N = 2$ (see Appendix A and Refs. [13-15]). The latter situation is possible for exactly symmetric photonic circuit when $\omega_2 = \omega_1 + \omega_p$ and $\xi_{1,1} = \xi_{2,2}$ excluded in Eqs. (32), (33). Close to this condition, there exist two maxima, one of which may be equal to unity and another one is close to unity.

The larger is the difference between the couplings $\xi_{11}$ and $\xi_{22}$ the smaller is the value of $\delta_1$ which can satisfy the CIRT condition. The smallest possible value of coupling that enables the CIRT found from Eq. (33) is

$$\delta_1^{(res)} = \frac{\xi_{11}\xi_{22}}{2}. \tag{34}$$

Eq. (33) demonstrates that the considered system can function as both a tunable frequency beam splitter [30] and a simultaneous spatial and frequency beam splitter. Indeed, in the applications of our interest the coupling amplitude $\delta_1$ is proportional to the modulation amplitude, which can be tuned. Then, at the resonance condition, the inelastic transmission power $|S_{14}|^2$ to the output waveguide 2 varies from zero to unity when $\delta_1$ varies from zero to the value defined by Eq. (33). Simultaneously, the power of elastic transition back into the input waveguide $|S_{11}|^2 = 1 - |S_{14}|^2$ varies from unity to zero.

Characteristic plots of transmission power as a function of frequency for the interstate coupling satisfying Eq. (33) are shown in Fig. 5. Generally, for unequal couplings to waveguides and eigenfrequencies not satisfying the exact resonant condition $\omega_2 - \omega_1 - \omega_p = 0$, the inelastic transmission power $|S_{14}(\omega)|^2$ has two maxima, but only one of them is equal to unity (curve 1 in Fig. 5(b1)). The closer the couplings $\xi_{11}$ and $\xi_{22}$ are to each other, the larger the separation between these maxima is and the closer the smaller maximum becomes to unity (curve 2 in Fig. 5(b1)). The interplay of parameters may suppress or completely eliminate one of the maxima. Then the transmission power spectrum possesses a single maximum equal to unity (curve 3 in Fig. 5(b1)). The value of resonant coupling $\delta_1^{(res)}$ required to achieve the CIRT can be reduced by the reduction of one of the couplings to the waveguides, which leads to narrowing of the transmission peak (curve 4 in Fig. 5(b1)).

### B. Effect of losses and leakages

The material losses and leakages to other output channels reduce the inelastic transmission power $|S_{14}|^2$. For relatively small losses and leakages, $\gamma_1 + \xi_{12}^2 \ll \xi_{11}^2$ and $\gamma_2 + \xi_{21}^2 \ll \xi_{22}^2$, the reduction can be found from Eq. (10) by the perturbation theory (see Appendix B):

$$\max\left(|S_{1,4}|^2\right) = \left|S_{1,4}(\omega_1^{(res)})\right|^2 = 1 - \Delta, \quad \Delta = \frac{\gamma_1 + \xi_{12}^2}{\xi_{11}^2} + \frac{\gamma_2 + \xi_{21}^2}{\xi_{22}^2}. \tag{35}$$

Here we introduce the deviation $\Delta$ of the inelastic transmission amplitude from unity. Eq. (35) is also valid for $\xi_{11} = \xi_{22}$, $\omega_2 = \omega_1$, and arbitrary $\delta_1$. It follows from Eq. (35) that the contributions of losses and leakages from individual states to the reduction of inelastic transmission power are combined and summed. Simple calculations show that this effect occurs in the first-order perturbation over losses and leakages only. While the solid curves in Fig. 5(b1) correspond to the lossless inelastic transmission, the dashed curves of the same color correspond to the transmission with material losses $\gamma_1 = \gamma_2 = 0.1\xi_{22}^2$. To better understand the effect of material losses, Fig. 5(b2) shows the dependencies of the maximum inelastic transmission power on the dimensionless loss $\gamma_n/0.1\xi_{22}^2$ assuming $\gamma_1 = \gamma_2$. These dependencies are compared with the analytical approximation of for relatively small losses found from Eq. (35) (dashed curves). In turn, Fig. 5(b3) shows the dependence of resonant coupling $\delta_1^{(res)}$ and resonant frequency $\omega_1^{(res)}$ on the dimensionless eigenfrequency separation $(\omega_2 - \omega_1 - \omega_p)/\xi_{22}^2$ indicating their values corresponding to the curves shown in Fig. 5(b1) by circles. From Eq. (35), the effect of losses and leakages can be reduced for sufficiently large couplings to waveguides, $\xi_{1,1}^2$ and $\xi_{2,2}^2$. However, from Eq. (33), the latter condition requires proportionally larger interstate coupling $\delta_1^{(res)}$ which may not be available experimentally. On the other hand, the inelastic transmission resonance (curve 4 in Fig. 5(b1)) which corresponds to relatively small $\delta_1^{(res)}$ (green circles at curves 4 in Fig. 5(b1)) reduces its magnitude with growing losses much faster than the resonances at larger $\delta_1^{(res)}$ (curve 4 vs. other curves in Fig. 5(b2)).

### C. Inelastic Butterworth filter

It follows from Eq. (32) that if the couplings to waveguides, $\xi_{11}$ and $\xi_{22}$, are close to each other and the resonant condition $\omega_2 - \omega_1 - \omega_p = 0$ is not accurately satisfied, the CIRT may correspond to frequency that is situated far away from the frequencies $\omega_1$ and $\omega_2 - \omega_p$. The inelastic transmission spectra shown in Fig. 5(c1) illustrate this effect. For $\omega_2 - \omega_1 - \omega_p = 0$ and $\xi_{11} = \xi_{22}$, we have $\delta_1^{(res)} = \frac{1}{2}\xi_{22}^2$ and $\omega_1^{(res)} = \omega_1$ so that

$$\left|S_{1,4}(\omega)\right|^2 = \frac{\xi_{2,2}^8}{\xi_{2,2}^8 + 4(\omega - \omega_1)^4} \tag{36}$$

(curve 1 in Fig. 5(c1)), which is the second order Butterworth filter. However, even for very small deviations from this condition (still satisfying Eq. (32)), the resonance splits into two peaks with significant separation (curve 2 in Fig. 5(c1)). While the maximum of one of these peaks remains equal to unity, it is slightly smaller than unity for another one. Similar to Fig. 5(b1), the dashed curves in Fig. 5(c1) show the inelastic transmission spectra with losses $\gamma_1 = \gamma_2 = 0.1\xi_{22}^2$. Similar to Figs. 5(b2) and 5(b3), Figs. 5(c2) and 5(c3) show the dependencies of maximum inelastic transmission on material losses and the dependencies of resonance couplings and frequencies on the eigenfrequency separation.

### D. Singularity of automatic tuning towards CIRT

Both the interstate coupling $\delta_1$ and the eigenfrequency separation $\omega_2 - \omega_1 - \omega_p$ can be tuned by changing the power of modulation, its frequency $\omega_p$, as well as individual eigenfrequencies $\omega_n$. For this reason, experimentally, the modulation power can be set to *automatically satisfy* the condition of Eq. (33) for the interstate coupling. Then, as follows from Eqs. (32) and (33), the position of the resonance and, hence, the inelastic transmission amplitude experience a *singular behavior* as a function of modulation frequency $\omega_p$ if the system's parameters are close to the condition $\omega_2 - \omega_1 - \omega_p = 0$ and $\xi_{11} = \xi_{22}$. Very small deviations from this condition can cause large displacement of resonances as illustrated in Fig. 5. In particular, this figure demonstrates a fast transition from the transmission power spectrum with degenerated resonances defined by Eq. (36) (curve 1 in Fig. 5(c1)) to the power spectrum with strongly separated resonances (curve 2 in Fig. 5(c1)). Similar tunability can be achieved in stationary photonic circuits, where $\omega_p = 0$, by tuning the eigenfrequency difference $\omega_2 - \omega_1$ and coupling coefficient $\delta_1$ by nonlocal refractive index variations $\Delta n_0(r)$ and $\Delta n_1(r)$. It can be shown, however, that the noted singularity is absent for the independent tuning of the circuit parameters.

## V. PHOTONIC CIRCUITS WITH THREE RESONANT STATES

The tunability of all couplings $\delta_n$ introduced by variation of the modulation power may be limited by the experimentally feasible SDM $\Delta\omega_1^{(mod)}(\mathbf{r})$ (see Eqs. (14) and (15)). Nevertheless, in the absence of losses and leakages and for couplings having the same order, $\delta_n \sim \delta_m$ The condition of CIRT can be achieved by varying the total modulation power. Here we demonstrate this fact for a photonic circuit consisting of $N = 3$ eigenstates (Fig. 6(a)) with an arbitrarily predetermined ratio of couplings

$$\alpha = \frac{\delta_2}{\delta_1} \tag{37}$$

fixed by the distribution of $\Delta n(x, y, z)$ so that couplings $\delta_1$ and $\delta_2$ are varied in proportion to the total modulation power. As in the previous Section IV, we first ignore the effects of losses and leakages and determine the values of $\delta_1$ and the input frequency $\omega$ corresponding to the condition of CIRT, and then find the reduction of transmission power due to the presence of these effects.

### A. Lossless photonic circuits without leakages, inelastic Butterworth filter

In the absence of losses and leakages we set $\xi_{2,1} = \xi_{3,1} = \xi_{1,2} = \xi_{2,2} = 0$ and $\gamma_1 = \gamma_2 = \gamma_3 = 0$ and the condition of CIRT is given by Eq. (17). Solution of this equation yields two possible values of the resonant frequency (see Appendix C):

$$\omega_{1,2}^{(res)} = \frac{(\omega_3 - 2\omega_p)\xi_{1,1}^2 - \alpha^2 \omega_1 \xi_{3,2}^2 \pm \tfrac{1}{2}\xi_{1,1}\xi_{3,2}D^{1/2}}{\xi_{1,1}^2 - \alpha^2 \xi_{3,2}^2},$$
$$D = 4\alpha^2(\omega_3 - \omega_1 - 2\omega_p)^2 + (\xi_{1,1}^2 - \alpha^2 \xi_{3,2}^2)(\alpha^2 \xi_{1,1}^2 - \xi_{3,2}^2), \tag{38}$$

and two possible values of the corresponding resonant coupling parameter $\delta_1$

$$\delta_{1,2}^{(res)} = \left\{ \frac{\left(\omega_{1,2}^{(res)} - \omega_2 + \omega_p\right)\left[\left(\omega_{1,2}^{(res)} - \omega_1\right)\xi_{3,2}^2 - \left(\omega_{1,2}^{(res)} - \omega_3 + 2\omega_p\right)\xi_{1,1}^2\right]}{\xi_{3,2}^2 - \alpha^2 \xi_{1,1}^2} \right\}^{1/2} \tag{39}$$

expressed through the frequencies of uncoupled states, $\omega_1$, $\omega_2$, $\omega_3$, coupling to waveguides, $\xi_{1,1}$, $\xi_{3,2}$, and the ratio of interstate couplings $\alpha$. The photonic circuit parameters are assumed fixed. Under the conditions that $D \geq 0$ in Eq. (38) and that the expression in the curly brackets in Eq. (39) is positive, these two equations determine two values of modulation power when, in the absence of material losses and leakages, the magnitude of inelastic transmission amplitude can achieve unity. Remarkably, the value of the resonant frequencies $\omega_{1,2}^{(res)}$ does not depend on eigenfrequency $\omega_2$, while the corresponding couplings $\delta_{1,2}^{(res)}$ depend on $\omega_2$ linearly. Thus, for $\omega_{1,2}^{(res)}$ close to $\omega_2$, the required $\delta_{1,2}^{(res)}$ can be made sufficiently small (compare with the case $N = 2$ of Section IV).

It follows from Eqs. (38) and (39) that if $\xi_{3,2}^2 \cong \alpha^2 \xi_{1,1}^2$ then the separation of the resonant frequencies from the eigenfrequencies of uncoupled eigenstates as well as the values of the resonant coupling may grow infinitely unless the photonic circuit eigenfrequencies are perfectly aligned. This result is similar to that for $N = 2$ illustrated in Fig. 5. In the case of aligned eigenfrequencies, $\omega_1 = \omega_2 - \omega_p = \omega_3 - 2\omega_p$, Eqs. (38) and (39) are simplified to

$$\omega_{1,2}^{(res)} = \omega_1 \pm \tfrac{1}{2}\xi_{1,1}\xi_{3,2}\left(\frac{\alpha^2 \xi_{1,1}^2 - \xi_{3,2}^2}{\xi_{1,1}^2 - \alpha^2 \xi_{3,2}^2}\right)^{1/2} \tag{40}$$

$$\delta_{1,2}^{(res)} = \tfrac{1}{2}\xi_{1,1}\xi_{3,2}\left(\frac{\xi_{1,1}^2 - \xi_{3,2}^2}{\xi_{1,1}^2 - \alpha^2\xi_{3,2}^2}\right)^{1/2} \tag{41}$$

These equations can only be satisfied if

$$\min\left(\frac{\xi_{1,1}}{\xi_{3,2}}, \frac{\xi_{3,2}}{\xi_{1,1}}\right) < \alpha < \max\left(\frac{\xi_{1,1}}{\xi_{3,2}}, \frac{\xi_{3,2}}{\xi_{1,1}}\right). \tag{42}$$

From this inequality, we find that, as in the case of two resonant states considered in Section IV, the completely symmetric photonic circuit with $\xi_{1,1} = \xi_{3,2}$ can be realized for $\alpha = 1$ only, i.e., for equal interstate couplings and perfectly aligned photonic circuits.

### B. Inelastic transmission spectra, the effect of losses and leakages

The characteristic plots of inelastic transmission power $|S_{16}(\omega)|^2$ as a function of frequency for the resonant values of couplings $\delta_{1,2}^{(res)}$ defined by Eqs. (40) and (41) are shown in Fig. 6. Generally, as illustrated in Fig. 6(b1), curves 1, 2 and 3, these plots may have one, two, or three maxima though only one of them achieves unity. From Eq. (39), if the resonance frequency $\omega_{1,2}^{(res)}$ is close to the eigenfrequency $\omega_2$, the resonance coupling $\delta_2^{(res)}$ is very small and, in the absence of losses, the resonance peak of $|S_{1,6}(\omega)|^2$ can be very narrow, as illustrated by curve 3 in Fig. 6(b1). However, for smaller coupling $\delta_2^{(res)}$ the effect of losses (dashed curves of the same color, for which we assume $\gamma_1 = \gamma_2 = \gamma_3 = 0.1\xi_{22}^2$) is much stronger. This effect becomes clearer from Fig. 6(b2) which shows the dependence of the maximum inelastic transmission power on material losses where we assumed, again, equal losses for all eigenstates of the system, $\gamma_1 = \gamma_2 = \gamma_3$. Fig. 6(b3) shows the dependencies of resonant couplings and frequencies on the dimensionless eigenfrequency separation $(\omega_3 - \omega_2 - \omega_p)/\xi_{22}^2$.

The characteristic spectral plots of inelastic transmission power for the case of perfectly aligned eigenfrequencies, $\omega_1 = \omega_2 - \omega_p = \omega_3 - 2\omega_p$, are shown in Fig. 6(c1). Curve 1 in this figure (similar to curve 1 in Fig. 5(c1)) illustrates the degenerated dependence of transmission power on frequency near its maximum for the aligned circuit configuration when $\alpha = \frac{\xi_{3,2}}{\xi_{1,1}} = 1$, $\omega_{1,2}^{(res)} = \omega_1$, and $\delta_1^{(res)} = 2^{3/2}\xi_{3,2}^2$. Then we arrive at the inelastic Butterworth filter (see Appendix D):

$$|S_{1,6}(\omega)|^2 = \frac{\xi_{3,2}^{12}}{\xi_{3,2}^{12} + [2(\omega - \omega_1)]^6}. \tag{43}$$

For a small deviation from this configuration, curve 2 illustrates the splitting of the single resonance into three resonances, where two of them have a maximum equal to unity, and the third maximum is close to unity. Generally, for the aligned eigenfrequencies and away from the symmetric configuration, the inelastic transmission power has two symmetric maxima equal to unity, as seen in curves 3 and 4 in Fig. 6(c1).

As in the previous cases, dashed curves in this plot take into account losses $\gamma_1 = \gamma_2 = \gamma_3 = 0.1\xi_{22}^2$. For relatively small losses and leakages, the reduction of transmission power is found in Appendix B for the close to the aligned photonic circuit configuration when $\omega_1 \cong \omega_2 - \omega_p \cong \omega_3 - 2\omega_p$ and $\alpha = 1$. Then $\omega_{1,2}^{(res)} = \omega_1 \pm \delta_1^{(res)}$, $\delta_1^{(res)} \cong \tfrac{1}{2}\xi_{1,1}\xi_{3,2}$ and the inelastic transmission power is calculated from

$$\left|S_{1,6}(\omega_1 \pm \tfrac{1}{2}\xi_{1,1}\xi_{3,2})\right|^2 \cong 1-\Delta, \quad \Delta = \left[\frac{1}{\xi_{1,1}^2}\left(\gamma_1 + \xi_{1,2}^2\right) + \left(\frac{1}{\xi_{1,1}^2} + \frac{1}{\xi_{3,2}^2}\right)\left(\gamma_2 + \xi_{2,1}^2 + \xi_{2,2}^2\right) + \frac{1}{\xi_{3,2}^2}\left(\gamma_3 + \xi_{3,1}^2\right)\right]. \quad (44)$$

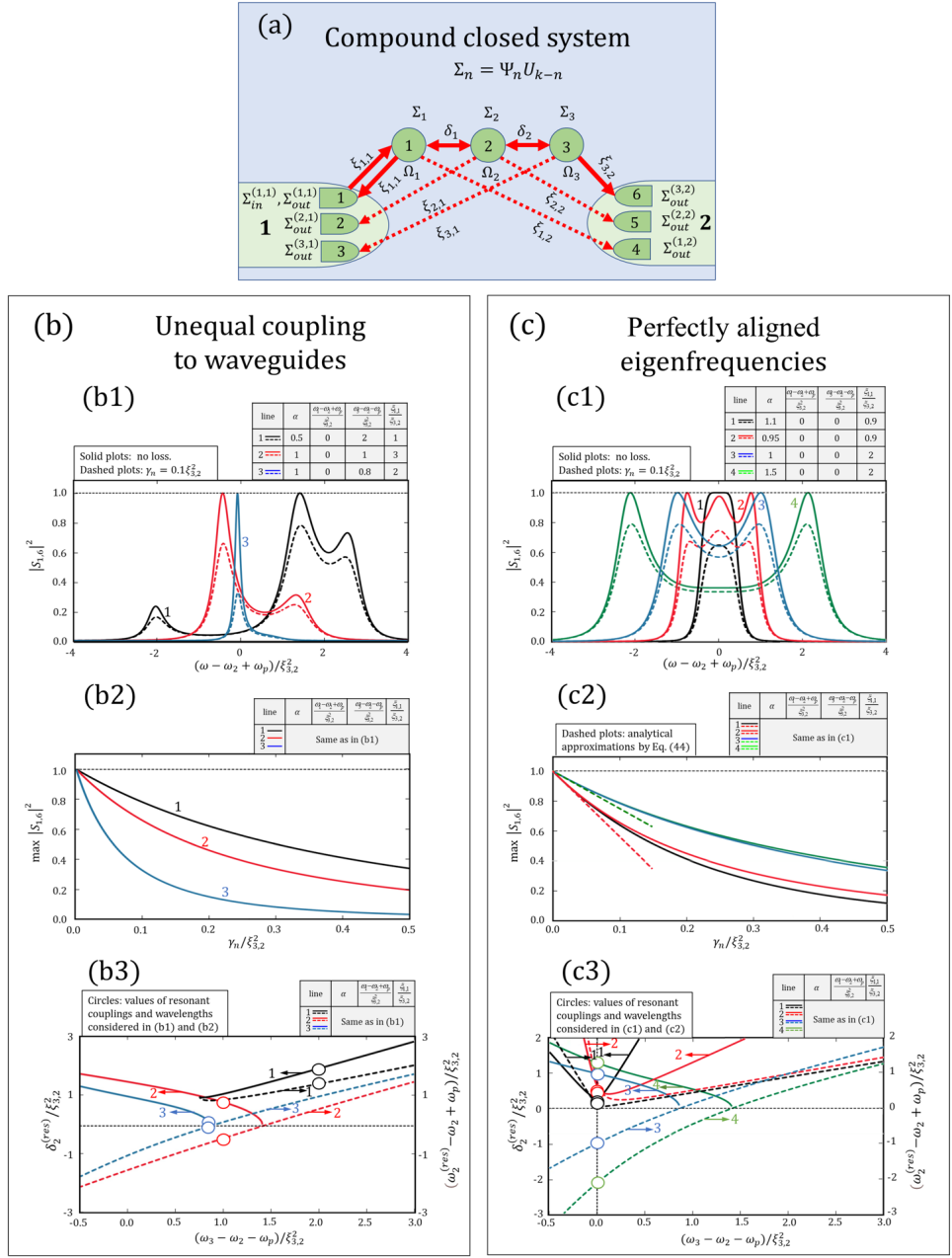

FIG. 6. Characteristic plots of transmission power as a function of frequency for the modulated system of three resonant states.

This equation shows that the contribution of losses and leakages of the intermediate state to the reduction of transmission power is qualitatively different from that of states directly coupled to waveguides. Fig. 6(c2) compares the maximum inelastic transmission power as a function of material losses for the photonic circuits with parameters used in Fig. 6(c1). Dashed lines in Fig. 6(c2) are calculated from Eq. (44). Finally, Fig. 6(c3) shows the dependencies of dimensionless resonant couplings and frequencies on the dimensionless eigenfrequency separation $(\omega_3 - \omega_2 - \omega_p)/\xi_{22}^2$ where, as previously, the values indicated by circles correspond to the curves in Fig. 6(c1) of the same color. As in the examples above, the smaller the resonant coupling, the smaller the material losses required to achieve an inelastic transmission amplitude close to unity.

### C. Resonant transparency enhanced by leakage and loss

At first sight, leakage and loss in a photonic circuit can only decrease its resonant transparency. Here we demonstrate that there exist situations where increasing leakage and loss can increase the resonant transparency of a circuit. As an example, we consider a Butterworth inelastic photonic circuit having two compound resonant states $\Sigma_1$ and $\Sigma_2$ with equal eigenstates, $\Omega_1 = \Omega_2$, zero material losses, $\gamma_1 = \gamma_2 = 0$, equal couplings to waveguides, $\xi_{1,1} = \xi_{2,2}$, and interstate coupling $\delta_1 = \xi_{2,2}^2/2$ (see Section IVC). This circuit experiences CIRT at $\Omega = \Omega_1$. Next, we add an auxiliary state $\Sigma_3$ with eigenfrequency $\Omega_3$ to this circuit as illustrated in Figs. 7(a) and 7(b). We introduce coupling $\delta_2$ between states $\Sigma_2$ and $\Sigma_3$, coupling $\xi_{3,2}$ of state $\Sigma_3$ to waveguide 2, and material loss $\gamma_3$ of $\Sigma_3$. Then, the transparency of this circuit at frequency $\Omega = \Omega_1$ is

$$|S_{1,5}(\Omega_1)|^2 = 1 - \Delta, \quad \Delta = \frac{4\delta_2^2 \left(\delta_2^2 + \xi_{1,1}^2 \gamma_3^{(tot)}\right)}{\left(2\delta_2^2 + \xi_{1,1}^2 \gamma_3^{(tot)}\right)^2 + 4(\Omega_3 - \Omega_1)^2 \xi_{1,1}^4}, \quad \gamma_3^{(tot)} = \gamma_3 + \xi_{3,2}^2, \tag{45}$$

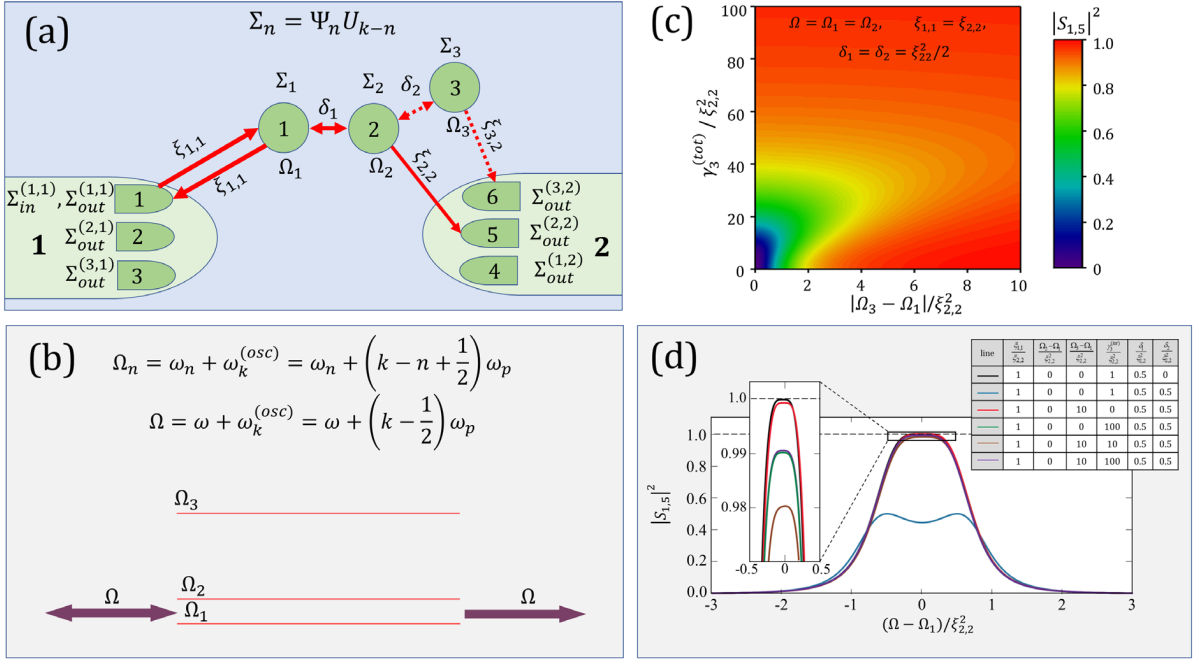

FIG. 7. (a) A Butterworth inelastic photonic circuit with compound resonant states $\Sigma_1$ and $\Sigma_2$ complemented by an auxiliary state $\Sigma_3$. (b) Configuration of compound eigenfrequencies $\Omega_1$, $\Omega_2$, and $\Omega_3$ of states $\Sigma_1$, $\Sigma_2$, and $\Sigma_3$. (c) Transmission power surface plot demonstrating the growth of transparency $|S_{1,5}|^2$ with increasing the loss and leakage of state $\Sigma_3$. (d) Transparency $|S_{1,5}|^2$ as a function of compound frequency $\Omega$ for characteristic values of parameters $\Omega_3 - \Omega_1$, $\delta_n$, $\xi_{1,1}$, and $\gamma_3^{(tot)}$ indicated in the table.

where we introduced the total loss $\gamma_3^{(tot)}$ of state $\Sigma_3$ which includes material loss $\gamma_3$ and leakage caused by coupling $\xi_{3,2}$ to waveguide 2 (compare with Eq. (43)). It can be shown that for any input frequency $\Omega$, the transmittance $|S_{1,5}(\Omega)|^2$ of the considered circuit depends on loss $\gamma_3$ and coupling $\xi_{3,2}$ through their combination in $\gamma_3^{(tot)}$ only. It follows from Eq. (43) that the effect of state $\Sigma_3$ becomes negligible not only for a relatively large deviation of the eigenfrequency $\Omega_3$ from the resonance $\Omega_1 = \Omega_2$ of the original two state circuit but also for a relatively large total loss $\gamma_3^{(tot)}$ of state $\Sigma_3$, which can be enhanced either by increasing its material loss $\gamma_3$ or coupling $\xi_{3,2}$ to waveguide 2. This result is visualized by the surface plot of $|S_{1,5}|^2$ in Fig. 7(c), which was calculated from Eq. (44). It is seen from this plot that the growth of leakage of state $\Sigma_3$ into waveguide 2 and/or its loss results in increasing of transparency $|S_{1,5}|^2$ approaching unity even for zero deviation of eigenfrequency $\Omega_3$ from the resonance $\Omega_1$ of the Butterworth filter. Complementary, Fig. 7(d) shows the behavior of transparency $|S_{1,5}|^2$ on frequency $\Omega$ for characteristic values of parameters $\Omega_3 - \Omega_1$, $\delta_n$, $\xi_{1,1}$, and $\gamma_3^{(tot)}$.

## VI. PHOTONIC CIRCUITS WITH *N* RESONANT STATES

Here, we describe properties of inelastic resonant transmission through time-modulated photonic circuits, which directly follow from the related properties of the elastic resonant circuits [13-15, 39]. Ideally, the photonic circuits of our interest are composed of states with equally spaced eigenfrequencies $\omega_n$, in our formulation – with equal compound eigenfrequencies $\Omega_n = \omega_n + \left(k - n + \frac{1}{2}\right)\omega_p = \Omega_1 = \omega_1 + \left(k - \frac{1}{2}\right)\omega_p$. In this Section we describe several cases of special interest. First, we consider photonic circuits experiencing the CIRT at the resonant frequency coinciding with $\omega_1$. Such circuits can be designed to arrive at the CIRT over a bandwidth rather than for a single resonance frequency. In analogy to the circuits enabling complete elastic transmission amplitude approaching unity over a bandwidth [13, 15] these circuits can be called *inelastic Butterworth filters*. Next, we assume that in addition to all equal compound eigenfrequencies $\Omega_n = \Omega_1$ all interstate couplings induced in the photonic circuits are equal too, i.e., $\delta_n = \delta_1$. In the latter case, a circuit with equal couplings to waveguides, $\xi_{1,1} = \xi_{N,2}$, (identical to the previously considered elastic resonant transmission case [13-15]), due to its spatial symmetry, a lossless photonic circuit composed of *N* resonant eigenstates can exhibit the *CIRT at N resonance frequencies.* Here we prove a less evident result for the case of unequal couplings, $\xi_{1,1} \neq \xi_{N,2}$, which, as shown in the previous Sections IV and V, may become of special importance in approaching the condition of the CIRT experimentally. We show that at $\xi_{1,1} \neq \xi_{N,2}$ a lossless photonic circuit composed of *N* resonant eigenstates with equally spaced eigenfrequencies $\omega_n$ and equal interstate couplings $\delta_n$ can exhibit the *CIRT at $N - 1$ resonance frequencies.* Finally, we discuss the effects of losses, leakages, and misalignment.

### A. Photonic circuits without leakages

To exhibit the inelastic transmission amplitude close to unity, photonic circuits should be designed to have negligible leakages from the intermediate states $2, 3, \ldots, N - 1$ into the waveguides. In the absence of leakages, the only nonzero elements of *S*-matrix are the reflection amplitude $S_{1,1}(\omega)$ and the transmission amplitude $S_{1,2N}(\omega)$, which are found from Eq. (10) as:

$$S_{1,1}(\omega) = 1 - \frac{i\xi_{1,1}^2 M_{1,1}(\Theta)}{\det(\Theta)}, \quad S_{1,2N}(\omega) = \frac{(-1)^N i\xi_{1,1}\xi_{N,2}\prod_{n=1}^{N-1}\delta_n}{\det(\Theta)},$$

$$\Theta = \Omega(\omega)\mathbf{I}_N - \mathbf{\Omega} + \frac{i}{2}(\mathbf{\Gamma} + \mathbf{\Xi}), \quad \mathbf{\Xi} = \begin{pmatrix} \xi_{1,1}^2 & 0 & \ldots & 0 \\ 0 & 0 & \ldots & 0 \\ \ldots & \ldots & \ldots & \ldots \\ 0 & 0 & \ldots & \xi_{N,2}^2 \end{pmatrix}.$$

(46)

Here $M_{n,m}(\Theta)$ is a minor of matrix $\Theta$. The expression for transmission amplitude $S_{1,2N}(\omega)$ in Eq. (46) replicates the expressions previously known for the elastic transmission [13, 15].

## B. Lossless photonic circuits with equal compound eigenfrequencies $\Omega_n$ and arbitrary interstate couplings $\delta_n$

First, we consider a lossless photonic circuit having eigenstates with equal compound eigenfrequencies $\Omega_n = \omega_n + \left(k - n + \frac{1}{2}\right)\omega_p = \Omega_1 = \omega_1 + \left(k - \frac{1}{2}\right)\omega_p$ and different interstate coupling parameters $\delta_n$. Calculations (see Appendix E) show that the CIRT of this circuit, $|S_{1,2N}(\omega)|^2 = 1$, happens at $\Omega_1^{(res)} = \Omega_1$, i.e., at light frequency $\omega = \omega_1^{(res)} = \omega_1$, and different conditions for even and odd $N$:

$$\xi_{1,1}\xi_{N,2}\delta_2\delta_4...\delta_{N-2} = 2\delta_1\delta_3...\delta_{N-1} \quad \text{(even } N\text{)}, \tag{47}$$

$$\xi_{1,1}\delta_2\delta_4...\delta_{N-1} = \xi_{N,2}\delta_1\delta_3...\delta_{N-2} \quad \text{(odd } N\text{)}. \tag{48}$$

For $N = 2$, Eq. (47) coincides with Eq. (33), and for $N = 3$ Eq. (48) coincides with the condition of $\omega_1^{(res)} = \omega_1$ following from Eq. (40).

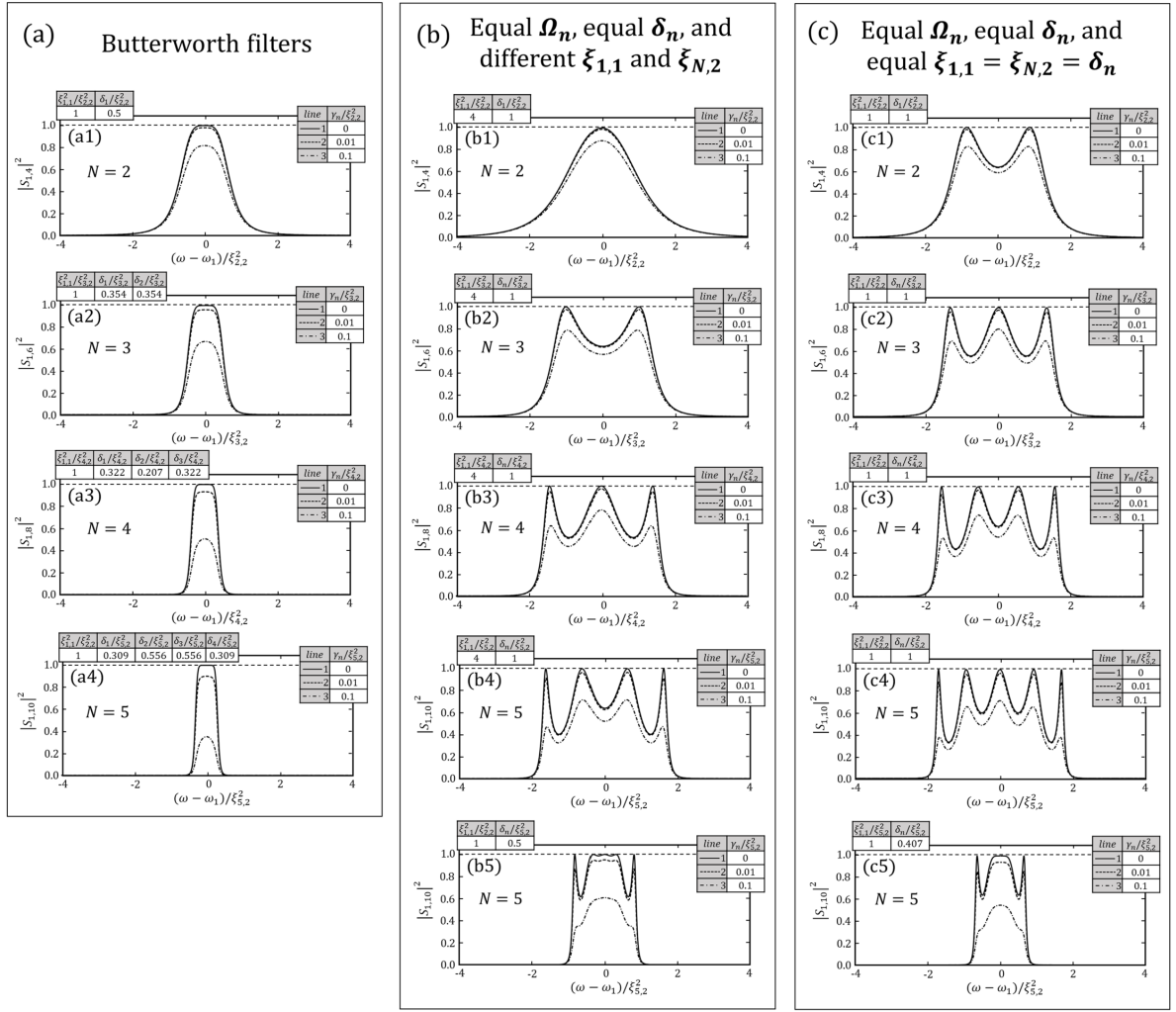

FIG. 8. The inelastic transmission spectra for photonic circuits composed of $N = 2, 3, 4,$ and $5$ localized states. (a1)-(a4) The Butterworth filters composed of unequally coupled localized states with symmetric couplings to waveguides, $\xi_{1,1} = \xi_{N,2}$. (b1)-(b5) Photonic circuits composed of equally coupled localized states with asymmetric couplings to waveguides, $\xi_{1,1} \neq \xi_{N,2}$. (c1)-(c5) Photonic circuits composed of equally coupled localized states with symmetric couplings to waveguides, $\xi_{1,1} = \xi_{N,2}$.

## C. Inelastic Butterworth filters

Using Eqs. (47) and (48) we can introduce an inelastic analogue of a Butterworth filter [15]. The inelastic second- and third-order Butterworth filters are described by Eqs. (36) and (43) of Sections IV and V. Generally, the transmission power of an $N^{th}$ order lossless inelastic Butterworth filter composed of $N$ resonantly coupled eigenstates can be written down as:

$$|S_{1,2N}(\omega)|^2 = \frac{\xi_{N,2}^{4N}}{\xi_{N,2}^{4N} + \rho_N [2(\omega - \omega_1)]^{2N}} \qquad (49)$$

where $\rho_N$ is a dimensionless parameter defined below.

Coupling parameters $\delta_n$, $\xi_{1,1}$ and $\xi_{N,2}$ in Eqs. (47) and (48) can be chosen so that the corresponding circuits become the inelastic Butterworth filters. For $N = 2$ and $N = 3$ these parameters and the expressions for the inelastic transmission power, Eqs. (36) and (43), are presented in Sections IV and V. More cumbersome calculations determine the Butterworth filters $|S_{1,2N}(\omega)|^2$ for $N = 4$ and $N = 5$ (see Appendix D). The parameters obtained are collected in Table 1 where the results for $N = 4$ coincides with that found in Ref. [15]. The inelastic transmission spectra for the Butterworth filters with $N = 2, 3, 4,$ and $5$ are shown in Fig. 8(a1)-(a4). In these plots, the decimal quantities of $\delta_n/\xi_{N,2}^2$, which exact values are collected in Table 1, are given. Solid lines describe the lossless filters, while dashed and dash-dotted lines include losses in each of coupled states equal to $0.01\xi_{N,2}^2$ and $0.1\xi_{N,2}^2$, respectively.

| N | $\xi_{11}/\xi_{N,2}$ | $\delta_1/\xi_{N,2}^2$ | $\delta_2/\xi_{N,2}^2$ | $\delta_3/\xi_{N,2}^2$ | $\delta_4/\xi_{N,2}^2$ | $\rho_N$ |
|---|---|---|---|---|---|---|
| 2 | 1 | $\frac{1}{2}$ | – | – | – | $\frac{1}{2}$ |
| 3 | 1 | $2^{-3/2}$ | $2^{-3/2}$ | – | – | 1 |
| 4 | 1 | $\frac{1}{2}(2^{1/2} - 1)^{1/2}$ | $\frac{1}{2}(2^{1/2} - 1)$ | $\frac{1}{2}(2^{1/2} - 1)^{1/2}$ | – | $3 \cdot 2^{\frac{1}{2}} + \frac{17}{4}$ |
| 5 | 1 | $\frac{1}{4}(5^{1/2} - 1)$ | $\frac{1}{2}\left(\frac{5^{1/2}}{2} - 1\right)^{1/2}$ | $\frac{1}{2}\left(\frac{5^{1/2}}{2} - 1\right)^{1/2}$ | $\frac{1}{4}(5^{1/2} - 1)$ | $\frac{1}{2}\left(55 \cdot 5^{\frac{1}{2}} + 123\right)$ |

Table 1. Parameters of inelastic Butterworth filters

### 1. Photonic circuits with equal compound eigenfrequencies $\Omega_n$, equal interstate couplings $\delta_n$, and arbitrary couplings to waveguides $\xi_{1,1}$ and $\xi_{N,2}$

For lossless photonic circuits having eigenstates with equal compound eigenfrequencies $\Omega_n = \omega_n + \left(k - n + \frac{1}{2}\right)\omega_p = \Omega_1 = \omega_1 + \left(k - \frac{1}{2}\right)\omega_p$, equal losses $\gamma_n = \gamma$, equal coupling parameters $\delta_n = \delta$, and arbitrary couplings to waveguides $\xi_{1,1}$ and $\xi_{N,2}$, we find $\det(\Theta)$ and $M_{1,1}(\Theta)$ in Eq. (46):

$$\det(\Theta) = \left[\left(\omega - \omega_1 + \tfrac{i}{2}(\gamma + \xi_{1,1}^2)\right)\left(\omega - \omega_1 + \tfrac{i}{2}(\gamma + \xi_{N,2}^2)\right) - \delta^2\right]D_{N-2}$$
$$- \delta^2\left(\omega - \omega_1 + \tfrac{i}{2}(2\gamma + \xi_{1,1}^2 + \xi_{N,2}^2)\right)D_{N-3},$$
$$M_{1,1}(\Theta) = \left(\omega - \omega_1 + \tfrac{i}{2}(\gamma + \xi_{N,2}^2)\right)D_{N-2} - \delta^2 D_{N-3}, \qquad (50)$$
$$D_N = \frac{(\omega - \omega_1 + \tfrac{i}{2}\gamma)^N}{2^{N+1}R}\left[(1+R)^{N+1} - (1-R)^{N+1}\right], \quad R = \left(1 - \frac{4\delta}{\omega - \omega_1 + \tfrac{i}{2}\gamma}\right)^{1/2}.$$

It follows from Eqs. (46) and (50) that, for a losses circuit ($\gamma = 0$), there are only two possible conditions of CIRT (see Appendix A):

$$\delta = \frac{1}{2}\xi_{1,1}\xi_{N,2} \qquad (51)$$

$$\xi_{N,2}^2 = \xi_{1,1}^2 \qquad (52)$$

Similar to the elastic photonic circuits, lossless inelastic photonic circuits composed of $N$ states can be designed to be completely transparent at $N$ resonant frequencies. Here we show that, for different couplings to waveguides, the circuits described by Eq. (50) can still have $N - 1$ resonances exhibiting CIRT.

Similar to the elastic photonic circuits [13-15], lossless inelastic photonic circuits composed of $N$ states can be designed to be *completely transparent at $N - 1$ resonant frequencies*. This fact was discussed in Sections IV and V for the circuits composed of 2 and 3 states. Generally, we consider eigenstates with equal compound eigenfrequencies $\Omega_n = \omega_n + \left(k - n + \tfrac{1}{2}\right)\omega_p = \Omega_1 = \omega_1 + \left(k - \tfrac{1}{2}\right)\omega_p$, and equal interstate couplings $\delta_n = \delta$. Under the condition of Eq. (51), we find that the CIRT takes place at frequencies (see S6):

$$\omega_n^{(res)} = \omega_1 + 2\delta\cos\left(\frac{\pi n}{N}\right), \quad n = 1, 2, \dots N - 1. \qquad (53)$$

Plots (b1)-(b4) in Fig. 8 show representative inelastic transmission spectra for $N = 2, 3, 4$, and 5 calculated using Eqs. (46) and (50) at $\xi_{1,1}^2/\xi_{N,2}^2 = 4$ and $\delta/\xi_{N,2}^2 = 1$. Fig. 8(b5) shows the spectrum for $N = 5$ optimized to have a flatter behavior near the CIRT region and a minimum ratio $\gamma_n/\delta$. This spectrum corresponds to a smaller modulation-induced ratio $\delta/\xi_{N,2}^2 = 0.5$ which is easier to achieve experimentally.

### 2. *Photonic circuits with equal compound eigenfrequencies $\Omega_n$, equal interstate couplings $\delta_n$, and equal couplings to waveguides $\xi_{1,1} = \xi_{N,2}$*

A photonic circuit with equal compound eigenfrequencies $\Omega_n$ and equal interstate couplings $\delta_n$ is symmetric with respect to its center. Then, in the absence of coupling to the input-output waveguides, $\xi_{1,1} = \xi_{N,2} = 0$, each of the eigenstates of such a circuit will have the same symmetry or anti-symmetry. Due to this symmetry and provided that couplings to waveguides are equal, $\xi_{1,1} = \xi_{N,2}$, and small enough, Eqs. (46) and (50) describe such a circuit having $N$ resonances $\omega = \omega_n^{(res)}$, $n = 1, 2, \dots, N$, exhibiting CIRT that satisfy the equation (see Appendix A):

$$\frac{1}{R}\left\{\left[\left(\frac{1+R}{1-R}\right)^{N+1} - 1\right] + \frac{\xi_{11}^4}{4\delta^2}\left(\frac{1+R}{1-R}\right)\left[\left(\frac{1+R}{1-R}\right)^{N-1} - 1\right]\right\} = 0, \quad R = \sqrt{1 - \frac{4\delta^2}{\omega^2}}. \qquad (54)$$

For small $\xi_{N,2}^2/\delta \ll 1$, solutions of this equation are

$$\omega_n^{(res)} = \omega_1 + 2\delta \cos\left(\frac{\pi n}{N+1}\right), \quad n = 1, 2, \ldots N. \tag{55}$$

Plots (c1), (c2), (c3), and (c4) in Fig. 8 show representative inelastic transmission spectra for $N = 2, 3, 4$, and 5 calculated using Eqs. (46) and (50) at $\xi_{1,1} = \xi_{N,2}$ and $\delta/\xi_{N,2}^2 = 1$. Increasing the couplings to waveguides $\xi_{1,1} = \xi_{N,2}$ or, similarly, decreasing the interstate coupling $\delta$ beyond the thresholds depending on $N$ reduces the number of $\omega_n^{(res)}$ exhibiting CIRT. As an example, Fig. 8(c5) presents a transmission spectrum with flattered CITR at $\delta/\xi_{N,2}^2 = 0.407$ corresponding to minimum ratio $\gamma_n/\delta$.

### 3. Adjustment of eigenfrequencies by perturbation theory

Preliminary designed photonic circuit with known eigenstates $\Psi_m^{(0)}$ and eigenfrequencies $\omega_m^{(0)}$ can be adjusted by perturbation theory. Then, the requested shifts $\Delta\omega_m$ of $\omega_m^{(0)}$ are determined from Eqs. (14) and (15) through the introduced perturbation of refractive index $\Delta n_0(z)$. Assuming that photonic circuit consists of $N$ eigenstates, we can present $\Delta n_0(z)$ as a superposition of $N$ known functions $\Delta n_m^{(0)}(\mathbf{r})$, $m = 1, 2, \ldots, N$:

$$\Delta n_0(\mathbf{r}) = \sum_{m=1}^{N} a_k \Delta n_m^{(0)}(\mathbf{r}) \tag{56}$$

with unknown coefficients $a_k$. Substitution of Eq. (56) into Eqs. (14) and (15) yields:

$$\Delta\omega_m = \sum_{n=1}^{N} F_{m,n} a_n, \quad F_{m,n} = \frac{\omega}{n_0}\left\langle \Psi_n^{(0)} \left| \Delta n_m^{(0)} \right| \Psi_n^{(0)} \right\rangle. \tag{57}$$

From Eq. (57) we find coefficients $a_k$ as

$$a_k = \sum_{m=1}^{N} H_{k,m} \Delta\omega_m, \tag{58}$$

where matrix $H$ is the inverse of $F$, $H = F^{-1}$.

### 4. Determination of coupling parameters

We consider a photonic circuit modulated by refractive index $\Delta n_1(\mathbf{r})$ (see Eqs. (13)-(16)). The corresponding modulation induced couplings are defined by Eq. (16). Assuming that photonic circuit consists of $N$ eigenstates, we can present $\Delta n_1(z)$ as a superposition of $N-1$ known functions $\Delta n_{1,m}^{(0)}(z)$, $m = 1, 2, \ldots, N-1$:

$$\Delta n_1(\mathbf{r}) = \sum_{k=1}^{N-1} a_k \Delta n_{1,m}^{(0)}(\mathbf{r}) \tag{59}$$

with unknown coefficients $a_k$. Substitution of Eq. (59) into Eqs. (13)-(16) yields:

$$\delta_n = \sum_{m=1}^{N-1} F_{n,m} a_m, \quad F_{n,m} = \frac{\omega}{n_0}\left\langle \Psi_n^{(0)} \left| \Delta n_{1,m}^{(0)} \right| \Psi_{n+1}^{(0)} \right\rangle. \tag{60}$$

From Eq. (60) we find coefficients $a_k$ as

$$a_m = \sum_{n=1}^{N-1} H_{n,m}\delta_n, \tag{61}$$

where matrix $H$ is the inverse of $F$, $H = F^{-1}$. Thus, formally, the spatial distribution of the modulation can be determined for any predetermined couplings $\delta_n$. The major problem though consists in finding sufficiently smooth and relatively small function $\Delta n_1(\mathbf{r})$ possible to implement experimentally.

### D. Effect of losses and leakages

Fig. 8 compares the transmission spectra of lossless circuits exhibiting CIRT at resonant frequencies $\omega_n^{(res)}$ with the same circuits having material losses. Evidently, losses grow with the number of cascaded states required to arrive at the frequency conversion with increment $(N-1)\omega_p$. While the width of resonances decreases with decreasing of the interstate couplings $\delta_n$, and of couplings to waveguides $\xi_{1,1}$ and $\xi_{N,2}$, wider resonances in Fig. 8, are less affected by losses than narrower ones and therefore are preferable for applications of our interest. The reason of different losses for different $\omega_n^{(res)}$ can be explained by different spatial distribution of the corresponding resonant compound states described by Eq. (11). In the first order of perturbation theory, the effect of leakages from each eigenstate is similar to that of the material losses and can be combined in a total loss $\gamma_n^{(tot)}$ (see Eqs. (35) and (44) for $N = 2$ and $N = 3$ and Appendix B for $N = 3, 4, 5$ and $6$), so that

$$\gamma_n^{(tot)} = \gamma_n + \xi_{n,1}^2 + \xi_{n,2}^2, \quad 1 < n < N,$$
$$\gamma_1^{(tot)} = \gamma_1 + \xi_{1,2}^2, \quad \gamma_N^{(tot)} = \gamma_N + \xi_{N,1}^2. \tag{62}$$

For example, for a photonic circuit composed of $N = 4$ eigenstates, equal compound eigenfrequencies $\Omega_n = \omega_n + \left(k - n + \frac{1}{2}\right)\omega_p = \Omega_1 = \omega_1 + \left(k - \frac{1}{2}\right)\omega_p$, equal coupling parameters $\delta_n = \delta$, and arbitrary couplings to waveguides $\xi_{1,1}$ and $\xi_{N,2}$, Eq. (53) yields the resonant frequencies exhibiting CIRT as $\omega_1^{(res)} = \omega_1$ and $\omega_{2,3}^{(res)} = \omega_1 \pm \sqrt{2}\delta$. Then, as shown in Appendix B, in the first order of perturbation theory we have

$$\left|S_{1,8}(\omega_1^{(res)})\right|^2 \cong 1 - \left[\frac{1}{\xi_{1,1}^2}\left(\gamma_1^{(tot)} + \gamma_3^{(tot)}\right) + \frac{1}{\xi_{4,2}^2}\left(\gamma_2^{(tot)} + \gamma_4^{(tot)}\right)\right], \tag{63}$$

$$\left|S_{1,8}(\omega_{2,3}^{(res)})\right|^2 \cong 1 - \left[\frac{1}{\xi_{1,1}^2}\left(\gamma_1^{(tot)} + 2\gamma_2^{(tot)} + \gamma_3^{(tot)}\right) + \frac{1}{\xi_{4,2}^2}\left(\gamma_2^{(tot)} + 2\gamma_3^{(tot)} + \gamma_4^{(tot)}\right)\right] \tag{64}$$

where $\delta = \xi_{1,1}\xi_{4,2}/2$. It is seen that the contribution of losses and leakages is noticeably greater at resonances $\omega_{2,3}^{(res)}$ than at $\omega_1^{(res)}$, e.g., it is two times greater for equal losses, $\gamma_n^{(tot)} = \gamma_1^{(tot)}$. The latter result is in a reasonable agreement with the relation between transmission spectra shown in Fig. 8(b3). For larger $N$ this difference may become more significant. For example, at $N = 6$ the resonance frequencies found from Eq. (53) are $\omega_1^{(res)}$ (see Appendix B):

$$\left|S_{1,12}(\omega_1^{(res)})\right|^2 \cong 1 - \left[\frac{1}{\xi_{1,1}^2}\left(\gamma_1^{(tot)} + \gamma_3^{(tot)} + \gamma_5^{(tot)}\right) + \frac{1}{\xi_{6,2}^2}\left(\gamma_2^{(tot)} + \gamma_4^{(tot)} + \gamma_6^{(tot)}\right)\right], \tag{65}$$

$$\left|S_{1,12}(\omega_{2,3}^{(res)})\right|^2 \cong 1 - \left[\frac{1}{\xi_{1,1}^2}\left(\gamma_1^{(tot)} + \gamma_2^{(tot)} + \gamma_4^{(tot)} + \gamma_5^{(tot)}\right) + \frac{1}{\xi_{6,2}^2}\left(\gamma_2^{(tot)} + \gamma_3^{(tot)} + \gamma_5^{(tot)} + \gamma_6^{(tot)}\right)\right], \tag{66}$$

$$\left|S_{1,12}(\omega_{4,5}^{(res)})\right|^2 \cong 1 - \left[\frac{1}{\xi_{1,1}^2}\left(\gamma_1^{(tot)} + 3\gamma_2^{(tot)} + 4\gamma_3^{(tot)} + 3\gamma_4^{(tot)} + \gamma_5^{(tot)}\right) + \frac{1}{\xi_{6,2}^2}\left(\gamma_2^{(tot)} + 3\gamma_3^{(tot)} + 4\gamma_4^{(tot)} + 3\gamma_5^{(tot)} + \gamma_6^{(tot)}\right)\right], \quad (67)$$

so that, for equal total losses $\gamma_n^{(tot)} = \gamma_1^{(tot)}$, the attenuation of CIRT at $\omega_{4,5}^{(res)}$ is 4 times greater than at $\omega_1^{(res)}$. From Eqs. (63-67), we suggest that for $N$ resonant states and similar couplings to waveguides, $\xi_{1,1} \sim \xi_{N,2} \sim \xi$, and similar total losses, $\gamma_n^{(tot)} \sim \gamma^{(tot)}$, the estimate of the smallest deviation of the circuit's transparency from CIRT can be found as

$$\left|S_{1,12}(\omega_1^{(res)})\right|^2 \cong 1 - \Delta, \quad \Delta \sim \frac{N\gamma^{(tot)}}{\xi^2}. \quad (68)$$

### VII. MODELS OF PHOTONIC CIRCUITS WITH $N = 2$ RESONANT STATES

In this Section, we consider simplest models of low loss photonic circuits composed of $N = 2$ resonant states which may exhibit the CIRT. These models include two coupled microresonators (Fig. 9(a)), four coupled microresonators (Fig. 9(b)) and a SNAP bottle microresonator (Fig. 9(c)). The general theory of such photonic circuits is presented in Section IV. The transmission spectra of these circuits corresponding to equal couplings between the microresonators and waveguides and the relation between these coupling corresponding to a Butterworth filter replicating Figs. 8(a1) and (c1) are shown in Figs. 9(e1) and (e2). Possible qualitatively different spatial configurations of these photonic circuits are described as follows.

#### A. Coupled microresonators

To enable modulation induced inelastic transitions between states of a ring resonator, its FSR should be sufficiently small and, consequently, the resonator should have large mm-scale dimensions (see, e.g., [30]). Securing the inelastic transitions between only two states of such resonators is problematic since the spacing of their multiple eigenfrequencies becomes effectively equal with increasing the resonator dimensions. This problem can be solved by ring resonators fabricated with multiple mode waveguides which can possess pairs of eigenstates with close eigenfrequencies [29]. However, spatial separation of the output light with converted frequency in such resonators is problematic. Alternatively, this problem can be solved by a circuit of two coupled microresonators (Fig. 9(a)) considered in this Section. We discuss the design of such a circuit approaching CIRT and optimize the SDM to arrive at the maximum possible modulation induced coupling.

##### 1. Photonic circuits comprising two coupled microresonators

A photonic circuit consisting of two weakly coupled microresonators, MR$_1$ and MR$_2$, shown in Fig. 9(a) can act as a frequency beam splitter [30] and also can perform simultaneous spatial and frequency beam splitting. This circuit can exhibit close to complete frequency conversion with the limitation set by Eq. (35). Its resonant eigenstates, $\Psi_1(\mathbf{r})$ and $\Psi_2(\mathbf{r})$, are the linear combinations of two normalized states of uncoupled microresonators, $\Psi_1^{(0)}(\mathbf{r})$ and $\Psi_2^{(0)}(\mathbf{r})$. In order to minimize coupling $\xi_{12}$ of $\Psi_1(\mathbf{r})$ to waveguide WG$_2$ and coupling $\xi_{21}$ of $\Psi_2(\mathbf{r})$ to waveguide WG$_1$ (see Eq. (33)), we have to assume that the spacing between eigenfrequencies of uncoupled MR$_1$ and MR$_2$ $\omega_2^{(0)} - \omega_1^{(0)}$ is much greater than the original coupling between them $\delta_1^{(0)}$. Then the eigenfrequencies and eigenstates of the coupled ring microresonators are found from Eqs. (1)-(3) as:

$$\begin{pmatrix} \omega_1 \\ \omega_2 \end{pmatrix} \cong \begin{pmatrix} \omega_1^{(0)} - \varepsilon^2 \left( \omega_2^{(0)} - \omega_1^{(0)} \right) \\ \omega_1^{(0)} + \varepsilon^2 \left( \omega_2^{(0)} - \omega_1^{(0)} \right) \end{pmatrix},$$

$$\begin{pmatrix} \Psi_1(\mathbf{r}) \\ \Psi_2(\mathbf{r}) \end{pmatrix} \cong \mathbf{C} \begin{pmatrix} \Psi_1^{(0)}(\mathbf{r}) \\ \Psi_2^{(0)}(\mathbf{r}) \end{pmatrix}, \quad \mathbf{C} = \begin{pmatrix} 1 & \varepsilon^2 \\ -\varepsilon^2 & 1 \end{pmatrix}, \tag{69}$$

$$\varepsilon = \frac{\delta_1^{(0)}}{\omega_2^{(0)} - \omega_1^{(0)}} \cong \frac{\delta_1^{(0)}}{\omega_p} \ll 1.$$

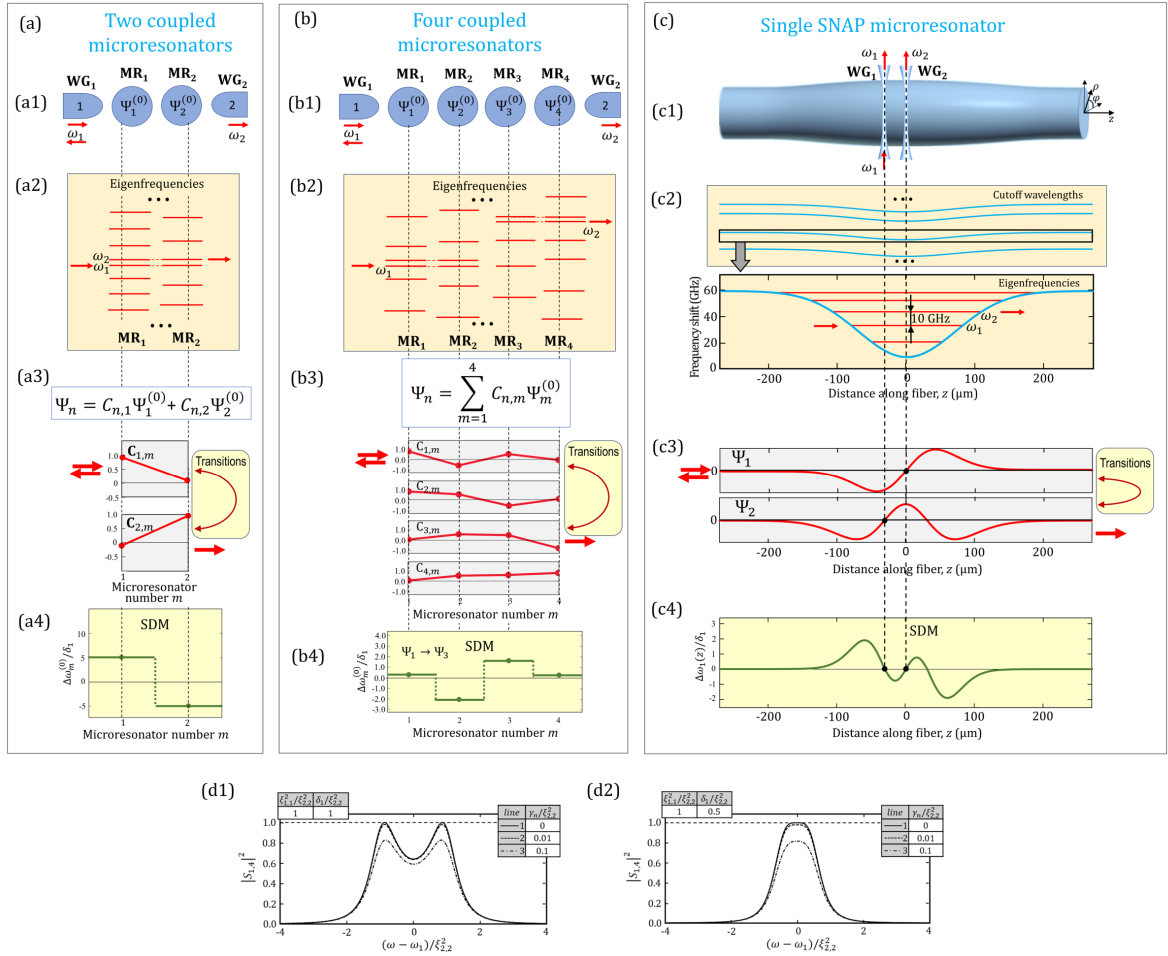

FIG. 9. Optimized photonic circuits enabling CIRT through two resonant states composed of (a) two coupled microresonators, (b) Four coupled microresonators, and (c) single SNAP bottle microresonator. (d1) Transmission through a photonic circuit having 2 eigenstates and equal modulation induced couplings between them and waveguides, $\delta_1 = \Delta \xi_{11} = \Delta \xi_{22}$. (d2) Transmission through a Butterworth photonic circuit having 2 eigenstates and equal modulation induced couplings to waveguides, $\Delta \xi_{11} = \Delta \xi_{22} = 2\delta_1$. Plots (d1) and (d2) are identical to plots (c1) and (a1) in Fig. 8.

Assuming that MR$_1$ and MR$_2$ are similar and taking into account that couplings $\xi_{nk}$ are proportional to $\langle \Psi_n(\mathbf{r}) | \Psi_{out}^{(n,k)}(\mathbf{r}) \rangle$, we find from Eq. (69) that $\xi_{12}/\xi_{11} = \xi_{21}/\xi_{22} = \varepsilon^2$. Consequently, the deviation of the maximum inelastic transmission power from unity defined by Eq. (35) caused by leakage into the output channels is $2\varepsilon^4$ and, thus, rapidly vanish with $\varepsilon$. Contrarily, the contribution of the remaining terms in Eq. (35) proportional to material losses $\gamma_n$ are inverse proportional to $\varepsilon^2$. Indeed, it follows from Eqs. (16) and (69) that the modulation induced coupling between states $\Psi_1(r)$ and $\Psi_2(r)$ is $\delta_1 \sim \varepsilon^2 \Delta n_1 \omega/n_0$, i.e., is proportional to $\varepsilon^2$. Setting $\gamma_1 = \gamma_2 = \gamma$, we find from Eq. (35) that the CIRT deviation from unity is $\Delta \sim 2\varepsilon^{-2} n_0/\Delta nQ + 2\varepsilon^4$. We minimize the latter expression over $\varepsilon$ and find $\Delta \sim 3 \cdot 2^{1/3}(n_0/(Q\Delta n))^{2/3}$ for $\varepsilon \sim (n_0/(2Q\Delta n))^{1/3}$. Assuming, as in the previous example, that $\xi_{11}^2 \sim \xi_{22}^2 \sim \delta_1$, $Q \lesssim 10^7$, and $\Delta n/n_0 \lesssim 10^{-4}$, we find $\varepsilon^2 \gtrsim 0.1$ and $\Delta \gtrsim 0.04$. From Eq. (69), the spatial distributions of states $\Psi_n(\mathbf{r})$ along microresonators are characterized by vectors $\mathbf{C}_n$, the $n^{\text{th}}$ columns of matrix $\mathbf{C} = \{C_{n,m}\}$. These distributions are shown in Fig. 9(a3) for $\varepsilon^2 = 0.1$.

In the absence of waveguide 2, the leakage channels are absent since then the inelastically transmitted light is reflected back into WG$_1$. Then, the frequency beam splitting (rather than both spatial and frequency beam splitting) can be realized. Assuming that in this case the ring resonators are symmetric with $\omega_2^{(0)} = \omega_1^{(0)}$, we have

$$\Psi_{1,2}(\mathbf{r}) = 2^{-1/2}\left(\Psi_1^{(0)}(\mathbf{r}) \pm \Psi_2^{(0)}(\mathbf{r})\right), \quad \omega_{1,2} \cong \omega_1^{(0)} \mp \delta_1^{(0)}. \tag{70}$$

Then we find from Eqs. (14)-(16) that $\delta_1 \sim \omega \Delta n/n_0$ and $\Delta \sim n_0/\Delta nQ$. For example, setting $\Delta n/n_0 \cong \delta_1/\omega \cong 1.5 \cdot 10^{-5}$, and $Q \cong 0.8 \cdot 10^6$, similar to parameters used in Ref. [30], we find for the maximum inelastic transmission power $1 - \Delta \cong 0.8$, in good agreement with experiment [30].

### 2. *Optimization of modulation induced coupling for the circuit of two coupled microresonators*

For coupled microresonators described by Eq. (69), the optimal spatial distribution of refractive index modulation $\Delta n_1(\mathbf{r})$ is found from Eqs. (14), (18), (69), and (70) as:

$$\Delta n_1(\mathbf{r}) = \frac{\delta_1 \omega}{\varepsilon^2 n_0} \frac{\left|\Psi_1^{(0)}(\mathbf{r})\right|^2 - \left|\Psi_2^{(0)}(\mathbf{r})\right|^2}{\left\langle\left(\Psi_1^{(0)}\right)^2 \middle| \left(\Psi_1^{(0)}\right)^2\right\rangle + \left\langle\left(\Psi_2^{(0)}\right)^2 \middle| \left(\Psi_2^{(0)}\right)^2\right\rangle}. \tag{71}$$

From the same equations, we find for the optimal induced frequency shifts at MR$_1$ and MR$_2$:

$$\Delta\omega_{1,n}^{(0)} = \frac{\delta_1}{2\varepsilon^2}\begin{cases} 1 & n=1 \\ -1 & n=2 \end{cases}. \tag{72}$$

It follows from Eq. (71), that at microresonator $n$ the distribution of $\Delta n_1(\mathbf{r})$ is proportional to $\left|\Psi_n^{(0)}(\mathbf{r})\right|^2$. In particular, for ring resonators, assuming that $\Delta n_1(\mathbf{r})$ is uniform within their waveguide cross-sections, we find that the optimal distribution of $\Delta n_1$ is uniform and has opposite signs in MR$_1$ and MR$_2$, in good agreement with the experimental design [30]. This result is noticeably different from the optimized nonuniform SDM in a single ring resonator designed for the optical frequency comb generation [40].

From Eq. (72), the values of $\Delta\omega_{1,n}^{(0)}$, and thus the required modulation amplitude, grow with the reduction of coupling between MR$_1$ and MR$_2$ as $\varepsilon^{-2}$. It is shown in the previous Section VIIA1 that $\varepsilon^2 = 0.1$ yields a quite small leakage of $2\varepsilon^4 = 0.02$. The bottom plot in Fig. 9(a4) shows the dependence of $\Delta\omega_{1,n}^{(0)}/\delta_1$ on microresonator number $n$ found for $\varepsilon^2 = 0.1$. From Eq. (72), the amplitude of this distribution is $\varepsilon^{-2}/2 = 5$ suggesting that the considered circuit can exhibit a close to CIRT provided that the required microresonator frequency shifts $\Delta\omega_{1,n}^{(0)}$ induced by modulation is at least five times greater that the resulting coupling $\delta_1$ between MR$_1$ and MR$_2$.

### 3. *A circuit of four coupled microresonators with small leakage and large spatial overlap between eigenstates*

As shown in the previous Section VIIA2, reduction of leakages for a circuit of two coupled microresonators necessarily leads to the enhancement of the amplitude of modulation required to arrive at CIRT realized with spatially different input and output waveguides. This problem can be solved by photonic circuits composed of more than two microresonators. As an example, here we consider a circuit of four microresonators, $MR_1$, $MR_2$, $MR_3$, and $MR_4$, shown in Fig. 9(b1). The values of eigenfrequencies of these microresonators and their couplings are defined by matrix $\mathbf{\Omega}^{(0)}$ from Eq. (1):

$$\mathbf{\Omega}^{(0)} = \begin{pmatrix} \omega_1^{(0)} - \delta_1^{(0)} & \delta_0^{(0)} & 0 & 0 \\ \delta_0^{(0)} & \omega_1^{(0)} & \delta_1^{(0)} & 0 \\ 0 & \delta_1^{(0)} & \omega_1^{(0)} & \delta_0^{(0)} \\ 0 & 0 & \delta_0 & \omega_1^{(0)} + \delta_1^{(0)} \end{pmatrix} \tag{73}$$

The idea behind the selected parameters of this matrix is as follows. We assume that coupling $\delta_0^{(0)}$ is small. Setting $\delta_0^{(0)} = 0$ we find the joint eigenfrequencies of $MR_2$ and $MR_3$ outlined in Eq. (73) as $\omega_1^{(0)} - \delta_1^{(0)}$ and $\omega_1^{(0)} + \delta_1^{(0)}$, which respectfully coincide with the eigenfrequencies of $MR_1$ and $MR_4$. Then, similar to consideration in previous Section VIIA2, we find that for small coupling $\delta_0^{(0)}$, penetration of joint state of $MR_2$ and $MR_3$ with frequency $\omega_1^{(0)} - \delta_1^{(0)}$ into waveguide $WG_2$ as well as penetration of joint state of $MR_2$ and $MR_3$ with frequency $\omega_1^{(0)} + \delta_1^{(0)}$ into waveguide $WG_1$ is small. In the first order in $\delta_0^{(0)}$, we find the eigenfrequencies $\omega_n$ of this circuit as

$$\begin{pmatrix} \omega_1 \\ \omega_2 \\ \omega_3 \\ \omega_4 \end{pmatrix} \cong \begin{pmatrix} \omega_1^{(0)} - \delta_1^{(0)} - 2^{-1/2} \delta_0^{(0)} \\ \omega_1^{(0)} - \delta_1^{(0)} + 2^{-1/2} \delta_0^{(0)} \\ \omega_1^{(0)} + \delta_1^{(0)} - 2^{-1/2} \delta_0^{(0)} \\ \omega_1^{(0)} + \delta_1^{(0)} + 2^{-1/2} \delta_0^{(0)} \end{pmatrix}, \quad \left(\frac{\delta_0^{(0)}}{\delta_1^{(0)}}\right)^2 \ll 1. \tag{74}$$

Respectfully, using Eqs. (2) and (3), we express normalized eigenstates $\Psi_n$ of this circuit through the normalized eigenstates $\Psi_n^{(0)}$ of the standing along microresonators as

$$\Psi_n(\mathbf{r}) = \sum_{m=1}^{4} C_{n,m} \Psi_m^{(0)}(\mathbf{r}), \quad n = 1, 2, 3, 4,. \tag{75}$$

where in the first order in $\delta_0^{(0)}$

$$\mathbf{C} = \begin{pmatrix} \frac{\sqrt{2}}{2} - \frac{\delta_0^{(0)}}{8\delta_1^{(0)}} & -\frac{1}{2} - \frac{\sqrt{2}\delta_0^{(0)}}{16\delta_1^{(0)}} & \frac{1}{2} - \frac{3\sqrt{2}\delta_0^{(0)}}{16\delta_1^{(0)}} & -\frac{\delta_0^{(0)}}{4\delta_1^{(0)}} \\ \frac{\sqrt{2}}{2} + \frac{\delta_0^{(0)}}{8\delta_1^{(0)}} & \frac{1}{2} - \frac{\sqrt{2}\delta_0^{(0)}}{16\delta_1^{(0)}} & -\frac{1}{2} - \frac{3\sqrt{2}\delta_0^{(0)}}{16\delta_1^{(0)}} & \frac{\delta_0^{(0)}}{4\delta_1^{(0)}} \\ \frac{\delta_0^{(0)}}{4\delta_1^{(0)}} & \frac{1}{2} + \frac{5\sqrt{2}\delta_0^{(0)}}{16\delta_1^{(0)}} & \frac{1}{2} + \frac{\sqrt{2}\delta_0^{(0)}}{16\delta_1^{(0)}} & -\frac{\sqrt{2}}{2} - \frac{3\delta_0^{(0)}}{8\delta_1^{(0)}} \\ \frac{\delta_0^{(0)}}{4\delta_1^{(0)}} & \frac{1}{2} - \frac{5\sqrt{2}\delta_0^{(0)}}{16\delta_1^{(0)}} & \frac{1}{2} - \frac{\sqrt{2}\delta_0^{(0)}}{16\delta_1^{(0)}} & \frac{\sqrt{2}}{2} - \frac{3\delta_0^{(0)}}{8\delta_1^{(0)}} \end{pmatrix}, \quad \left(\frac{\delta_0^{(0)}}{\delta_1^{(0)}}\right)^2 \ll 1. \tag{76}$$

Fig. 9(b3) shows the plots of $C_{n,m}$ for states $\Psi_n$ as a function of position $m$ at microresonator $MR_m$ for $\delta_0^{(0)}/\delta_1^{(0)} = 0.2$ when Eqs. (74) and (76) are found to have better than 1% accuracy. The relative leakages for these circuit parameters are found to be less than 1% as well. In this case, the circuit eigenfrequencies are:

$$\begin{pmatrix} \omega_1 \\ \omega_2 \\ \omega_3 \\ \omega_4 \end{pmatrix} \cong \begin{pmatrix} \omega_1^{(0)} - 1.15\delta_1^{(0)} \\ \omega_1^{(0)} - 0.87\delta_1^{(0)} \\ \omega_1^{(0)} + 0.87\delta_1^{(0)} \\ \omega_1^{(0)} + 1.15\delta_1^{(0)} \end{pmatrix}. \tag{77}$$

It is seen from this equation and Fig. 9(b3) that there are several pairs of eigenstates with minimal leakages and, at the same time, large spatial overlap, which can ensure CIRT with up-conversion by $\omega_p \cong 2.3\delta_1^{(0)}$, $\omega_p \cong 2.02\delta_1^{(0)}$, and $\omega_p \cong 1.74\delta_1^{(0)}$.

*4. Optimization of modulation induced coupling for the circuit of four coupled microresonators*

Optimization of the modulation induced coupling can be performed following Section IIIH similar to that considered for two coupled microresonators in Section VIIA2. As shown in Section IIIH the SDM with the smallest possible amplitude should be proportional to the product $\Psi_m(\mathbf{r})\Psi_n(\mathbf{r})$ of eigenstates $\Psi_m(\mathbf{r})$ and $\Psi_n(\mathbf{r})$ participating in resonant inelastic transmission. Then, from Eqs. (14), (25) and (75) we find for the optimal SDM:

$$\Delta n_1(\mathbf{r}) = \delta_1 \frac{n_0}{\omega} \frac{\sum_{k=1}^{4} C_{m,k} C_{n,k} \left|\Psi_k^{(0)}(\mathbf{r})\right|^2}{\sum_{k=1}^{4} C_{m,k}^2 C_{n,k}^2 \left\langle \left(\Psi_k^{(0)}\right)^2 \left|\left(\Psi_k^{(0)}\right)^2\right.\right\rangle}. \tag{78}$$

In a particular case, when $\Delta n_1(\mathbf{r})$ is constant and equal to $\Delta\omega_k^{(0)}$ within microresonator $k$, this equation is replaced by equation

$$\Delta\omega_k^{(0)} = \delta_1 \frac{C_{m,k} C_{n,k}}{\sum_{q=1}^{4} C_{m,q}^2 C_{n,q}^2} \tag{79}$$

generalizing Eq. (72). Fig. 9(b4) shows the plot of optimal $\Delta\omega_n^{(0)}$ as a function of microresonator number $n$ calculated for inelastic transition through states $\Psi_1(\mathbf{r})$ and $\Psi_3(\mathbf{r})$ defined by coefficients $C_{1,n}$ and $C_{3,n}$ plotted in Fig. 9(b3). Comparison of Fig 9(b4) and Fig. 9(a4) shows that, to generate similar coupling with similar small eigenstate leakages, the four microresonator circuit requires modulation with 2.5 smaller amplitude.

### B. A SNAP bottle microresonator

SNAP microresonators are microresonators fabricated at the surface of an optical fiber by nanoscale variation of its effective radius which can be introduced with an exceptional subangstrom precision by different methods including fabrication of stationary microresonators [33, 34, 42-44] and microresonators with tunable free spectral range [45-47]. A SNAP bottle microresonator considered in this Section having states $\Psi_1(\mathbf{r})$ and $\Psi_2(\mathbf{r})$ subject to modulation induced coupling is shown in Fig. 9(c). In contrast to circuits of two and four coupled microresonators considered in previous Sections (Figs. 9(a) and (b)), the leakages of states $\Psi_1(\mathbf{r})$ and $\Psi_2(\mathbf{r})$ to the input-output waveguides can be completely eliminated while their spatial overlap remain significant as required for the effective modulation induced coupling.

### 1. Design of a SNAP bottle microresonator

The localized eigenstates in SNAP microresonators are formed by whispering gallery modes (WGMs) whose frequencies $\omega$ are close to the cutoff frequencies $\omega_{lp}^{(cut)}(z)$ of the optical fiber numerated by their azimuthal and radial quantum numbers $l$ and $p$. Due to the dramatically small variation of the effective fiber radius, the spatial variation of eigenstates of a SNAP microresonator circuit can be separated in cylindrical coordinates $(z, \rho, \varphi)$ (Fig. 9(c1)) as $\Psi_{lpn}(\mathbf{r}) = \exp(il\varphi)\Phi_{lpn}(z)\Lambda_{lp}(\rho)$. Here $q$ is the axial quantum number. Under the condition that the eigenfrequencies $\omega_q$ of a SNAP resonator are close to the cutoff frequency, $|\omega_{lpn} - \omega_{lp}^{(cut)}(z)| \ll \omega_{lpn}$, the eigenstate axial dependence $\Phi_{lpn}(z)$ is determined by the Schrödinger-type wave equation [33],

$$\frac{\chi^2}{2}\frac{\partial^2 \Phi_n}{\partial z^2} + \left(\omega_n - \omega^{(cut)}(z)\right)\Phi_n = 0, \quad \chi = \frac{c}{n_0\sqrt{\omega_n}}, \tag{80}$$

where $c$ is the speed of light and $n_0$ is the fiber refractive index. Here and below, we consider the propagation of WGMs near a single cutoff frequency $\omega^{(cut)}(z)$ only and, for brevity, the quantum numbers $l$ and $p$ are omitted. The effect of parametric modulation of SNAP microresonators is enhanced due to their dramatically small variation of effective radius which is proportional to the variation of $\omega^{(cut)}(z)$ [33]. It can be achieved by pumping with modulated light at a frequency separated from frequency $\omega$, as well as using highly nonlinear and structured fibers. The corresponding modulation induced interstate couplings and eigenfrequency shifts are calculated from Eqs. (14)-(16).

A model of a silica SNAP bottle microresonator with the characteristic axial dimension of 200 μm considered here is shown in Fig. 9(c1). We present eigenstates $\Psi_1(\mathbf{r})$ and $\Psi_2(\mathbf{r})$ in the cylindrical coordinates as $\Psi_n(\mathbf{r}) = \exp(il\varphi)\Phi_n(z)\Lambda(\rho)$ where $\Phi_1(z)$ and $\Phi_2(z)$ correspond to the second and third axial states of this microresonator with eigenfrequency spacing equal to 10 GHz at telecommunication frequency $\omega$ =193 THz. In our model, *the leakages* reducing the inelastic transmission amplitude *are completely eliminated* by placing the input-output waveguide WG$_1$ at the node of $\Phi_2(z)$ and output WG$_2$ at the node of $\Phi_1(z)$ as shown in Fig. 9(c3). Therefore, the deviation of inelastic transmission amplitude from unity can be found using the noted above estimate $\Delta \sim n_0/(\Delta n Q)$. Thus, at a high effective index modulation amplitude $\Delta n/n_0 \sim 10^{-4}$ and microresonator Q-factor $Q \sim 10^8$ [48] the deviation of the inelastic transparency from CIRT can be as small as $\Delta \sim 10^{-4}$.

### 2. Optimization of modulation induced coupling

As in previous Sections, we optimize the modulation induced coupling $\Delta n_1(\mathbf{r})$ rescaling it to the frequency variation $\Delta\omega_1(\mathbf{r})$ using Eq. (71) and assuming that it depends on the axial coordinate of the SNAP microresonator only, $\Delta\omega_1(\mathbf{r}) = \Delta\omega_1(z)$. Using the same arguments as in Section VII, we look for the optimal modulation induced coupling in the form $\Delta\omega_1(z) = A \cdot \Phi_1(z)\Phi_2(z)$ where functions $\Phi_1(z)$ and $\Phi_2(z)$ are normalized. Then, Eq. (71) is transformed to:

$$\Delta\omega_1(z) = \delta_1 \frac{\Phi_1(z)\Phi_2(z)}{\left\langle (\Phi_1)^2 \big| (\Phi_1)^2 \right\rangle}. \tag{81}$$

In our numerical modeling, we use the calculated distributions $\Phi_1(z)$ and $\Phi_2(z)$ shown in Fig. 9(c3). The resulting figure of merit spatial dependence $\Delta\omega_1(z)/\delta_1$ is shown in Fig. 9(c4). It is seen that the amplitude of optimized $\Delta\omega_1(z)$ is reasonably comparable with the value of $\delta_1$.

## VIII.   MODELS OF PHOTONIC CIRCUITS WITH $N > 2$ RESONANT STATES

In this Section, we consider realizations of photonic circuits enabling CIRT through $N > 2$ resonant eigenstates. Fig. 3(c) shows the general configuration of compound states $\Sigma_1(\mathbf{r}), \Sigma_2(\mathbf{r}), \ldots, \Sigma_N(\mathbf{r})$ and Figs. 10(a), (b), (c), and (d) illustrate possible spatial distributions of light described by eigenstates $\Psi_1(\mathbf{r}), \Psi_2(\mathbf{r}), \ldots, \Psi_N(\mathbf{r})$ corresponding to these compound states. These functions are assumed to have eigenfrequencies $\omega_1, \omega_2, \ldots, \omega_N$ with accurately constant FSR $\omega_{n+1} - \omega_n \cong \Delta\omega$. We suggest that these $N$ states form a closed group where the interstate transitions induced by parametric modulation with frequency $\omega_p \cong \Delta\omega$ are possible. Examples of such groups include a racetrack microresonator (Fig. 2(b)) a Kac series of coupled microresonators (Fig. 10(a)), a SNAP semi-parabolic microresonator (Fig. 10(b)), coupled microresonators with linearly shifted eigenfrequencies (Fig. 10(c)), and a tilted series of coupled SNAP microresonators (Fig. 10(d)) are considered below.

### A. Photonic circuits incorporating a racetrack microresonator

A possible configuration of a photonic circuit similar to that demonstrated in Ref. [30] is illustrated in Fig. 2(b). This photonic circuit includes a racetrack (ring) resonator MR$_1$ which dimensions are sufficiently large to have the accurately constant FSR $\Delta\omega$ for a series of consequent eigenfrequencies, $\ldots, \omega_0, \omega_1, \ldots, \omega_N, \omega_{N+1}, \ldots$. In Fig. 2(b), the design of Ref. [30] is complemented by the output waveguide 2, WG$_2$, to enable the simultaneous spatial and frequency beam splitting in addition to the frequency beam splitting considered in [30]. To interrupt the FSR uniformity of this series keeping only $N$ of them, the authors of Ref. [30] added the second smaller microresonator MR$_2$ coupled to MR$_1$ and having FSR equal to $(N+2)\Delta\omega$. The MR$_2$ eigenfrequencies $\omega'_m$ and $\omega'_{m+1}$ coincided with $\omega_0$ and $\omega_{N+1}$, respectively (in Fig. 2(b), we have $N=4$). Coupling of MR$_1$ with MR$_2$ splits and shifts $\omega_0$ and $\omega_{N+1}$ and, thus, bound the modulation induced transitions within the series $\omega_1, \ldots, \omega_N$. To avoid leakages to the input-output waveguides, we also have to ensure that, in the selected group $\Psi_1(\mathbf{r}), \Psi_2(\mathbf{r}), \ldots, \Psi_N(\mathbf{r})$, the state $\Psi_1(\mathbf{r})$ is the only state coupled to WG$_1$, and the state $\Psi_N(\boldsymbol{r})$ is the only state coupled to WG$_2$. To this end, following the idea of Ref. [30], we add microresonators MR$_3$ and MR$_4$ which indirectly couple MR$_1$ to WG$_1$ and MR$_1$ to WG$_2$, respectively, and have eigenfrequencies $\omega'_1$ and $\omega'_2$ coinciding with $\omega_1$ and $\omega_N$. This approach allows us to spatially separate the intermediate states from the waveguides and make the leakages from them small.

#### *1. Conditions of CIRT*

In the experiment [30], modulation of MR$_1$ was introduced by capacitors adjacent to the MR$_1$ waveguide. Then, the modulation induced coupling between intermediate states $\Psi_1(\mathbf{r}), \Psi_2(\mathbf{r}), \ldots, \Psi_N(\mathbf{r})$ localized inside MR$_1$ are close to each other, $\delta_n = \delta_1$, while couplings between $\Psi_1(\mathbf{r})$ and $\Psi_2(\mathbf{r})$ and between $\Psi_{N-1}(\mathbf{r})$ and $\Psi_N(\mathbf{r})$ can be made equal to $\delta$ by the appropriate design of SDM. For equal couplings to waveguides, $\xi_{1,1} = \xi_{N,2}$ (as in the case of experiment [30] where WG$_1$ and WG$_2$ coincide), the CIRT of the photonic circuit is described by the theory of Section VI and illustrated in Figs. 8(b1)-(b5). For small material losses and leakages with $\gamma_n^{(tot)} \sim 0.01$ GHz, which requires $Q \gtrsim 10^7$, and couplings $\delta \sim \xi_{1,1}^2 \sim \xi_{N,2}^2 \sim 1$ GHz, it follows from Eq. (67) that the deviation from CIRT can be as small as $\Delta \sim 0.01 N$. However, for characteristic $Q \sim 10^6$ [30] the smallest deviation is estimated as $\Delta \sim 0.1 N$.

In order to have the FSR $\Delta\omega$ of MR$_1$ coinciding with the experimentally achievable modulation frequencies $\omega_p \lesssim 10$ GHz, the characteristic dimensions of this resonator should be $R \gtrsim c/n_0\Delta\omega \sim 1$ cm. We will see in the following Sections that dimensions of photonic circuit can be significantly reduced by employing coupled microresonator circuits and SNAP microresonators with equally spaced eigenfrequencies.

#### *2. Optimization of modulation induced coupling*

Here we suggest that the modulation is applied to the central ring resonator MR$_1$ only (Fig. 2(b1)). We consider the eigenfunctions of this resonator with the same transverse quantum numbers and different axial quantum numbers $n$ expressed as $\Psi_n(\mathbf{r}) = L^{-1/2}\exp(2\pi i n z/L)\Lambda(x,y)$, where $L$ is the ring circumference length and coordinates $z$ and $(x,y)$ are the axial and transverse coordinates, respectfully. Function $\Lambda(x,y)$ is assumed normalized, so that $\Psi_n(\mathbf{r})$ is normalized as well. The problem of optimization of the SDM in this case was solved in Ref. [40] under the assumption that this distribution is independent of transverse coordinates $(x, y)$. Here we assume that all the modulation induced couplings between adjacent states $\Psi_n(\mathbf{r})$ and $\Psi_{n+1}(\mathbf{r})$ are the same, $\delta_n = \delta_{10}$. Then, as follows from Ref. [40] (see Section 5 and Appendices J and K of this paper), the optimized SDM which maximizes $|\delta_{10}|$ is $\Delta n_1(z) = A \cdot \cos((2\pi z/L) + \varphi_0)$ with arbitrary phase $\varphi_0$. As previously, it is instructive to introduce the SDM in frequency units, $\Delta\omega_1(z) = (\omega/n_0)\Delta n_1(z)$ so that the dimensionless ratio $\Delta\omega_1(z)/\delta_{10}$ may serve as a figure of merit of the modulation efficiency. Then we have from Eq. (16)

$$\Delta\omega_1(z) = 2\delta_{10} \cos\left(\frac{2\pi z}{L} + \varphi_0\right) \tag{82}$$

Thus, the maximum of the required optimized amplitude of modulation is only two times greater than $\delta_{10}$.

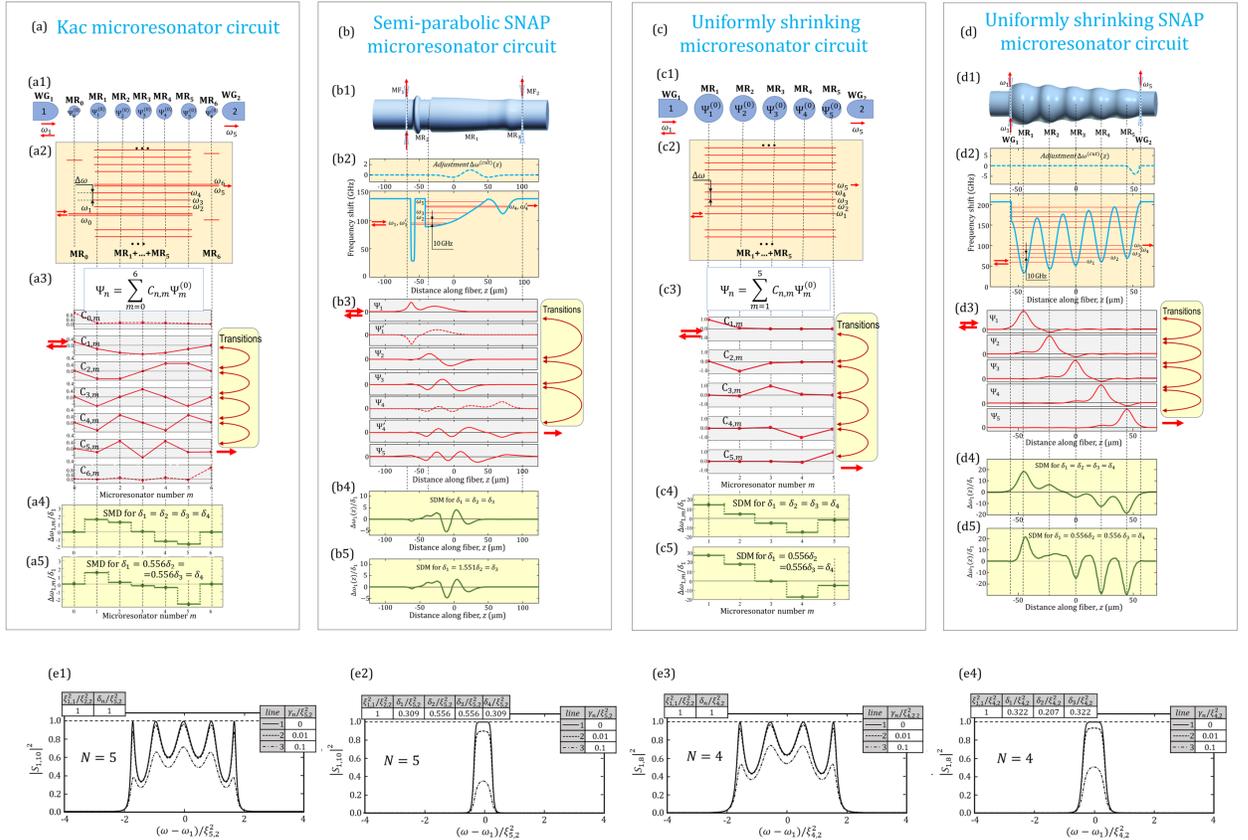

FIG. 10. Optimized photonic circuits enabling CIRT through multiple resonant states. (a) Kac microresonator circuit. (b) Semi-parabolic SNAP microresonator circuit. (c) Uniformly shrinking microresonator circuit. (d) Uniformly shrinking SNAP microresonator circuit. (e1) and (e3) Transmission through a photonic circuit having 5 and 4 eigenstates and equal modulation induced couplings between them and waveguides, $\delta_n = \Delta\xi_{11} = \Delta\xi_{22}$. (e2) and (e4) Transmission through a Butterworth photonic circuit having 5 and 4 eigenstates and equal modulation induced couplings to waveguides, $\Delta\xi_{11} = \Delta\xi_{22}$. Plots (e1), (e2), (e3), and (e4) are identical to plots (c4), (a4), (c3), and (a3) in Fig. 8, respectively.

## B. A Kac microresonator circuit

An alternative to a relatively large racetrack (ring) resonator with locally uniform FSR is a circuit of much smaller coupled microresonators which possesses a series of equally spaced eigenfrequencies illustrated in Fig. 10(a1). These can be ring, photonic crystal, as well as SNAP microresonators. An example of such circuit is a circuit of equal microresonators with varying coupling parameters $\delta_n^{(0)} = \frac{1}{2}\Delta\omega\sqrt{n(N-n)/(N-1)}$ (Kac microresonator circuit) which has $N$ eigenfrequencies with equidistant spacing $\Delta\omega$ [49-51] and, in addition, microscopic dimensions. In particular, a class of microresonator circuits with constant FSR composed of states with eigenfrequencies $\omega_n$ and couplings $\delta_n$ varying with $n$ can be determined in the semiclassical approximation [52]. Generally, the eigenstates of such circuits are delocalized, and the intermediate states may leak into the input-output waveguides. To suppress this leakage, two additional microresonators placed between the photonic circuit and waveguides (MR$_1$ and MR$_6$ in Fig. 10(a1), similar to MR$_3$ and MR$_4$ in Fig. 9(b1)) can be added.

### 1. Design of the microresonator circuit

Figs. 10(a3), (a4), and (a5) present the results of modelling of a photonic circuit illustrated in Fig. 10(a1), which includes a Kac series of $N = 5$ microresonators MR$_1$, ..., MR$_5$ coupled to waveguides WG$_1$ and WG$_2$ through microresonators MR$_0$ and MR$_6$ to suppress leakages. The eigenstates of this circuit are $\Psi_0(\mathbf{r}), \Psi_1(\mathbf{r})$, ..., $\Psi_6(\mathbf{r})$ among which only five $\Psi_1(\mathbf{r}), \Psi_2(\mathbf{r})$, ..., $\Psi_5(\mathbf{r})$ are designed to have equally spaced eigenfrequencies. Microresonators MR$_1$, ..., MR$_5$ standing along possess eigenstates $\Psi_1^{(0)}(\mathbf{r}), ..., \Psi_5^{(0)}(\mathbf{r})$ having equal eigenfrequencies $\omega_n = \omega_1$. Microresonators MF$_0$ and MF$_6$ have different eigenfrequencies which are determined by the condition of suitable coupling of state $\Psi_1(\mathbf{r})$ to waveguide WG$_1$ and state $\Psi_5(\mathbf{r})$ to waveguide WG$_2$ and smallest possible leakages. Consequently, Matrix $\boldsymbol{\Omega}^{(0)}$ in Eq. (1) is chosen as

$$\boldsymbol{\Omega}^{(0)} = \begin{pmatrix} \omega_1 - 2.05\delta_1^{(0)} & 0.05\delta_1^{(0)} & 0 & 0 & 0 & 0 & 0 \\ 0.05\delta_1^{(0)} & \omega_1 & \delta_1^{(0)} & 0 & 0 & 0 & 0 \\ 0 & \delta_1^{(0)} & \omega_1 & \delta_2^{(0)} & 0 & 0 & 0 \\ 0 & 0 & \delta_2^{(0)} & \omega_1 & \delta_3^{(0)} & 0 & 0 \\ 0 & 0 & 0 & \delta_3^{(0)} & \omega_1 & \delta_4^{(0)} & 0 \\ 0 & 0 & 0 & 0 & \delta_4^{(0)} & \omega_1 & 0.05\delta_1^{(0)} \\ 0 & 0 & 0 & 0 & 0 & 0.05\delta_1^{(0)} & \omega_1 + 2.05\delta_1^{(0)} \end{pmatrix}, \quad \delta_n^{(0)} = \frac{\Delta\omega}{2}\sqrt{n(5-n)}. \quad (83)$$

Assumed uncoupled from MF$_0$ and MF$_6$, the Kac circuit, which is determined by the $5 \times 5$ matrix outlined by a rectangle in Eq. (83), has five eigenfrequencies $\omega_1 - 2\Delta\omega$, $\omega_1 - \Delta\omega$, $\omega_1$, $\omega_1 + \Delta\omega$, and $\omega_1 + 2\Delta\omega$ with FSR equal to $\Delta\omega$. The eigenfrequencies $\omega_0 = \omega_1 - 2.05\Delta\omega$ and $\omega_6 = \omega_1 + 2.05\Delta\omega$ of microresonators MF$_0$ and MF$_6$ are chosen close to the edge eigenfrequencies $\omega_1 - 2\Delta\omega$ and $\omega_1 + 2\Delta\omega$ of the Kac circuit, which are supposed to be close to the input and output frequencies $\omega$ and $\omega + 4\omega_p$ of the designed photonic circuit. The eigenstates $\Psi_n(\mathbf{r})$ of this photonic circuit uncoupled from waveguides are determined from Eq. (2) as

$$\Psi_n(\mathbf{r}) = \sum_{m=0}^{N-1} C_{n,m} \Psi_n^{(0)}(\mathbf{r}), \quad n = 0,1,...6. \quad (84)$$

The spatial distributions of states $\Psi_n(\mathbf{r})$ along microresonators, which are characterized by vectors $\mathbf{C}_n$ (the $n^{\text{th}}$ columns of matrix $\mathbf{C} = \{C_{n,m}\}$), are plotted in Fig. 10(a3). The spacings between eigenfrequencies of the designed circuit indicated in Fig. 10(a2) show that the addition of microresonators MF$_0$ and MF$_6$ changes their constant FSR by less than 0.3%, while the separation between eigenfrequencies of MF$_0$ and MF$_1$ as well as MF$_5$ and MF$_6$ equal to $0.057\Delta\omega$ is sufficient to suppress the unwanted elastic and inelastic transitions to states $\Psi_0(\mathbf{r})$ and $\Psi_6(\mathbf{r})$. Comparison of coefficients $C_{n,0}$ with $C_{1,0}$ and $C_{n,6}$ with $C_{5,6}$ show that the relative leakages of states are around 1% or less.

### 2. *Optimization of modulation induced coupling*

We assume here that the modulation is introduced inside microresonators MF$_1$, ..., MF$_5$ only. Then, the modulation-induced effective frequency spatial distribution $\Delta\omega_1(r)$ (see Eq. (15)) is a sum of its parts localized in each of these microresonators,

$$\Delta\omega_1(\mathbf{r}) = \sum_{n=1}^{5} \Delta\omega_{1,n}^{(0)}(\mathbf{r}). \tag{85}$$

Eq. (85) allows us to optimize the modulation induced coupling as follows. We introduce the partial frequency shifts defined as

$$\Delta\omega_{1,n} = \left\langle \Psi_n^{(0)} \left| \Delta\omega_{1,n}^{(0)} \right| \Psi_n^{(0)} \right\rangle. \tag{86}$$

Then, from Eqs. (16), (85), and (86), the modulation-induced couplings between states $\Psi_n$ are determined as

$$\delta_n = \sum_{m=1}^{5} C_{n,m} C_{n+1,m} \Delta\omega_{1,m}^{(0)}, \quad n = 1,2,3,4. \tag{87}$$

For predetermined couplings $\delta_n$, Eq. (87) has 4 equations for 5 unknown parameters $\Delta\omega_{1,m}$. We determine all these parameters by the additional condition of the smallest possible magnitudes of $\Delta\omega_{1,m}$, which minimize the sum $\sum_{m=1}^{5} |\Delta\omega_{1,m}|^2$.

The plot of $\Delta\omega_{1,m}$ as a function of microresonator number $m$ determined under the condition of equal modulation induced couplings $\delta_1 = \delta_2 = \delta_3 = \delta_4$, which corresponds to CIRT with the transmission spectrum shown in Fig. 10(e1) (identical to that in Fig. 8(c4)), is presented in Fig. 10(a4). Due to the stronger spatial overlap of eigenstates $\Psi_n(\mathbf{r})$ than that for two weakly coupled states considered in Section VIIA (Fig. 9(a) and (b)), in spite of a larger number of states and additional conditions to $\delta_n$, the minimization of modulation magnitude is now performed more effectively and the magnitudes of all $\Delta\omega_{1,m}$ are noticeably closer to $\delta_1$ than in Fig. 9(b). From Fig. 10(a4), modulation induced coupling $\delta_2^{(0)} \sim 1$ GHz can be introduced by the frequency modulation $\Delta\omega_{1,m}$ of less than 2 GHz amplitude, a realistic value for lithium niobate waveguides [30]. Provided that $\gamma_n \sim 20$ MHz (corresponding to the microresonator Q-factor $\sim 10^7$ at $\omega \sim 200$ THz), and $\delta_n \sim (\xi_{01})^2 \sim (\xi'_{62})^2 \sim 1$ GHz, we estimate from Eq. (68) the smallest possible deviation of inelastic resonant transparency of the designed photonic circuit from CIRT as $\Delta \sim 0.05$.

Complementary, Fig. 10(a5) shows the optimized dependency of $\Delta\omega_{1,m}$ on the resonator number $m$ for a Butterworth filter composed of 5 microresonators. The corresponding CIRT spectrum is shown in Fig. 10(e2) (identical to that in Fig. 8(a4)). The smallest possible deviation of inelastic resonant transparency of the designed photonic circuit from CIRT in this case remains similar to the previous case with $\Delta \sim 0.05$.

### C. A photonic circuit comprising a SNAP semi-parabolic microresonator

As another alternative, $N$ eigenstates with constant FSR can be also realized in a SNAP bottle microresonator (see e.g., [33, 53, 54]). A photonic circuit comprising a semi-parabolic microresonator which potentially enables CERT is illustrated in Fig. 10(b1). The blue curve in Fig. 10(b2) shows the cutoff frequency $\omega^{(cut)}(z)$ of such a photonic circuit having $N = 4$ eigenstates with equally separated eigenfrequencies designed using Eq. (80). The FSR of these states is $\Delta\omega \cong 10$ GHz and the circuit length along the fiber axis is less than 200 μm. In addition to the semi-parabolic microresonator MR$_1$, this circuit includes two microresonators, MR$_2$ and MR$_3$, which act similarly to the auxiliary microresonators MR$_3$ and MR$_4$ in Fig. 2(b) and MR$_0$ and MR$_6$ in Fig. 10(a1), ensuring the suppression of leakages into the waveguides.

### 1. Design of the microresonator circuit

In our design, we first introduce a standing along semi-parabolic microresonator MR$_1$ and microresonators MR$_2$ and MR$_3$ having single eigenfrequencies coinciding, respectfully, with $\omega_1$ and $\omega_4$ of MR$_1$. The FSR of this microresonator is close to constant with small deviations caused by its finite shape (in comparison to infinite semi-parabolic microresonator which has exactly constant FSR). Moving MR$_2$ and MR$_3$ towards MR$_1$ splits $\omega_1$ and $\omega_4$ into eigenfrequencies $\omega_1$, $\omega_1'$ and $\omega_4$, $\omega_4'$ (Fig. 10(b2)). Simultaneously, coupling to MR$_2$ and MR$_3$ perturbs the eigenfrequencies of MR$_1$. To compensate for this perturbation of MR$_1$, we design the parabolic cutoff frequency variation by $\Delta\omega^{(cut)}(z)$ using the perturbation theory of Section VIC3 and arrive at equally spaced eigenfrequencies $\omega_1$, $\omega_2$, $\omega_3$, and $\omega_4'$. In particular, using Eq. (15) and the rescaling relation $\Delta n(z) = n_0 \Delta\omega(z)/\omega_0$, we present the required perturbation of the cutoff frequency variation as a linear combination of harmonic functions, $\Delta\omega^{(cut)}(z) = \sum_{m=1}^{N} a_m \cos(mbz)$ (see Appendix F). We find coefficients $a_m$ in the latter expression using Eq. (56)-(58) targeting equal separation between eigenfrequencies $\omega_1$, $\omega_2$, $\omega_3$, and $\omega_4'$. Finally, we perform fine adjustment of these frequencies by small displacements of MR$_2$ and MR$_3$. The determined profile of $\Delta\omega^{(cut)}(z)$ is shown in Fig. 10(b2).

Figs. 10(b2) and (b3) show the eigenfrequencies and spatial distribution of all seven eigenstates of the designed photonic circuit. The achieved FSR between eigenfrequencies $\omega_1, \omega_2, \omega_3$, and $\omega_4'$ is equal to 10 GHz with better than 0.2% accuracy, while the relative leakages from the states not directly adjacent to waveguides 1 and 2 are around 1% or less. We suggest that the leakages as well as FSR deviations of 9 MHz and 18 MHz from 10 GHz indicated in Fig. 10(b2) can be further reduced by a more accurate optimization of the model. Furthermore, provided the material losses $\gamma_n$ are small enough, these deviations can be compensated by the adjustment of modulation-induced interstate couplings similar to the adjustment performed in Appendix G for a photonic circuit comprised of $N = 3$ eigenstates. It is seen from Fig. 10(b3) that the eigenfrequencies of states $\Psi_1'(z)$, $\Psi_4(z)$ and $\Psi_5(z)$ are shifted from the sequence of equally spaced sequence $\omega_1, \omega_2, \omega_3$, and $\omega_4'$ by more than 0.5 GHz. For this reason, modulation induced coupling to these states for sufficiently small material losses can be ignored.

### 2. Optimization of modulation induced coupling

The theory presented in Section IIIH allows us to determine the optimized SDM for given coupling parameters $\delta_n$. We consider a SNAP photonic circuit modulated by refractive index $\Delta n_1(z)$ (see Eqs. (13)-(15)). The corresponding modulation induced cutoff frequency variation $\Delta\omega_1(z)$ is found by rescaling relation $\Delta\omega_1(z) = \omega_0 \Delta n_1(z)/n_0$.

As an example, we determine here $\Delta\omega_1(z)$ securing equal couplings $\delta_1 = \delta_2 = \delta_3$ corresponding to the CIRT spectrum shown in Fig. 10(e3) replicating Fig. 8(c3). We present $\Delta\omega_1(z)$ as a linear combination of basic functions $\Delta\omega_{1,k}^{(0)}(z) = \Delta\omega_0 \exp(-(z-z_0)^2/(v^2 k))$ and optimize parameters $z_0$ and $v$ to arrive at the smallest possible variation magnitude for a fixed value of $\Delta\omega_0$. The determined spatial distribution of $\Delta\omega_1(z)$ is shown in Fig. 10(b4). It is seen that the magnitude of $\Delta\omega_1(z)$ is still 10 times greater than $\delta_n$, i.e., for the introduction of coupling between states $\Psi_1, \Psi_2, \Psi_3$, and $\Psi_4'$ equal to $\delta_n = 100$ MHz we have to introduce the modulation with an amplitude of ~1 GHz. We suggest that employing a larger number of more flexible basic functions will allow to reduce the magnitude of $\Delta\omega_1(z)$ further. Provided that $\gamma_n \sim 2$ MHz (corresponding to the microresonator Q-factor ~ $10^8$ at $\omega \sim 200$ THz), and $\delta_n \sim (\xi_{11})^2 \sim (\xi_{42}')^2 \sim 100$ MHz, we estimate from Eq. (68) the smallest possible deviation of inelastic resonant transparency of the designed photonic circuit from CIRT as $\Delta \sim 0.05$.

As another example, Fig. 10(b5) shows the optimized cutoff frequency variation $\Delta\omega^{(cut)}(z)$ for the case when the modulation induced interstate couplings of the designed SNAP resonator circuit realize a Butterworth filter with the CIRT spectrum shown in Fig. 10(e4), identical to that in Fig. 8(a3). It is seen that this variation is similar in behavior and amplitude as in Fig. 10(b4) of the previous case.

### D. Photonic circuits incorporating coupled microresonators with linearly shifted eigenfrequencies

Strong spatial overlap of states $\Psi_n$ with equally spaced eigenfrequencies $\omega_n$ allows to achieve their effective inelastic resonant coupling at a minimum modulation amplitude. On the other hand, spatial expansion of these states leads to their leakages into input-output waveguides. As proposed in Ref. [30] and exploited in the previous Section VIIIC, this problem can be solved by two auxiliary microresonators which resonantly couple the first and last states, $\Psi_1$ and $\Psi_N$, respectively to the input and output waveguides and, at the same time, spatially separate other states from these waveguides. An alternative solution to this problem, though at the expense of reduction of spatial overlap between states $\Psi_n$, is proposed below. We design circuits of microresonators with varying dimensions, uniform FSR, and negligible leakages to waveguides without auxiliary microresonators.

### 1. Design of photonic circuits incorporating series of microresonators with uniformly shifted eigenfrequencies

Similar to the case of two weakly coupled microresonators considered in Section VIIA1, we can design a circuit of $N$ microresonators with couplings $\delta_n^{(0)}$ that are much smaller than a constant spacing of their eigenfrequencies $\Delta\omega = \left|\omega_{n+1}^{(0)} - \omega_n^{(0)}\right| \cong \omega_p$. This allows us to further simplify the design of coupled microresonator circuit illustrated in Figs. 2(d) and 10(c1).

To illustrate the idea of our design, we consider a circuit of 5 weakly coupled microresonators having $\delta_n^{(0)} \ll \left|\omega_{n+1}^{(0)} - \omega_n^{(0)}\right|$. We show that the FSR $\Delta\omega$ of the circuit can be made uniform by adjusting the eigenfrequencies of individual microresonators. In the example considered here, we assume that all couplings are equal, $\delta_n^{(0)} = \delta_1^{(0)}$, and make the FSR accurately uniform by setting the eigenfrequencies equal to $\omega_1 + n\Delta\omega$ for $n = 2,3,4$ and equal to $\omega_1 + \varepsilon_1$ and $\omega_1 + N\Delta\omega + \varepsilon_N$ for $n = 1$ and $n = N$, respectively. The eigenfrequencies of the circuit are defined from the equation

$$\det\left(\omega \mathbf{I} - \mathbf{\Omega}^{(0)}\right) = 0 . \tag{88}$$

Here $\mathbf{I}$ is the $5 \times 5$ unity matrix and

$$\mathbf{\Omega}^{(0)} = \begin{pmatrix} \omega_1^{(0)} + \varepsilon_1 & \delta_1^{(0)} & 0 & 0 & 0 \\ \delta_1^{(0)} & \omega_1^{(0)} + \Delta\omega & \delta_1^{(0)} & 0 & 0 \\ 0 & \delta_1^{(0)} & \omega_1^{(0)} + 2\Delta\omega & \delta_1^{(0)} & 0 \\ 0 & 0 & \delta_1^{(0)} & \omega_1^{(0)} + 3\Delta\omega & \delta_1^{(0)} \\ 0 & 0 & 0 & \delta_1^{(0)} & \omega_1^{(0)} + 4\Delta\omega + \varepsilon_5 \end{pmatrix}. \tag{89}$$

Expanding the determinant in Eq. (88) in series over $\delta_1^{(0)}$, $\varepsilon_1$, and $\varepsilon_N$ we find that the FSR of eigenfrequencies of this circuit have a constant FSR equal to $\Delta\omega$ with the accuracy $O\left(\left(\delta_1^{(0)}/\Delta\omega\right)^4\right)$ if

$$\varepsilon_1 = \frac{\left(\delta_1^{(0)}\right)^2}{\Delta\omega}, \quad \varepsilon_N = -\varepsilon_1. \tag{90}$$

For relatively week coupling $\delta_1^{(0)} = 0.1\Delta\omega$ and parameters $\varepsilon_1$ and $\varepsilon_N$ defined by Eq. (90), we find that, indeed, the eigenfrequencies of the designed photonic circuit have an accurately constant FSR:

$$\begin{pmatrix} \omega_1 \\ \omega_2 \\ \omega_3 \\ \omega_4 \\ \omega_5 \end{pmatrix} = \begin{pmatrix} \omega_1^{(0)} \\ \omega_1^{(0)} \\ \omega_1^{(0)} \\ \omega_1^{(0)} \\ \omega_1^{(0)} \end{pmatrix} + \begin{pmatrix} 0 \\ 1 \\ 2 \\ 3 \\ 4 \end{pmatrix} \Delta\omega + 10^{-5} \begin{pmatrix} 4.98 \\ -4.97 \\ 0 \\ 4.97 \\ -4.98 \end{pmatrix} \Delta\omega. \tag{91}$$

Following Eqs. (2) and (3), we express the eigenstates of this circuits $\Psi_n$ through the eigenstates of individual microresonators $\Psi_m^{(0)}$ as $\Psi_n = \sum_{m=1}^{5} C_{n,m} \Psi_m^{(0)}$ assuming $\delta_1^{(0)} = 0.1\Delta\omega$. The dependencies of $C_{n,m}$ for each of the states $\Psi_n$ on the microresonator number $m$, which characterize the distribution of magnitudes of $\Psi_n$ among microresonators, are shown on the plots of Fig. 10(c3). It is seen that the magnitudes of intermediate states $\Psi_n$ are small near waveguides 1 and 2 where their relative leakages are characterized by ratios $(C_{n,1}/C_{1,1})^2$ and $(C_{n,5}/C_{5,5})^2$, respectively. We find that the summarized leakages into each of the waveguides are equal to 1.01% only.

### 2. Optimization of modulation induced coupling

Here we consider the model designed in the previous Section and assume a relatively small coupling between microresonators $\delta_n^{(0)} = \delta_1^{(0)} = 0.1\Delta\omega$. As in the example of two coupled microresonators, the effect of the applied modulation is significantly reduced with decreasing of $\delta_1^{(0)}/\Delta\omega$. The modulation induced couplings between five states $\Psi_n$, $n = 1, 2, 3, 4, 5$, are determined by a linear system of equations, Eq. (24). Assuming that these couplings are equal, $\delta_n = \delta_1$, we have

$$\delta_1 = \sum_{m=1}^{5} C_{n,m} C_{n+1,m} \Delta\omega_{1,m}^{(0)}, \quad n = 1, 2, 3, 4. \tag{92}$$

The corresponding CIRT spectrum is shown in Fig. 10(e1) (identical to that in Fig. 8(c4)). Eq. (92) allows to express four of five frequency shifts $\Delta\omega_{1,m}^{(0)}$ through one of them, which in turn is determined by minimization of sum $\sum_{m=1}^{N} \left(\Delta\omega_{1,m}^{(0)}\right)^2$. The resulting dependence of optimum frequency shifts $\Delta\omega_{1,m}^{(0)}$ on the microresonator number $m$ is shown in Fig. 10(c4). It is seen that the required modulation amplitude is now an order of magnitude greater than that for a Kac photonic circuit shown in Fig. 10(a4). This is attributed to a relatively small coupling that was set here to suppress leakages.

In turn, Fig. 10(c5) shows the distribution of optimized frequency shifts $\Delta\omega_{1,m}^{(0)}$ for the case when the modulation induced interstate couplings realize a Butterworth filter with the CIRT spectrum shown in Fig. 10(e2) (identical to that shown in Fig. 8(a4)). It is seen that this variation is similar in behavior an amplitude as in Fig. 10(b4) of the previous case.

### E. Photonic circuits incorporating a tilted series of coupled SNAP microresonators

A photonic circuit, which is essentially similar to that considered in the previous Section VIIID, can be designed of coupled SNAP microresonators with tilted variation of radius and cutoff frequency $\omega^{(cut)}(z)$ as shown in Figs. 10(d1) and (d2). Here we design a SNAP circuit supporting $N = 5$ eigenstates $\Psi_n$ having eigenstates $\omega_n$ with accurately equal spacing and optimize the SDM required to approach the condition of CIRT.

### 1. Design of photonic circuit

The corresponding profile of $\omega^{(cut)}(z)$ is modelled as a sum of harmonic and linear functions. Varying the tilt, period, and amplitude of spatial oscillations of $\omega^{(cut)}(z)$, we can vary the values of eigenfrequency spacing and interstate coupling of this photonic circuit. While a spatially infinite periodically oscillating and tilted cutoff frequency variation $\omega^{(cut)}(z)$ has series of eigenfrequencies with exactly constant FSR, this is not correct for a finite structure of five coupled microresonators shown in Fig. 10(d1). To make the FSR of this structure constant with a good accuracy, we tune the position of the right-hand side cut edge of this structure and also adjust the left-hand side with the cutoff frequency perturbation (dashed blue curve in Fig. 10(d2)). As a result, the FSR of $N = 5$ eigenfrequencies of the designed photonic circuit is made equal to 10 GHz with a relative precision better than $8 \cdot 10^{-4}$. Five eigenstates $\Psi_n$ of this photonic circuit are shown in Fig. 10(d3). It is seen that in our design we set the interstate coupling relatively small. Similar to the photonic circuit considered in Section VIIID, this allows us to reduce leakages on the expense of the value of modulation induced coupling.

### *2. Optimization of modulation induced coupling*

As in previously in Section VIII, we consider the cases of equal modulation-induced couplings, $\delta_n = \delta_1$, and unequal coupling forming a Butterworth filter, whose CIRT spectra are shown in Figs. 10(e1) and (e2), respectively. Following the approach described above, we look for the optimized SDM in the form of linear combination of $\Psi_n \Psi_{n+1}$. The optimized spatial distributions of modulation for these cases are shown in Figs. 10(d4) and (d5). It is seen that amplitudes of these distributions are comparable with that for photonic circuit constructed of coupled microresonators with uniformly shifted eigenfrequencies (Figs. 10(c4) and (c5)) and is significantly greater than that for photonic circuits constructed of semi-parabolic SNAP microresonator (Fig. 10(b4) and (b5)) and Kac photonic circuit (Fig. 10(a4) and (a5)).

## IX. SUMMARY

In this work, we have developed a general theoretical framework describing the complete inelastic resonant transparency (CIRT) of time-modulated photonic circuits. By mapping the inelastic propagation of light in temporally modulated resonant systems to an equivalent elastic scattering problem in an extended configuration space, we have shown that the Mahaux–Weidenmüller formalism can be directly applied to describe light transmission through these circuits. This formulation establishes that, in the absence of losses, the S-matrix of a time-modulated resonant photonic circuit remains unitary, ensuring that light can experience frequency conversion with unity transmission amplitude—i.e., complete inelastic transparency.

The developed theory predicts that CIRT can be achieved by tuning only two experimentally accessible parameters: the modulation-induced interstate coupling and the input light frequency, while other circuit parameters may remain fixed. Using this principle, we have identified and optimized a broad class of resonant photonic circuits, which include coupled microrings, racetrack resonators, Kac-type chains, and SNAP bottle and semi-parabolic microresonators. In these circuits, light cascades coherently through $N$ equally spaced eigenstates and exits at a different frequency but nearly unchanged amplitude. The explicit CIRT conditions were derived for systems comprising two, three, and an arbitrary number of resonant states, and the corresponding transmission spectra were analyzed. For each configuration, we have evaluated the effects of losses and leakages, demonstrating that these effects scale linearly with the total loss and grow with the number of resonant states, thus defining the ultimate limits of achievable transparency.

We have established quantitative criteria for the optimization of the spatial distribution of modulation (SDM) that minimizes the modulation amplitude required to achieve a given coupling strength. The developed approach allows one to design modulation profiles achievable experimentally. For the considered SNAP bottle microresonator and coupled resonator circuits, the optimized SDM provides near-CIRT performance with modulation amplitudes several times smaller than those of unoptimized systems, confirming the robustness of the proposed method.

From a physical standpoint, the demonstrated possibility of complete inelastic transparency reveals a remarkable analogy between elastic and inelastic transport in resonant photonic systems. While complete elastic transparency corresponds to the perfect transmission of light without frequency change, the CIRT extends this concept to perfectly coherent frequency conversion within a microscopic structure. This opens the path to a broad class of miniature photonic devices incorporating frequency-shifting resonant circuits.

Future directions include the experimental realization of the presented designs in integrated photonic platforms such as lithium niobate, silicon nitride, and SNAP fiber devices. Incorporating strong electro-optic or acousto-optic modulation at microwave to gigahertz frequencies will allow observation of near-CIRT conditions with measurable power transfer between optical frequency channels. Further theoretical work can extend the present model to include higher-order modulation effects, nonlinear frequency mixing, and multimode or vector resonances. Another promising avenue is the use of dynamically tunable or multiple concurrent modulation frequencies to engineer complex coupling networks, enabling programmable and reconfigurable CIRT-based photonic processors.

Overall, the concept of complete inelastic transparency unifies and generalizes the phenomena of resonant tunneling, coherent frequency conversion, and transparent scattering. It provides a powerful foundation for designing compact, low loss, and tunable photonic systems where energy transfer between optical frequencies proceeds with the same perfection as elastic transmission in stationary resonant structures.

## DATA AVAILABILITY

The data that support the findings of this article are not publicly available. The data are available from the author upon reasonable request.

## ACKNOWLEDGMENTS

The author is grateful to A. A. Fotiadi and S. K. Turitsyn for useful discussions of these results. This research was supported by the Engineering and Physical Sciences Research Council (EPSRC) grants EP/W002868/1 and EP/X03772X/1 and by the Leverhulme Trust grant RPG-2022-014.

## APPENDIX A: CONDITION OF CIRT FOR PHOTONIC CIRCUIT WITH EQUAL COMPOUND EIGENFREQUENCIES AND EQUAL INTERSTATE COUPLING

Here we consider lossless photonic circuits setting $\gamma = 0$. From Eqs. (17) and (46), the condition of CIRT of reads:

$$\det(\Theta) - i\xi_{1,1}^2 M_{1,1}(\Theta) = 0. \tag{A.1}$$

Substitution of the expressions for $\det(\Theta)$ and $M_{1,1}(\Theta)$ from Eq. (50) into this equation yields:

$$\left[\left(\omega - \omega_1 + \tfrac{i}{2}\xi_{1,1}^2\right)\left(\omega - \omega_1 + \tfrac{i}{2}\xi_{N,2}^2\right) - \delta^2\right] D_{N-2} - \delta^2 \left(\omega - \omega_1 + \tfrac{i}{2}(\xi_{N,2}^2 - \xi_{1,1}^2)\right) D_{N-3} = 0. \tag{A.2}$$

Since $D_n$ are real, the ratio of the factors in front of $D_{N-2}$ and $D_{N-3}$ in Eq. (A.2) should be real as well. The latter condition is equivalent to

$$\left(\xi_{N,2}^2 - \xi_{1,1}^2\right)\left(\delta^2 - \frac{1}{4}\xi_{1,1}^2 \xi_{N,2}^2\right) = 0. \tag{A.3}$$

We find from Eq. (A.3) that there are only two possible conditions of CIRT which are given by the following equations:

$$\delta = \frac{1}{2}\xi_{1,1}\xi_{N,2}, \tag{A.4}$$

$$\xi_{N,2}^2 = \xi_{1,1}^2. \tag{A.5}$$

Eq. (A.5) is the condition of complete symmetry of the compound circuit when it can have $N$ resonances exhibiting CIRT. Under the condition of Eq. (A.4), the circuit is not symmetric though still has $N-1$ resonances exhibiting CIRT.

First, assuming the validity of Eq. (A.4) and using the expression for $D_n$ from Eq. (50) we rewrite Eq. (A.2) in the form:

$$\frac{1}{R}\left[\left(\frac{1+R}{1-R}\right)^N - 1\right] = 0, \quad R = \sqrt{1-\frac{4\delta^2}{\omega^2}}. \tag{A.6}$$

Solutions of this equation only exist if $R$ is imaginary, i.e., if $|\omega - \omega_1| < 2\delta$. Then this equation has exactly $N-1$ solutions

$$\omega_n^{(res)} = \omega_1 + 2\delta \cos\left(\frac{\pi n}{N}\right), \quad n = 1, 2, \ldots N-1. \tag{A.7}$$

Note that the values $n = 0$ and $n = N$ corresponding to $R = 0$ are excluded in this equation since the lefthand side of Eq. (A.6) does not tend to zero for $R \to 0$.

Next, under the validity of Eq. (A.5), we rewrite Eq. (A.2) in the form

$$\frac{1}{R}\left\{\left[\left(\frac{1+R}{1-R}\right)^{N+1} - 1\right] + \frac{\xi_{11}^4}{4\delta^2}\left(\frac{1+R}{1-R}\right)\left[\left(\frac{1+R}{1-R}\right)^{N-1} - 1\right]\right\} = 0, \quad R = \sqrt{1-\frac{4\delta^2}{\omega^2}}. \tag{A.8}$$

Unlike Eq. (A.6), this equation does not have analytical solutions. However, for relatively small coupling it has exactly $N$ solutions. Indeed, in the zero order over $\xi_{1,1}^4/\delta^2$ Eq. (A.8) becomes

$$\frac{1}{R}\left[\left(\frac{1+R}{1-R}\right)^{N+1} - 1\right] = 0. \tag{A.9}$$

$N$ solutions of this equation are

$$\omega_n^{(res)} = \omega_1 + 2\delta \cos\left(\frac{\pi n}{N+1}\right), \quad n = 1, 2, \ldots N. \tag{A.10}$$

Here, again, we have to exclude $n = 0$ and $n = N+1$. Therefore, for relatively small $\xi_{1,1}^4/\delta^2$ Eq. (A.8) will continue to have $N$ solutions corresponding to CIRT.

### APPENDIX B: THE EFFECT OF LOSSES AND LEAKAGES IN A PERFECTLY ALIGNED MODEL

In the presence of losses and leakages, Eq. (S1.1) is transformed into

$$\mathbf{S}(\omega) = \mathbf{I}_{2N} - i\mathbf{\Theta}(\omega), \quad \mathbf{\Theta}(\omega) = \xi^{\dagger}\left[\Delta\mathbf{\Omega}(\omega) + \frac{i}{2}\left(\mathbf{\Gamma} + \xi\xi^{\dagger}\right)\right]^{-1}\xi,$$

$$\Delta\mathbf{\Omega} = \begin{Bmatrix} \omega - \omega_1 & \delta & 0 & \cdots & 0 & 0 \\ \delta & \omega - \omega_1 & \delta & \cdots & 0 & 0 \\ 0 & \delta & \omega - \omega_1 & \cdots & 0 & 0 \\ & & \cdots & & \cdots & \\ 0 & 0 & 0 & \cdots & \omega - \omega_1 & \delta \\ 0 & 0 & 0 & \cdots & \delta & \omega - \omega_1 \end{Bmatrix}, \quad \mathbf{\Gamma} = \begin{Bmatrix} \gamma_1 & 0 & \cdots & 0 & 0 \\ 0 & \gamma_2 & \cdots & 0 & 0 \\ 0 & 0 & \cdots & 0 & 0 \\ \cdots & \cdots & & \cdots & \cdots \\ 0 & 0 & \cdots & \gamma_{n-1} & 0 \\ 0 & 0 & \cdots & 0 & \gamma_N \end{Bmatrix},$$

$$\xi = \begin{Bmatrix} \xi_{1,1} & 0 & 0 & \cdots & 0 & \xi_{1,2} & 0 & 0 & \cdots & 0 \\ 0 & \xi_{2,1} & 0 & \cdots & 0 & 0 & \xi_{2,2} & 0 & \cdots & 0 \\ 0 & 0 & \xi_{3,1} & \cdots & 0 & 0 & 0 & 0 & \cdots & 0 \\ & & & & \cdots & & & & & \\ 0 & 0 & 0 & \cdots & \xi_{N,1} & 0 & 0 & 0 & \cdots & \xi_{N,2} \end{Bmatrix}. \tag{B.1}$$

Here we determine the perturbation of the CIRT due to the relatively small material losses $\gamma_n$, $n = 1,2,\ldots,N$, and leakages $\xi_{n,1}$, $n = 2, 3, \ldots, N$, and $\xi_{n,2}$, $n = 1, 2, \ldots, N-1$. Generally, in the first order in losses and second order in leakages, and for comparable couplings to waveguides, $\xi_{1,1} \sim \xi_{N,2}$, we have at the frequencies $\omega_m^{(res)}$ defined by Eqs. (S1.3) and (S1.4):

$$\left|S_{1,2N}(\omega_m^{(res)})\right|^2 \cong 1 - \frac{1}{\xi_{1,1}^2}\left(\sum_{n=1}^{N} F_{m,n}^{(1,\gamma)}\gamma_n + \sum_{n=2}^{N} F_{m,n}^{(1,\xi_1)}\xi_{n,1}^2 + \sum_{n=1}^{N-1} F_{m,n}^{(1,\xi_2)}\xi_{n,2}^2\right) \\ + \frac{1}{\xi_{N,2}^2}\left(\sum_{n=1}^{N} F_{m,n}^{(2,\gamma)}\gamma_n + \sum_{n=2}^{N} F_{m,n}^{(2,\xi_1)}\xi_{n,1}^2 + \sum_{n=1}^{N-1} F_{m,n}^{(2,\xi_2)}\xi_{n,2}^2\right) \tag{B.2}$$

where $F_{m,n}^{(k,\gamma,\xi_j)} \geq 0$ are constants independent of the system's parameters. As examples, here we present the actual form of this expression for $N = 2, 3, 4, 5$ and 6.

For $N = 2$ there exist a single resonance frequency $\omega_1^{(res)} = \omega_1$ (see Eqs. (S1.3) and (S1.4))

$$\left|S_{1,4}(\omega_1)\right|^2 \cong 1 - \left[\frac{1}{\xi_{1,1}^2}\left(\gamma_1 + \xi_{1,2}^2\right) + \frac{1}{\xi_{2,2}^2}\left(\gamma_2 + \xi_{2,1}^2\right)\right] \tag{B.3}$$

For $N = 3$ and resonant frequencies defined by Eqs. (S1.3) and (S1.6) $\omega_{1,2}^{(res)} = \omega_1 \pm \delta$, Eq. (B.2) becomes:

$$\left|S_{1,6}(\omega_1 \pm \delta)\right|^2 \cong 1 - \left\{\frac{1}{\xi_{1,1}^2}\left[\left(\gamma_1 + \xi_{1,2}^2\right) + \left(\gamma_2 + \xi_{2,1}^2 + \xi_{2,2}^2\right)\right] + \frac{1}{\xi_{3,2}^2}\left[\left(\gamma_2 + \xi_{2,1}^2 + \xi_{2,2}^2\right) + \left(\gamma_3 + \xi_{3,1}^2\right)\right]\right\}. \tag{B.4}$$

For $N = 4$, the resonant frequencies defined by Eq. (S1.3) and (S1.7) are $\omega_1^{(res)} = \omega_1$ and $\omega_{2,3}^{(res)} = \omega_1 \pm \sqrt{2}\delta$. Then we have

$$|S_{1,8}(\omega_1)|^2 \cong 1 - \left\{ \frac{1}{\xi_{1,1}^2}\left[(\gamma_1 + \xi_{1,2}^2) + (\gamma_3 + \xi_{3,1}^2 + \xi_{3,2}^2)\right] + \frac{1}{\xi_{4,2}^2}\left[(\gamma_2 + \xi_{2,1}^2 + \xi_{2,2}^2) + (\gamma_4 + \xi_{4,1}^2)\right] \right\}, \tag{B.5}$$

$$|S_{1,8}(\omega_1 \pm \sqrt{2}\delta)|^2 \cong 1 - \left\{ \frac{1}{\xi_{1,1}^2}\left[(\gamma_1 + \xi_{1,2}^2) + 2(\gamma_2 + \xi_{2,1}^2 + \xi_{2,2}^2) + (\gamma_3 + \xi_{3,1}^2 + \xi_{3,2}^2)\right] \right.$$
$$\left. + \frac{1}{\xi_{4,2}^2}\left[(\gamma_2 + \xi_{2,1}^2 + \xi_{2,2}^2) + 2(\gamma_3 + \xi_{3,1}^2 + \xi_{3,2}^2) + (\gamma_4 + \xi_{4,1}^2)\right] \right\}. \tag{B.6}$$

For $N = 5$, the resonant frequencies defined by Eq. (S1.3) and (S1.8) are $\omega_{1,2,3,4}^{(res)} = \omega_1 \pm \frac{1}{2}(1 \pm \sqrt{5})\delta$. Then we have

$$\left|S_{1,10}\left(\omega_1 \pm \frac{1}{2}(1-\sqrt{5})\delta\right)\right|^2 \cong 1 -$$
$$\frac{1}{\xi_{1,1}^2}\left\{\left[(\gamma_1 + \xi_{1,2}^2) + \frac{1}{2}(3-\sqrt{5})(\gamma_2 + \xi_{2,1}^2 + \xi_{2,2}^2) + \frac{1}{2}(3-\sqrt{5})(\gamma_3 + \xi_{3,1}^2 + \xi_{3,2}^2) + (\gamma_4 + \xi_{4,1}^2 + \xi_{4,2}^2)\right]\right.$$
$$\left. + \frac{1}{\xi_{5,2}^2}\left[(\gamma_2 + \xi_{2,1}^2 + \xi_{2,2}^2) + \frac{1}{2}(3-\sqrt{5})(\gamma_3 + \xi_{3,1}^2 + \xi_{3,2}^2) + \frac{1}{2}(3-\sqrt{5})(\gamma_4 + \xi_{4,1}^2 + \xi_{4,2}^2) + (\gamma_5 + \xi_{5,1}^2)\right]\right\} \tag{B.7}$$

$$\left|S_{1,10}\left(\omega_1 \pm \frac{1}{2}(1+\sqrt{5})\delta\right)\right|^2 \cong 1 -$$
$$\left\{\frac{1}{\xi_{1,1}^2}\left[(\gamma_1 + \xi_{1,2}^2) + \frac{1}{2}(3+\sqrt{5})(\gamma_2 + \xi_{2,1}^2 + \xi_{2,2}^2) + \frac{1}{2}(3+\sqrt{5})(\gamma_3 + \xi_{3,1}^2 + \xi_{3,2}^2) + (\gamma_4 + \xi_{4,1}^2 + \xi_{4,2}^2)\right]\right.$$
$$\left. + \frac{1}{\xi_{5,2}^2}\left[(\gamma_2 + \xi_{2,1}^2 + \xi_{2,2}^2) + \frac{1}{2}(3+\sqrt{5})(\gamma_3 + \xi_{3,1}^2 + \xi_{3,2}^2) + \frac{1}{2}(3+\sqrt{5})(\gamma_4 + \xi_{4,1}^2 + \xi_{4,2}^2) + (\gamma_5 + \xi_{5,1}^2)\right]\right\} \tag{B.8}$$

For $N = 6$, the resonant frequencies defined by Eq. (S1.3) and (S1.9) are $\omega_1^{(res)} = \omega_1$, $\omega_{2,3}^{(res)} = \omega_1 \pm \delta$, and $\omega_{4,5}^{(res)} = \omega_1 \pm \sqrt{3}\delta$. Then we have

$$|S_{1,12}(\omega_1)|^2 \cong 1 - \left\{ \frac{1}{\xi_{1,1}^2}\left[(\gamma_1 + \xi_{1,2}^2) + (\gamma_3 + \xi_{3,1}^2 + \xi_{3,2}^2) + (\gamma_5 + \xi_{5,1}^2 + \xi_{5,2}^2)\right] \right.$$
$$\left. + \frac{1}{\xi_{6,2}^2}\left[(\gamma_2 + \xi_{2,1}^2 + \xi_{2,2}^2) + (\gamma_4 + \xi_{4,1}^2 + \xi_{4,2}^2) + (\gamma_6 + \xi_{6,1}^2)\right] \right\} \tag{B.9}$$

$$|S_{1,12}(\omega_1 \pm \delta)|^2 \cong 1 - \left\{ \frac{1}{\xi_{1,1}^2}\left[(\gamma_1 + \xi_{1,2}^2) + (\gamma_2 + \xi_{2,1}^2 + \xi_{2,2}^2) + (\gamma_4 + \xi_{4,1}^2 + \xi_{4,2}^2) + (\gamma_5 + \xi_{5,1}^2 + \xi_{5,2}^2)\right] \right.$$
$$\left. + \frac{1}{\xi_{6,2}^2}\left[(\gamma_2 + \xi_{2,1}^2 + \xi_{2,2}^2) + (\gamma_3 + \xi_{3,1}^2 + \xi_{3,2}^2) + (\gamma_5 + \xi_{5,1}^2 + \xi_{5,2}^2) + (\gamma_6 + \xi_{6,1}^2)\right] \right\}$$
$$\tag{B.10}$$

$$\left|S_{1,12}(\omega_1 \pm \sqrt{3}\delta)\right|^2 \cong 1 -$$

$$\left\{ \frac{1}{\xi_{1,1}^2}\left[\left(\gamma_1 + \xi_{1,2}^2\right) + 3\left(\gamma_2 + \xi_{2,1}^2 + \xi_{2,2}^2\right) + 4\left(\gamma_3 + \xi_{3,1}^2 + \xi_{3,2}^2\right) + 3\left(\gamma_4 + \xi_{4,1}^2 + \xi_{4,2}^2\right) + \left(\gamma_5 + \xi_{5,1}^2 + \xi_{5,2}^2\right)\right] \right. \quad \text{(B.11)}$$

$$\left. + \frac{1}{\xi_{6,2}^2}\left[\left(\gamma_2 + \xi_{2,1}^2 + \xi_{2,2}^2\right) + 3\left(\gamma_3 + \xi_{3,1}^2 + \xi_{3,2}^2\right) + 4\left(\gamma_4 + \xi_{4,1}^2 + \xi_{4,2}^2\right) + 3\left(\gamma_5 + \xi_{5,1}^2 + \xi_{5,2}^2\right) + \left(\gamma_6 + \xi_{6,1}^2\right)\right] \right\}$$

It is seen from the comparison of these expressions that the contribution of individual leakages may grow with the number of eigenstates participating in the frequency conversion process.

## APPENDIX C: A RESONANT CIRCUIT WITH THREE EIGENSTATES

Here we analyze the inelastic transparency of a photonic circuit having $N = 3$ resonant eigenstates. The S-matrix of this circuit is described by the MW equation:

$$\mathbf{S}(\omega) = \mathbf{I}_6 - i\boldsymbol{\Theta}(\omega), \quad \boldsymbol{\Theta}(\omega) = \boldsymbol{\xi}^\dagger \left[\Omega(\omega)\mathbf{I}_3 - \boldsymbol{\Omega} + \frac{i}{2}\left(\boldsymbol{\Gamma} + \boldsymbol{\xi}\boldsymbol{\xi}^\dagger\right)\right]^{-1} \boldsymbol{\xi},$$

$$\boldsymbol{\Omega} = \begin{Bmatrix} \Omega_1 & \delta_1 & 0 \\ \delta_1 & \Omega_2 & \delta_2 \\ 0 & \delta_2 & \Omega_3 \end{Bmatrix} \quad \boldsymbol{\Gamma} = \begin{Bmatrix} \gamma_1 & 0 & 0 \\ 0 & \gamma_2 & 0 \\ 0 & 0 & \gamma_3 \end{Bmatrix},$$

(C.1)

$$\Omega(\omega) = \omega + \left(k - \tfrac{1}{2}\right)\omega_p, \quad \Omega_n = \omega_n + \left(k - n + \tfrac{1}{2}\right)\omega_p,$$

$$\boldsymbol{\xi} = \begin{Bmatrix} \xi_{1,1} & 0 & 0 & \xi_{1,2} & 0 & 0 \\ & \xi_{2,1} & 0 & 0 & \xi_{2,2} & 0 \\ 0 & 0 & \xi_{3,1} & 0 & 0 & \xi_{3,2} \end{Bmatrix}.$$

First, we assume that the leakages and losses are absent, $\xi_{2,1} = \xi_{3,1} = \xi_{1,2} = \xi_{2,2} = 0$ and $\gamma_1 = \gamma_2 = \gamma_3 = 0$. Then, the matrix to be inverted in Eq. (C.1) becomes

$$\Omega(\omega)\mathbf{I}_3 - \boldsymbol{\Omega} + \frac{i}{2}\left(\boldsymbol{\Gamma} + \boldsymbol{\xi}\boldsymbol{\xi}^\dagger\right) = \begin{Bmatrix} \omega - \omega_1 - \tfrac{i}{2}\xi_{11}^2 & \delta_1 & 0 \\ \delta_1 & \omega - \omega_2 + \omega_p & \delta_2 \\ 0 & \delta_2 & \omega - \omega_3 + 2\omega_p - \tfrac{i}{2}\xi_{32}^2 \end{Bmatrix}. \quad \text{(C.2)}$$

In our consideration we assume that the interstate couplings $\delta_1$ and $\delta_2$ are varied by the modulation power proportionally so that

$$\delta_2 = \alpha\delta_1, \quad \text{(C.3)}$$

where $\delta_1$ can be varied and $\alpha$ is a constant. We are looking for the conditions when the CIRT can be achieved by variation of the light frequency $\omega$ and modulation power (proportional to $\delta_1^2$) only assuming that other parameters of the system are given.

In the absence of losses, we have $S_{0,1} = S_{0,2} = S_{0,3} = S_{0,4} = 0$ so that $|S_{0,0}|^2 + |S_{0,5}|^2 = 1$. Therefore, the condition of CIRT $|S_{0,5}|^2 = 1$ is equivalent to $S_{0,0} = 0$ (see Eq. (17) of the main text). The actual complex-valued expression for $S_{0,0}$ includes the numerator and denominator. The condition that the numerator of this expression is equal to zero is reduced to two equations for the real system's parameters:

$$4\delta_1^2(\omega - \omega_3 + 2\omega_p) + 4(\omega - \omega_1)(\omega - \omega_2 + \omega_p)^2 - (\omega - \omega_2 + \omega_p)\xi_{1,1}^2\xi_{3,2}^2 \\ - 4(\omega - \omega_1)(\omega - \omega_2 + \omega_p)(\omega - \omega_3 + 2\omega_p) = 0 \tag{C.4}$$

$$\delta_1^2(\xi_{3,2}^2 - \alpha^2\xi_{1,1}^2) + (\omega - \omega_2 + \omega_p)\left[(\omega - \omega_3 + 2\omega_p)\xi_{1,1}^2 - (\omega - \omega_1)\xi_{3,2}^2\right] = 0 \tag{C.5}$$

Substitution of $\delta_1^2$ from Eq. (C.5) into Eq. (C.4) yields the equation for $\omega$:

$$(\omega - \omega_2 + \omega_p)\Big\{4(\xi_{1,1}^2 - \alpha^2\xi_{3,2}^2)\omega^2 - 8\left[(\omega_3 - 2\omega_p)\xi_{1,1}^2 - \alpha^2\omega_1\xi_{3,2}^2\right]\omega \\ + 4(\omega_3 - 2\omega_p)^2\xi_{1,1}^2 - 4\alpha^2(\omega - \omega_1)^2\xi_{3,2}^2 + \xi_{1,1}^2\xi_{3,2}^4 - \alpha^2\xi_{1,1}^4\xi_{3,2}^2\Big\} = 0 \tag{C.6}$$

This equation has three solutions. We exclude the simplest one $\omega^{(2)} = \omega_2 - \omega_p$ since this condition leads to zeroing of the denominator of the expression for $S_{0,0}$ and non-zero total value of $S_{0,0}$. The roots $\omega^{(1,3)}$ of the remaining quadratic equation for $\omega$, determine two the resonant frequencies, $\omega_{1,2}^{(res)}$, and corresponding coupling parameter $\delta_{1,2}^{(res)}$ presented by Eqs. (38) and (39) of the main text:

$$\omega^{(1,3)} = \frac{(\omega_3 - 2\omega_p)\xi_{1,1}^2 - \alpha^2\omega_1\xi_{3,2}^2 \pm \frac{1}{2}\xi_{1,1}\xi_{3,2}D^{1/2}}{\xi_{1,1}^2 - \alpha^2\xi_{3,2}^2}, \tag{C.7}$$

$$D = 4\alpha^2(\omega_3 - \omega_1 - 2\omega_p)^2 + (\xi_{1,1}^2 - \alpha^2\xi_{3,2}^2)(\alpha^2\xi_{1,1}^2 - \xi_{3,2}^2),$$

which determine the corresponding coupling parameters $\delta_1^{(2,3)}$ as

$$\left(\delta_1^{(1,3)}\right)^2 = \frac{(\omega^{(1,3)} - \omega_2 + \omega_p)\left[(\omega^{(1,3)} - \omega_1)\xi_{3,2}^2 - (\omega^{(1,3)} - \omega_3 + 2\omega_p)\xi_{1,1}^2\right]}{\xi_{3,2}^2 - \alpha^2\xi_{1,1}^2}. \tag{C.8}$$

Under the conditions that $D \geq 0$ in Eq. (C.7) and that the right-hand side of Eq. (C.8) is positive, these two equations determine two values of parametric modulation power when, in the absence of material losses and leakages, the magnitude of inelastic transmission amplitude can achieve unity.

In the case of the equal couplings between eigenstates, $\alpha = 1$, and their perfectly aligned eigenvalues, $\omega_1 = \omega_2 - \omega_p = \omega_3 - 2\omega_p$, we return back to the case considered in Appendix A since then we find from Eqs. (C.7) and (C.8)

$$\omega^{(1,3)} = \omega_1 \pm \tfrac{1}{2}\xi_{1,1}\xi_{3,2}$$
$$\left(\delta_1^{(1,3)}\right)^2 = \frac{\xi_{1,1}^2\xi_{3,2}^2}{4} \tag{C.9}$$

## APPENDIX D: CALCULATION OF THE BUTTERWORTH FILTER PARAMETERS

To calculate the Butterworth filter parameters, we assume that all the eigenfrequencies of uncouple compound states are equal, $\Omega_n = \omega_n + \left(k - n + \frac{1}{2}\right)\omega_p = \Omega_1 = \omega_1 + \left(k - \frac{1}{2}\right)\omega_p$. Then, the expression for the inelastic transmission power $|S_{1,2N}|^2$ of a Butterworth filter in Eq. (46) can be written down in the form

$$|S_{1,2N}|^2 = \frac{\xi_{N,2}^{4N}}{\xi_{N,2}^{4N} + \rho_N \left[2(\Omega - \Omega_1)\right]^{2N}} \tag{D.1}$$

identical to Eq. (49). To find the parameters of photonic circuit having this spectrum, we successively expanded the expression for the transmission power $|S_{1,8}|^2$ found from Eq. (1) into series of $\Omega - \Omega_1$ using Mathcad. Here we illustrate our approach for the case $N = 4$ when, as follows from Eq. (D.1), this expansion should have the form

$$|S_{1,8}|^2 = 1 - \frac{\rho_4}{\xi_{4,2}^{16}} 2^8 (\Omega - \Omega_1)^8 + O\left((\Omega - \Omega_1)^{16}\right) \tag{D.2}$$

For $N = 4$, the condition of CIRT given by Eq. (47) becomes

$$\xi_{1,1} = \frac{2\delta_1 \delta_3}{\delta_2 \xi_{4,2}}. \tag{D.3}$$

Substituting this expression for $\xi_{1,1}$ into the expression for $|S_{1,8}|^2$ and expanding it into series over $\Omega - \Omega_1$ up to the second order, we have

$$|S_{1,8}|^2 = 1 - \frac{\left(\delta_1^2 \delta_2^2 \xi_{4,2}^4 - 4\delta_1^2 \delta_2^2 \delta_3^2 - 4\delta_1^2 \delta_3^4 + \delta_2^4 \xi_{4,2}^4\right)^2}{16\delta_1^4 \delta_2^4 \delta_3^4 \xi_{4,2}^4}(\Omega - \Omega_1)^2 + O\left((\Omega - \Omega_1)^4\right) \tag{D.4}$$

Zeroing the term $\sim (\Omega - \Omega_1)^2$ in this expression yields:

$$\delta_1 = \delta_2^2 \xi_{4,2}^2 \left(4\delta_2^2 \delta_3^2 + 4\delta_3^4 - \delta_2^2 \xi_{4,2}^4\right)^{-1/2}. \tag{D.5}$$

After substituting the expressions for $\xi_{1,1}$ and $\delta_1$ from Eqs. (D.3) and (D.5) into the expression for $|S_{1,2N}|^2$ we find

$$|S_{1,2N}|^2 = 1 - \frac{\left(2\delta_2^4 + 4\delta_2^2 \delta_3^2 - \delta_2^4 \xi_{4,2}^4 + 2\delta_3^4\right)^2}{\delta_2^8 \xi_{4,2}^8}(\Omega - \Omega_1)^4 + O\left((\Omega - \Omega_1)^6\right). \tag{D.6}$$

Zeroing the term $\sim (\Omega - \Omega_1)^4$ in this expression yields:

$$\xi_{4,2} = 2^{1/4} \left(\frac{\delta_2^2 + \delta_3^2}{\delta_2}\right)^{1/2}. \tag{D.7}$$

After substituting the expressions for $\xi_{4,2}$, $\xi_{1,1}$ and $\delta_1$ from Eqs. (D.3), (D.5) and (D.7) into the

-50-

expression for $|S_{1,2N}|^2$ we find

$$|S_{1,2N}|^2 = 1 - \frac{\left(\delta_2^4 + 2\delta_2^2\delta_3^2 - \delta_3^4\right)^2}{8\delta_3^4\delta_2^6\left(\delta_2^2 + \delta_3^2\right)^2}(\Omega - \Omega_1)^6 + O\left((\Omega - \Omega_1)^8\right). \tag{D.8}$$

Zeroing the term $\sim (\Omega - \Omega_1)^6$ in this expression yields:

$$\delta_2 = \left(2^{1/2} - 1\right)^{1/2} \delta_3. \tag{D.9}$$

After substituting the expressions for $\delta_2$, $\xi_{4,2}$, $\xi_{1,1}$ and $\delta_1$ from Eqs. (D.3), (D.5), (D.7), and (D.9) into the expression for $|S_{1,2N}|^2$ we find

$$|S_{1,2N}|^2 = 1 - \frac{1}{4\delta_3^8}(\Omega - \Omega_1)^8 + O\left((\Omega - \Omega_1)^{16}\right). \tag{D.10}$$

Finally, combining Eqs. (D.3), (D.5), (D.7), and (D.9) we find the expressions for the photonic circuit parameters at $N = 4$ summarized in Table 1.

## APPENDIX E: PHOTONIC CIRCUITS COMPOSED OF COMPOUND STATES WITH EQUAL EIGENFREQUENCIES

For the case of compound states with equal eigenfrequencies, $\Omega_n = \omega_n + \left(k - n + \frac{1}{2}\right)\omega_p = \Omega_1 = \omega_1 + \left(k - \frac{1}{2}\right)\omega_p$, and the same input frequency, the determinant in Eq. (46) is

$$\det\begin{pmatrix} \frac{i}{2}\xi_{1,1} & \delta_1 & \cdots & 0 & 0 \\ \delta_1 & 0 & \cdots & 0 & 0 \\ \cdots & \cdots & \cdots & \cdots & \cdots \\ 0 & 0 & \cdots & 0 & \delta_{N-1} \\ 0 & 0 & \cdots & \delta_{N-1} & \frac{i}{2}\xi_{2,N} \end{pmatrix} = \begin{cases} \frac{i}{2}\left(\xi_{1,1}^2\delta_2^2\delta_4^2\ldots\delta_{N-1}^2 + \xi_{N,2}^2\delta_1^2\delta_3^2\ldots\delta_N^2\right) & \text{odd } N \\ \frac{1}{4}\xi_{1,1}^2\xi_{N,2}^2\delta_2^2\delta_4^2\ldots\delta_{N-2}^2 + \delta_1^2\delta_3^2\ldots\delta_{N-1}^2 & \text{even } N \end{cases} \tag{E.1}$$

Eqs. (47) and (48) immediately follow from Eq. (46) for $S_{1,2N}(\omega)$ and (E.1).

## APPENDIX F: ADJUSTMENT OF EIGENFREQUENCIES BY PERTURBATION THEORY

Preliminary designed photonic circuit with known eigenstates $\Psi_m^{(0)}$ and eigenfrequencies $\omega_m^{(0)}$ can be adjusted by perturbation theory. As an example, for a SNAP photonic circuit, the perturbation of refractive index $\Delta n(z)$ is related to the perturbation of the cutoff frequency $\Delta\omega(z)$ by the rescaling relation $\Delta n(z) = n_0\Delta\omega(z)/\omega_0$. The requested shifts $\Delta\omega_m$ of $\omega_m^{(0)}$ are determined through $\Delta\omega(z)$ in the first order of the perturbation theory as

$$\Delta\omega_m = \left\langle \Psi_m^{(0)} \middle| \Delta\omega \middle| \Psi_m^{(0)} \right\rangle \quad (F.1)$$

Assuming that photonic circuit consists of $N$ eigenstates, we can present $\Delta\omega(z)$ as a superposition of $N$ known functions $\Delta\omega_k^{(0)}(z)$, $k = 1,2,\ldots,N$:

$$\Delta\omega(z) = \sum_{k=1}^{N} a_k \Delta\omega_k^{(0)}(z) \quad (F.2)$$

with unknown coefficients $a_k$. Substitution of Eq. (F.2) into Eq. (F.1) yields:

$$\Delta\omega_m = \sum_{k=1}^{N} F_{m,k} a_k, \quad F_{m,k} = \left\langle \Psi_m^{(0)} \middle| \Delta\omega_k^{(0)} \middle| \Psi_m^{(0)} \right\rangle. \quad (F.3)$$

From Eq. (F.3) we find coefficients $a_k$ as

$$a_k = \sum_{m=1}^{N} H_{k,m} \Delta\omega_m, \quad (F.4)$$

where matrix $H$ is the inverse of $F$, $H = F^{-1}$.

In Section VIIIC, we used this approach to design a SNAP photonic circuit comprising a semi-parabolic microresonator. In our preliminary design, the FSR of four eigenstates of this circuit was not accurately constant. An additional problem with our design was that two of seven eigenstates of the circuit had close eigenfrequencies so that the conventional first order perturbation theory fails. However, since a major part of all eigenstates were localized in the semi-parabolic microresonator, we applied the theory described above ignoring the rest of the circuit. We presented Eq. (80) for a semi-parabolic microresonator as

$$\frac{\chi^2}{2} \frac{\partial^2 \Phi_{lpq}}{\partial z^2} + \left(\omega - \alpha z^2 \theta(z) + \Delta\omega(z)\theta(z)\right) \Phi_{lpq} = 0, \quad \chi = \frac{c}{n_0 \sqrt{\omega}}, \quad (F.5)$$

and rewrite this equation in the dimensionless form:

$$\frac{\partial^2 \Psi}{\partial x^2} + \left(\nu - x^2 \theta(x) + \Delta\nu(x)\theta(x)\right) \Psi = 0,$$

$$x = \frac{(2\alpha)^{1/4}}{\chi^{1/2}}, \quad \nu = \frac{2\omega}{\chi(2\alpha)^{1/2}}, \quad \Delta\nu(x) = \frac{2\omega}{\chi(2\alpha)^{1/2}} \Delta\omega(z), \quad (F.6)$$

Here $\theta(x)$ is the Heaviside function and $\Delta\nu(x)$ is the rescaled adjustment function $\Delta\omega(z)$ to be found. At $\Delta\nu(x) = 0$, the normalized localized solutions of Eq. (F.6) are expressed through the Hermite polynomials $H_m(x)$:

$$\Psi_m^{(0)}(x) = \frac{\pi^{-1/4}}{\sqrt{2^m m!}} \exp\left(-\frac{x^2}{2}\right) H_m(x) \quad (F.7)$$

In Section VIIIC, we adjust our preliminary design of SNAP photonic circuit as described above by presenting $\Delta\nu(x)$ in the form

$$\Delta \nu(x) = \sum_{k=1}^{N} a_k \cos(kx) \tag{F.8}$$

Then

$$\left\langle \Psi_m^{(0)} \left| \cos(kx) \right| \Psi_m^{(0)} \right\rangle = \exp\left(-\frac{k^2}{4}\right) L_m\left(\frac{k^2}{2}\right), \tag{F.9}$$

where $L_m(y)$ is the Laguerre polynomial.

### APPENDIX G: ADJUSTING THE DEVIATIONS FROM THE PERFECTLY ALIGNED MODEL FOR $N = 2$ AND 3.

The conditions of CIRT introduced in Appendix A may be challenging to achieve experimentally. However, violation of these conditions caused by deviations of the resonance eigenfrequencies and coupling parameters from those defined in Appendix A can be compensated by adjusting the values of resonance frequency $\omega_n^{(res)}$ and modulation power. Here we present the values of the adjusted resonance frequency and modulation-induced coupling found by the perturbation theory for $N = 2$ and 3.

For the photonic circuit possessing $N = 2$ eigenstates, we introduce the perturbed eigenfrequencies $\omega_1 + \Delta\omega_1$ and $\omega_1 + \omega_p + \Delta\omega_1$. We assume $|\Delta\omega_1|, |\Delta\omega_2| \ll \xi_{1,1}^2, \xi_{2,2}^2$, so that the separation between these eigenfrequencies deviates from $\omega_p$ by a relatively small value. Then we find that in the second order in $|\Delta\omega_1|$, $|\Delta\omega_2|$, $|\Delta\omega|$, where $\Delta\omega = \omega - \omega_1^{(res)}$ and $\omega_1^{(res)} = \omega_1$, the reduction of the inelastic resonant amplitude is determined by the equation:

$$\left| S_{1,4}(\omega_1^{(res)} + \Delta\omega) \right|^2 \cong 1 - \left( \frac{1}{\xi_{1,1}^2}(\Delta\omega - \Delta\omega_1) + \frac{1}{\xi_{2,2}^2}(\Delta\omega - \Delta\omega_2) \right)^2 \tag{G.1}$$

From Eq. (G.1), we find the new resonant frequency corresponding to the CIRT:

$$\Delta\omega_1^{(res)} = \frac{\Delta\omega_1 \xi_{22}^2 - \Delta\omega_2 \xi_{11}^2}{\xi_{22}^2 - \xi_{22}^2} \tag{G.2}$$

For the photonic circuit with $N = 3$ eigenstates, we introduce the perturbed eigenfrequencies $\omega_1 + \Delta\omega_1$, $\omega_1 + \omega_p + \Delta\omega_1$, and $\omega_1 + 2\omega_p + \Delta\omega_1$ and, perturbed couplings $\delta_1 + \Delta\delta_1$ and $\delta_1 + \Delta\delta_2$. To apply the perturbation theory, we assume that $|\Delta\omega_1|, |\Delta\omega_2|, |\Delta\omega_3|, |\Delta\delta_1|, |\Delta\delta_2| \ll \xi_{1,1}^2, \xi_{3,2}^2$. Then in a small neighborhood $\Delta\omega = \omega - \omega_1^{(res)}$ of the original resonance frequency $\omega_1^{(res)} = \omega_1 + \delta_1$ and neighborhood $\Delta\delta_0 = \delta - \delta_1$ of the original coupling $\delta_1$ (see Eqs. (S1.3) and (S1.6)) we find

$$\begin{aligned}
&\left| S_{1,4}(\omega_1^{(res)} + \Delta\omega) \right|^2 \cong 1 - \Xi^{-1}\left[ (2\Xi\Delta\omega - \sigma_1)^2 + \sigma_2^2 \right] \\
&\Xi = 2\left( \xi_{1,1}^4 - \xi_{1,1}^2 \xi_{3,2}^2 + \xi_{1,1}^4 \right) \\
&\sigma_1 = (4\Delta\delta_0 + 4\Delta\delta_2 - 2\Delta\omega_2 - 2\Delta\omega_3)\xi_{1,1}^4 + (2\Delta\omega_1 + 2\Delta\omega_3)\xi_{1,1}^2 \xi_{3,2}^2 + (4\Delta\delta_0 + 4\Delta\delta_1 - 2\Delta\omega_1 - 2\Delta\omega_2)\xi_{3,2}^4 \\
&\sigma_2 = 2\xi_{11}\xi_{32}\left[ (2\Delta\delta_0 + 2\Delta\delta_1 - \Delta\omega_2 + \Delta\omega_3)\xi_{1,1}^2 - (2\Delta\delta_0 + 2\Delta\delta_1 2 + \Delta\omega_1 - \Delta\omega_2)\xi_{3,2}^2 \right]
\end{aligned} \tag{G.3}$$

We determine the perturbed resonant frequency $\omega_1^{(res)} + \Delta\omega_1^{(res)}$ and perturbed coupling $\delta_1 + \Delta\delta_0$ by zeroing the deviation of $|S_{1,4}|^2$ from unity in Eq. (G.3), i.e., by setting

$$\begin{aligned} 2\Xi\Delta\omega &= \sigma_1 \\ \sigma_2 &= 0 \end{aligned} \quad (G.4)$$

From these equations, we find:

$$\begin{aligned} \Delta\omega_1^{(res)} &= \frac{(\Delta\delta_2 - \Delta\delta_1 - \Delta\omega_3)\xi_{1,1}^2 + (\Delta\delta_2 - \Delta\delta_1 + \Delta\omega_1)\xi_{1,1}^2}{\xi_{3,2}^2 - \xi_{1,1}^2} \\ \Delta\delta_0 &= \frac{(\Delta\omega_2 - \Delta\omega_3 - 2\Delta\delta_1)\xi_{1,1}^2 + (\Delta\omega_1 - \Delta\omega_2 + 2\Delta\delta_2)\xi_{1,1}^2}{\xi_{3,2}^2 - \xi_{1,1}^2} \end{aligned} \quad (G.5)$$